# La secrète harmonie du désordre aléatoire dévoilée dans la feuille de papier

Essai sur une forme d'espace moteur en milieu stochastique

Jacques Silvy

**Table**

# Préface

*« Sans la croyance qu'il est possible de saisir la réalité avec nos constructions théoriques, sans la croyance en l'harmonie interne de notre monde, il ne pourrait y avoir de science ».*
*Albert Einstein et Léopold Infeld - L'évolution des idées en physique.*

Cette étude concerne les propriétés des ensembles d'éléments associés dans une structure qui résulte de processus multiples où intervient le hasard. Il est possible de caractériser ces ensembles par une configuration géométrique spatiale qui synthétise les corrélations statistiques entre leurs variables descriptives. Lorsque l'ensemble est contraint dans un champ de forces sa réponse est prévisible en étudiant l'interaction entre le champ et cette configuration. La réponse ainsi quantifiée est holiste et harmonieuse. Elle est qualifiée de secrète car bien qu'immanente au sein de l'ensemble elle ne coule pas de source et des constructions théoriques sont nécessaires pour la mettre en évidence.

Les caractéristiques des ensembles désordonnés aléatoires nous sont connues par des lois probabilistes et sont d'autant mieux définies que le nombre d'éléments identifiés dans ces ensembles est important. Les grands ensembles forment des univers lesquels peuvent être de différentes natures, par exemple : matérielle, sociétale. Citons les univers physiques qui s'étendent des univers macroscopiques lointains tels ceux du cosmos aux univers microscopiques rapprochés des particules dans la matière condensée en incluant la classe intermédiaire des univers mésoscopiques qui existent dans les matériaux dans notre proche environnement. Cette dernière classe qu'il est plus facile d'appréhender par nos sens est prise comme exemple dans cette étude où il est typiquement fait référence à un matériau qui est structuré de manière non ordonnée, aléatoire: la feuille de papier.

Je montre dans cette étude comment un ensemble constitué par des fibres et dont la formation s'est effectuée par la conjonction d'une multiplicité de processus complexes où intervient le hasard, présente une réponse globalement émergente, prévisible et harmonieuse, vis-à-vis des sollicitations auxquelles il est soumis. Réciproquement on doit se poser la question de l'existence d'un principe organisationnel compatible avec la formation d'un ensemble homogène à structure désordonnée aléatoire, lorsqu'une réponse harmonieuse vis-à-vis de ses sollicitations est identifiée dans son comportement. Le qualificatif d'harmonieux exprime d'une part le caractère équilibré, perceptible par nos sens, des formes révélées par l'association des parties de l'ensemble et d'autre part son comportement vis-à-vis des sollicitations qui se formalise en fonction d'expressions mathématiques harmoniques elliptiques.

Le lecteur pourra penser que les développements présentés dans les exemples cités paraissent éloignés du contexte des objectifs annoncés. L'auteur a choisi à dessein l'étude de la feuille de papier afin de présenter un ensemble ayant une structure complexe à caractère stochastique existant dans un matériau à portée de main pour chacun. Des raisonnements à l'échelle mésoscopique, intermédiaire entre celles des domaines macroscopique et microscopique, facilitent par ailleurs l'établissement de liens entre ces extrêmes ce qui est une des préoccupations de la physique contemporaine.

Le vocabulaire utilisé est celui du langage commun et le plus souvent adapté à la classe des matériaux ; il est transposable dans d'autres domaines d'intérêt. Des notes annexes ainsi qu'une bibliographie qui fait référence à des travaux réalisés suivant les concepts présentés ou proches du domaine étudié complètent le texte.

# I

## Des univers alentours, à portée de main, c'est le cas de la feuille de papier

I-1. Les matériaux à structure désordonnée aléatoire.

A la différence des ensembles d'éléments dont la structure est ordonnée comme c'est le cas pour les matériaux cristallins, il n'est pas possible de prévoir de manière certaine dans les matériaux à structure désordonnée la position relative des éléments qui constituent leur texture [1]. Dans la plupart des cas les paramètres géométriques qui caractérisent la texture de ces ensembles vérifient des lois de probabilité et des méthodes statistiques permettent de les quantifier avec un certain niveau de vraisemblance.
Dans la nature, de nombreux matériaux ont une texture qui vérifie ces critères, par exemple : les sols, les roches magasins d'hydrocarbures constituées par l'assemblages de grains minéraux, les amas fibreux en forme de pelotes rejetées sur les grèves par le mouvement des vagues, certains matériaux inertes ou la matière vivante dont la structure est cellulaire ou alvéolaire. Ces éléments ont une morphologie qui peut être caractérisée de manière globale par les valeurs moyennées de paramètres tels que : la porosité, la surface spécifique, la taille et la forme des pores.
Des matériaux tels que les panneaux de fibres ou de copeaux de bois utilisés dans la construction, certains matériaux en feuilles renforcés par des fibres organiques ou métalliques, les produits textiles non tissés, le papier et le carton à base de fibres végétales, sont également des assemblages de particules dont la texture une fois consolidée est non ordonnée aléatoire.
Le verre sous ses différentes formes de réalisations : transparentes ou diffusantes, la plupart des films de polymères sont des matériaux dont la morphologie à l'échelle micellaire ou moléculaire est non ordonnée. La réalisation de ces matériaux d'origine naturelle ou artificielle résulte de processus multiples et complexes ou intervient le hasard[2] ce qui rend stochastique l'ordonnancement de leurs éléments lorsque nous les examinons dans des conditions qui permettent de les caractériser.

I-2. La feuille de papier.

Envisageons le cas du papier, un matériau communément répandu de par ses nombreuses utilisations. Bien qu'omniprésent dans notre environnement peu de personnes s'interrogent cependant sur sa structure complexe [3] ainsi que son processus de fabrication. Dans une feuille de papier de format A4, de surface 1/16 de m2, de dimensions : 29,7 cm en longueur, 21,0 cm en largeur et 60 microns d'épaisseur, de la qualité standard utilisée pour l'impression, le nombre de fibres végétales qui sont le principal constituant du papier, est d'environ 44,7 millions soit 11.952 fibres par mm³ de la feuille [4]. Cette quantité de fibres varie suivant la nature du végétal utilisé ainsi que de l'épaisseur et de la texture de la feuille, paramètres qui sont fonctions des usages du papier. La multiplicité des usages justifie l'appellation d' « univers fibreux » pour caractériser l'étendue du domaines des utilisations du papier compte tenu de ses propriétés[5].
Sur les machines à papier les plus modernes et suivant les sortes de papiers il peut suffir d'à peine 0,3 sec pour que 75 milliards de fibres dispersées par des micro turbulences intenses dans une suspension aqueuse, soient réparties sur une surface de : 10 m x 10 m, en réalisant une feuille homogène d'un seul tenant et de consistance moyenne [6].

- Les chiffres en exposant renvoient aux notes complémentaires répertoriées dans l'appendice

La texture de la feuille de papier est ainsi formée de manière quasi instantanée, en continu, sur la table de fabrication de la machine à papier, voir par exemple la figure 5.

En examinant la feuille à l'aide d'un microscope, nous identifions les éléments fibreux dispersés dans la feuille. Le nombre de fibres par unité de volume de la feuille, la longueur moyenne des cordes dans les pores entre les interfaces des parois des fibres, la direction des fibres dans le plan ainsi que dans l'épaisseur de la feuille définissent de manière statistique la taille et la forme moyenne des pores de la feuille à une échelle sub-millimétrique [7], voir les photos : sur la figure 1 d'un papier buvard, photo réalisée au microscope électronique à balayage par Ana Paula Gomes à l'Universidade da Beira Interior à Covilha (Portugal), et figure 2, photo d'un papier pour condensateur électrique, réalisée au microscope électronique en transmission par Henri Chanzy, Directeur de recherche au Centre d'Etudes et de Recherches sur les Macromolécules Végétales (C.E.R.M.A.V) à Grenoble.

La suspension fibreuse à son arrivée sur la machine à papier a la consistance d'une soupe faiblement concentrée, avec des micros agrégats de fibres, de taille millimétrique, qui forment des grumeaux appelés flocs en terminologie papetière. La suspension s'écoule dans la caisse de distribution de la pâte, située en tête de la machine à papier, avec un régime de micros turbulences intenses destiné à rompre les flocs et à distribuer la pâte de manière homogène sur la toile de filtration ou s'effectue la formation de la feuille. L'opération est réalisée à l'aide d'un répartiteur hydraulique dimensionné en fonction de la largeur de la toile et profilé en forme de venturi convergent avec à sa sortie des règles métalliques soigneusement rectifiées et dont l'espacement qui est de l'ordre du centimètre est contrôlé par des asservissements qui mettent en œuvre une haute technologie. Un brusque changement d'état de la pâte à papier, de liquide à l'état semi solide, s'effectue au moment de l'impact de la suspension sur la toile sans fin de filtration, entraînée en défilement continu. Cette transition de phase s'accompagne de modifications de la structure fibreuse qui d'isotrope en moyenne dans la suspension aqueuse, devient anisotrope dans le matelas fibreux couché sur la toile de filtration.

La formation de la feuille est complétée au cours de son transport sur la machine à papier, par compression du matelas fibreux au contact de feutres qui absorbent l'eau puis par séchage en évaporant l'eau au contact de la surface polie de cylindres chauffés. Tout au long de son parcours sur la machine, dans les presses et dans la sécherie, la feuille est soumise à des contraintes de tension dans le plan de la feuille et de compression suivant son épaisseur à l'aide de toiles perméables sans fin qui assurent la planéité de la feuille jusqu'à son bobinage en fin de machine.

La vitesse d'entraînement de la feuille sur la machine à papier est de l'ordre de 100 Km/h pour les sortes communes destinées à l'impression, voire supérieure : 120 Km/h pour les papiers de faible grammage[8], par exemple 14 g/m², destinés à des utilisations d'essuyage. L'ensemble des opérations de la fabrication est contrôlé par des opérateurs, spécialistes du procédé papetier, assistés par un ordinateur central qui assure la régularité de la fabrication et qui minimise les casses de la feuille.

La structure de la feuille se reproduit ainsi en continu sur la machine à papier sur une largeur, on dit une laize en terminologie papetière, qui peut atteindre 10 mètres pour les sortes courantes et une épaisseur en moyenne inférieure au dixième de millimètre. Compte tenu de ce processus, un papier fabriqué suivant les règles de l'art est un matériau dont la texture est homogène lorsqu'elle est évaluée à l'échelle des dimensions nécessaires pour sa mise en œuvre et qui recouvrent une étendue de l'ordre du millimètre, voire inférieure dans le cas de l'impression, jusqu'à plusieurs mètres lorsque le papier est utilisé pour la fabrication de produits manufacturés destinés à l'emballage et à la construction dans l'habitat et le génie civil.

En fonction des réglages de la machine à papier des gradients de vitesse sont formés dans l'épaisseur de la veine liquide de la suspension fibreuse pendant le court instant où s'effectue la formation de la feuille. Les fibres dont la longueur est voisine du millimètre et la largeur voisine de la dizaine de micromètre sont déposées sur la toile de filtration avec leur grand axe situé en moyenne dans un plan parallèle au plan de la toile d'où la structure feuilletée et lisse en surface de la feuille de papier [9]. Sous l'effet des forces induites par le gradient de vitesse dans l'épaisseur de la feuille lors de sa formation , l'orientation des fibres est privilégiée à divers degrés dans le sens de fabrication de la machine, on dit le sens marche. Ainsi les directions du sens marche et du sens travers qui lui est perpendiculaire dans le plan de la feuille et celle de l'épaisseur perpendiculaire au plan de la feuille constituent trois directions d'axes principaux privilégiés pour la feuille. Cette configuration existe déjà à l'intérieur des flocs à une échelle inférieure au millimètre, voir par exemple sur la figure 11 l'analyse de l'orientation des fibres par diffusion d'un rayon laser dans un papier d'emballage ainsi que figures : 12 et 13 l'analyse d'un papier journal par des tomographiques aux rayons X, réalisée à l'E.S.R.F de Grenoble. Cette structure confère à la feuille des caractéristiques structurales spécifiques qui sont à l'origine des propriétés fonctionnelles du papier pour ses différents usages [10].

Pour les utilisations courantes l'épaisseur d'une feuille de papier est inférieure au dixième de millimètre avec une régularité de l'ordre de quelques microns. Pour les cartons l'épaisseur de la feuille peut cependant être supérieure au millimètre. En un point de la surface de la feuille la courbure est en moyenne nulle lorsqu'elle est évaluée sur des distances de l'ordre du millimètre [11]. D'où les usages du papier compte tenu de sa planéité pour l'impression, l'écriture, la reprographie voire le calage de précision des pièces mécaniques. Tout autre est la forme de l'interface des fibres et des pores à l'échelle submillimétrique en deçà de la surface de la feuille. De point en point sur les interfaces de la texture, la courbure est de manière aléatoire positive ou négative, définissant des formes d'une grande complexité. C'est ainsi qu'apparaît la texture d'une feuille de papier et qui échappe à nos observations au cours de ses utilisations communes.

L'utilisation du papier comme support de l'écriture est apparue au troisième siècle avant notre ère si l'on en juge par l'existence de documents écrits sur des feuilles de papier datant de cette époque et trouvés dans la grande muraille de Chine. Dès le premier siècle de notre ère les Chinois ont utilisé pour l'écriture des supports textiles constitués d'une trame non tissée formée à partir de débris de fils de soie mis en suspension dans l'eau et consolidée par séchage à l'air sur un support de filtration textile mis en tension sur des cadres [12]. Puis vint le développement de la fabrication artisanale du papier à partir de fibres végétales en provenance de plantes annuelles et d'écorces d'arbres tels Broussonetia Papirifera et Murientis Papirifera, des matières premières moins onéreuses que la soie et plus faciles à se procurer en grande quantité. Cette invention du papier en utilisant des fibres d'origine végétale est due à Tsai-Lun, un superintendant de la maison impériale de l'empereur Hoti [13] qui l'a réalisée en l'an 105.

On peut penser que cette innovation est l'appropriation par l'homme d'un matériau existant dans la nature et réalisé par les guêpes cartonnières de l'espèce poliste Bernhardis Cornis qui construisent leur nid avec des fibres végétales collectées dans leur proche environnement. La texture d'une feuille de papier est à s'y méprendre identique à celle de la feuille qui constitue l'enveloppe du nid de cette sorte de guêpes. On peut s'en rendre compte en comparant au microscope la texture de ces deux matériaux, voir les photos au microscope électronique à balayage réalisées au laboratoire de microscopie de l'E.F.P.G, par Christian Voillot et Martine Rueff, Docteurs Ingénieurs, spécialistes du procédé papetier et enseignants chercheurs à Grenoble INP-Pagora. La photo de la figure 3, représente le « papier mâché » des guêpes, et celle de la figure 4, un papier d'emballage tel qu'il est fabriqué industriellement de nos jours. La texture de ces deux types de feuilles est constituée dans un cas comme dans l'autre de fibres de bois résineux [14]. L'existence sur notre planète du matériau papier pourrait ainsi se situer à l'époque de l'apparition des insectes vespidés. En s'appropriant ce matériau qui existe dans la nature depuis plusieurs dizaines de millions d'années, l'homme à conféré au papier ses lettres de noblesse en confirmant sa pérennité pour un développement durable à long terme.

La matière première du papier provient en majorité des fibres végétales dont la paroi est constituée de faisceaux micro fibrillaires de cellulose. Cette macromolécule organique, la plus abondante sur la terre, est produite dans les végétaux par un processus bioénergétique : la photosynthèse, résultat de l'effet de l'énergie solaire sur le dioxyde de carbone et l'eau [15]. Il existe également une génération de cellulose par un processus d'origine bactérienne.

Depuis les premières réalisations artisanales du papier, la texture de base de la feuille a été peu modifiée, même après l'invention de la machine à papier par Nicolas Robert en 1799 qui transforma le procédé de fabrication du papier à la main, feuille après feuille, en un procédé continu entièrement automatisé de nos jours [16]. A titre d'exemple la fabrication d'un papier destiné à l'impression du journal est réalisée sur des machines dont la production est en moyenne de 1000 tonnes en 24 heures ce qui représente en une heure une feuille de papier de 100 kilomètres de long, de 10 mètres de large et de 52 micromètres d'épaisseur, enroulée en continu, en bobines, au bout de la machine à papier.

I-3. Les multiples usages du papier.

C'est la forme des micros espaces qui sont réalisés par les fibres dans la texture de la feuille qui confère au papier son adéquation pour ses usages multiples. Au cours de la fabrication de la feuille l'élimination de l'eau par évaporation dans les micro pores de la texture fibreuse, développe des forces de tension dues à la capillarité qui correspondent à des pressions de plusieurs centaines d'atmosphères et qui s'exercent sur les parois des fibres cellulosiques en établissant des liaisons physico-chimiques du type pont hydrogène entre les fibres. Cette liaison a une forte résilience et permet de réaliser une texture cohérente d'un seul tenant avec une multiplicité de pores communiquant avec l'extérieur en conférant à la feuille de papier une grande porosité alliée à une grande légèreté compte tenu de sa faible épaisseur. Pour un papier d'impression, la porosité c'est-à-dire la proportion en volume des pores de la feuille représente en moyenne 50 % qui se reparti aux interfaces proches de la surface et à l'intérieur de la feuille. La taille des pores, <<g (θ)>>, évaluée en tant que longueur moyenne statistique des cordes entre les interfaces des pores et des fibres à l'intérieur de la feuille est en moyenne de l'ordre de quelques micromètres ; elle est variable suivant les sortes de papiers. Ces valeurs de porosité et de taille des pores [17] correspondent à une surface développée à l'intérieur de la feuille qui est plus de dix fois supérieure à celle des faces de la feuille.

Les parois des fibres végétales ont une structure lamellaire, en feuillets constitués de faisceaux de micro fibrilles cellulosiques en majorité cristallines et orientées dans la paroi des lamelles en formant une structure hélicoïdale avec un espacement entre les feuillets de quelques dizaines de nanomètres. Suivant les procédés de fabrication de la pâte et du papier les interfaces dans les parois des fibres sont accessibles à l'échelle nanométrique contribuant ainsi à l'augmentation de la surface spécifique de la feuille de papier qui peut atteindre des valeurs de 50 m2/g.

Ces différentes caractéristiques physiques, de porosité, de surface spécifique externe et interne, de taille des pores sont à l'origine des propriétés spécifiques du papier : sa perméabilité pour les fluides, ses propriétés de filtration, d'absorption des liquides, de rétention des particules, de fixation des encres, de barrière antibactérienne mise à profit pour la conservation du matériel chirurgical après stérilisation dans des poches en papier, ses propriétés mécaniques de rigidité ou de souplesse en fonction de l'épaisseur et du volume massique de la feuille, ses propriétés optiques d'opacité, de blancheur, ses propriétés d'isolation thermique, de conduction des ions en tant que membrane séparatrice dans les générateurs électrochimiques et de forte capacité électrique lorsque la feuille de papier est densifiée à l'extrême pour son utilisation dans des condensateurs électriques d'une grande fiabilité[18].

Il n'apparaît pas possible de représenter par une expression mathématique la texture d'une feuille de papier qui soit destinée à un usage spécifique tant la forme de l'espace constitué entre les interfaces des pores et des fibres à l'intérieur de la feuille est complexe. Des images de coupes micro tomographiques, voir par exemple la figure 12, obtenues par l'absorption de rayons X de haute énergie et par contraste de phase, au synchrotron de l'ESRF à Grenoble[19], permettent d'observer sans la modifier la texture interne des feuilles de papier à une échelle microscopique qui correspond à des volumes d'analyse de quelques centièmes de mm³. Ces images évoquent celles d'hyperespaces projetés en trois dimensions.

C'est ainsi que Charles.H. Hinton, Professeur à l'université de Princeton, spécialiste des espaces multidimensionnels, disait « avoir puisé dans la trame grossière de l'enchevêtrement des fibres de misérables papiers » la source de son inspiration pour ses recherches. Ce chercheur, mathématicien et physicien pragmatique, écrivait en 1885 dans son essai « Many Dimensions » : « Regardant ces mêmes journaux et les examinant de plus en plus précisément au microscope je découvrais à profusion des formes d'une variété et d'une ampleur inouïes qui sur le moment, surpassèrent tout ce que mes rêves les plus fous avaient pu concevoir car même là dans ces lettres illisibles et ce papier chiffonné, se trouve, si vous regardez bien, l'espace lui même, avec l'infinité de toutes ses formes possibles ».

J'ai fait ce rappel sur le papier en tant qu'exemple d'un matériau qui est palpable par tous et dont les propriétés dépendent de sa texture stochastique. Comment caractériser un grand nombre d'ensembles d'objets présents dans notre environnement et dont la structure est désordonnée aléatoire? Hermann Weyl, écrivait en 1920: « quels sont les caractères qui doivent être explicités afin de pouvoir caractériser, avec un degré arbitraire de précision et d'une manière conceptuelle, un objet particulier se trouvant dans la portion continûment étendue de la réalité qui est en question ?». Cet éminent mathématicien et physicien estimait que « l'on n'entrait pas dans l'espace comme dans une maison locative, l'espace étant la forme des phénomènes qui se présentent à la conscience ». Ces propos viennent en corollaire de la réflexion d'Anaxagore qui cinq siècles avant notre ère estimait que : « les phénomènes sont la vue des choses cachées ». Henri Poincaré quand à lui a conclu que « le véritable espace est l'espace moteur… ». Effectivement le mouvement est nécessaire pour que le hasard intervienne dans la structuration d'un ensemble désordonné aléatoire.

## II

**Un concept de caractérisation des ensembles à texture désordonnée aléatoire : le pore équivalent**

<u>II-1. Homogénéisation de la courbure spatiale dans les ensembles à texture désordonnée aléatoire.</u>

Comme nous l'avons remarqué plus haut s'il est possible de décrire par des lois déterministes le comportement d'un ensemble d'éléments dont la texture est ordonnée comme c'est le cas par exemple pour un matériau cristallin, les mêmes raisonnements ne sont pas applicables dans le cas d'un ensemble désordonné dans lequel la position des éléments résulte de processus où intervient le hasard. L'étude de l'assemblage des éléments dans ce type d'ensemble s'effectue par l'évaluation statistique de paramètres choisis pour caractériser sa structure. Ces paramètres sont définis par leur valeur moyenne associée à leur variance dans l'étendue d'un domaine considéré comme homogène, en mesure de représenter l'ensemble avec un certain degré de confiance. L'étendue de ce domaine élémentaire représentatif est fonction des sollicitations auxquelles l'ensemble est soumis.

Dans la plus part des méthodes d'analyse du comportement des ensembles désordonnés, par exemple par les méthodes dites d'homogénéisation, des opérateurs mathématiques sont appliqués dans le champ d'analyse défini à grande échelle auxquels sont affectées les valeurs de paramètres qui caractérisent l'ensemble analysé à petite échelle dans le domaine élémentaire représentatif. Les hypothèses de calcul sont ensuite validées après ajustement des valeurs de ces paramètres, compte tenu de la réponse physique de l'ensemble à grande échelle. Les méthodes d'homogénéisation sont utilisées avec succès dans différentes technologies et permettent d'importantes applications dans la pratique. Cependant compte tenu de l'effet des moyennes répétées aux différents niveaux d'analyse, les méthodes d'homogénéisation masquent la traçabilité de la réponse émergeante de l'ensemble ce qui rend difficile l'interprétation physique des phénomènes. Par une approche différente, basée sur la configuration géométrique de l'ensemble, il est possible de formuler une représentation homogénéisée adaptée à l'étude d'un ensemble et à son comportement dans des champs de contraintes de natures diverses.

Envisageons par exemple un matériau polyphasé dont on étudie le comportement lorsqu'il est sollicité dans un champ de forces. Une propriété des plus significatives pour cette étude conjointement à la nature des éléments qui composent le matériau est celle de la géométrie de l'interface entre les phases : celle des pores et des fibres par exemple dans une feuille de papier ou des grains dans un matériau tel qu'un alliage métallique ou dans un sol tel un grès. En dehors des points singuliers anguleux on peut caractériser la géométrie de cette interface en la rectifiant à la limite de résolution près par des éléments rectilignes, $dL_\theta$, ou des micros surfaces planes sous la forme de micro parallélogrammes, suivant que la structure est développée dans un plan ou en trois dimensions. L'orientation des éléments rectifiés est repérée en orientant leur normale vers une même phase par exemple celle des pores dans le cas d'un milieu poreux.

Les interfaces échantillonnées au hasard dans l'ensemble, sont regroupées en fonction de leur orientation, $\theta$, dans un même intervalle angulaire en cumulant leur longueur ou leur aire, sans tenir compte des caractéristiques de contiguïté des éléments rectifiés dans la texture . L'échantillonnage est répété dans le champ d'analyse et les résultats sont comparés à l'intérieur de chaque intervalle angulaire, en cumulant chaque fois les nouvelles valeurs d'avec les précédentes et en les rapportant à la totalité des valeurs des éléments décomptées toutes les orientations étant confondues, ceci jusqu'à l'obtention d'une répartition considérée comme stationnaire à la précision des mesures près. La courbe ou la surface gauche obtenue à la limite par lissage de cette répartition représente la densité de la probabilité en orientation des interfaces pondérée en longueur, $\Psi(\theta)$, ou en surface, des quantités qui sont sans dimension. Dans le cas de structures planes ou analysées sur des coupes ou des tranches fines, des expressions algébriques polynomiales conviennent, avec un ou deux paramètres d'ajustement, pour représenter, $\Psi(\theta)$, en fonction de l'orientation des interfaces des éléments dans la texture, ce qui facilite les comparaisons entre les matériaux, voir par exemple au paragraphe III-2 ci après.

La représentation de la densité de probabilité en orientation des interfaces s'effectue dans un système de coordonnées cartésiennes ou polaires mettant en évidence les anisotropies éventuelles des matériaux en fonction de l'orientation. Dans un réseau fibreux plan chaque élément d'une fibre rectifiée jouxte deux pores adjacents de part et d'autre de l'élément, délimitant ainsi deux interfaces qui ont la même direction mais dont l'orientation est opposée. La densité de probabilité en orientation des interfaces, $\Psi(\theta)$, est donc périodique, ses valeurs se reproduisant à un intervalle de $\pi$ radians. En représentation polaire la courbe indicatrice de $\Psi(\theta)$ est centrosymétrique. De manière générale les ensembles d'éléments désordonnés aléatoires homogènes sont centrosymétriques. A chaque interface qui structure l'ensemble suivant une orientation correspond une interface suivant la même direction mais d'orientation opposée. La courbe indicatrice de la densité de probabilité en orientation, $\Psi(\theta)$, représentée en coordonnées polaires est parfois dénommée : rose des directions. Cette figure, à l'exception des ensembles isotropes, n'a pas de similitude de forme d'avec une figure qui pourrait être détectée par un examen visuel de la texture, voire même imaginée par la pensée, contrairement à la configuration du pore équivalent qui est définie çi après.

II-2. Le concept du « pore équivalent ».

La répartition en orientation des interfaces des éléments rectifiés, pondérée par leur longueur ou leur aire, lorsqu'elle n'est pas évaluée de manière relative, a la dimension d'une longueur ou d'une aire rapportée à un intervalle d'angle. Cette quantité peut s'interpréter comme étant le rayon de courbure d'une figure qui enveloppe les interfaces de la texture de manière statistique et globale, en faisant abstraction des particularités géométriques dues à la contiguïté des éléments. Remarquons que la densité de probabilité en orientation définie au paragraphe précédent, $\Psi(\theta)$, peut suivant cette interprétation être considérée comme un rayon de courbure normé par rapport au périmètre de cette figure.

Compte tenu de la réalisation de l'échantillonnage il est possible d'évaluer le rayon de courbure dans des conditions telles que la totalité de la longueur ou de l'aire des interfaces de l'ensemble analysé soient prises en compte et de rapporter cette valeur soit à la surface soit au volume soit à la masse de l'ensemble en tenant compte de la masse surfacique ou de la masse volumique de la texture. Dans ces conditions la dimension du rayon de courbure normé est inversement proportionnelle à une longueur ou proportionnelle à une longueur ou une surface rapportées à la masse. Ce paramètre ainsi normé représente la courbure spécifique, surfacique ou massique dans le cas d'un ensemble plan ou la courbure spécifique, volumique ou massique dans le cas d'un ensemble en trois dimensions.

La courbe ou la surface sous-tendue par le rayon de courbure délimite d'un point de vue global un espace qui est statistiquement conforme à celui des interfaces discrétisées et rectifiées dans la texture. Cette figure que j'ai dénommée : le « pore équivalent », est un contour linéaire fermé dans le cas d'un ensemble plan ou une surface gauche en forme de boule simplement connexe dans le cas d'un ensemble en trois dimensions. D'un point de vue analytique la configuration du pore équivalent est celle de la fonction primitive du rayon de courbure représentatif de la densité de probabilité en orientation pondérée des éléments[20], une figure distincte de la rose des directions qui est la représentation moyennée des rayons de courbure en coordonnées polaires.

J'ai défini le concept du pore équivalent pour l'étude du papier et des textiles non tissés et plus généralement des matériaux fibreux dont les propriétés dépendent de la répartition en orientation des fibres qui enveloppent les pores dans leur texture[21].

L'appellation de « pore équivalent » se justifie du fait que dans un milieu poreux la configuration géométrique du pore équivalent lorsqu'il a une forme elliptique ou ellipsoïdale est semblable à celle du pore moyen défini selon l'usage par la valeur des cordes moyennes évaluées dans les différentes directions dans les pores entre les interfaces de la texture. Ainsi le pore équivalent et le pore moyen sont des concepts duals[22] ; voir la figure 8 ainsi que le paragraphe II-5 ci après. Le pore équivalent est parfois dénommé surface conforme équivalente ou profil équivalent conforme, appellations qui sont dérivées de celle de pore équivalent et mieux adaptées pour caractériser les états de surface des matériaux. Dans les analyses pétrographiques où l'on cherche à caractériser la forme moyenne des grains minéraux qui constituent les sols on définit leur forme équivalente suivant un concept différent mais relativement proche de celui du pore équivalent.

Le pore équivalent est un hologramme à la fois au sens étymologique et physique du terme. Les racines grecques de ce mot sont : « olos » qui signifie l'ensemble et « gramma » qui signifie l'épitaphe. Le pore équivalent représente en effet d'un point de vue global la configuration géométrique relative des éléments dans un ensemble. Son contour linéaire de dimension un ou sa surface de dimension deux permettent la résolution de problèmes qui se posent dans des espaces de dimension supérieure, dans un plan ou dans un volume. Cette propriété qui est une caractéristique des hologrammes justifie le qualificatif d'équivalent dans l'expression : pore équivalent [23].

II-3. Construction du pore équivalent par une transformation conforme.

Le concept de pore équivalent peut s'imaginer à partir d'une transformation conforme effectuée au niveau des éléments rectifiés dans la texture. Supposons un réseau fibreux plan constitué de filaments considérés comme des lignes de dimension un. Construisons une figure réalisée à partir de la translation de chacun des segments rectifiés des filaments qui tout en conservant leur longueur et leur direction par rapport à une direction de référence, les place à la queue leu leu à partir d'un point quelconque, choisi comme origine dans le plan, en les cumulant bout à bout une fois hiérarchisés en fonction de leur direction, voir la figure 7. La figure obtenue à la limite par le lissage de ce contour est une courbe dont la longueur est égale à la longueur cumulée des filaments dans le réseau étudié. Cette courbe en forme de demi boucle correspond à la moitié de la longueur des interfaces des filaments et des pores. Elle peut être refermée sur elle-même en la prolongeant d'autant par une symétrie centrale, les extrémités des segments rectifiés des filaments n'étant pas différenciées et leur orientation également répartie suivant, θ, et, θ + π, radians. La boucle ainsi construite représente à la limite de rectification près des éléments, le pore équivalent en tant qu'enveloppe des interfaces entre les filaments et les pores dans le réseau fibreux.

Cette construction du pore équivalent par une transformation conforme peut s'envisager également dans le cas d'un ensemble développé en trois dimensions après rectification de ses interfaces par des micros plans en forme de micro parallélogrammes. La hiérarchisation des micros plans suivant leur orientation, pondérée par l'aire cumulée de leurs surfaces, permet théoriquement de construire de proche en proche, une surface fermée et sans bords semblable à une boule dans l'espace et dont l'aire est égale à celle de l'interface entre les phases de la texture. Cette construction est certes plus complexe à réaliser que celle effectuée dans le cas d'un ensemble plan, même si elle est envisagée par la pensée. Le résultat cherché s'apparente à celui obtenu dans le jeu du cube de Erno Rubik dont le but est de regrouper sur chacune des faces d'un gros cube, des petits cubes unitaires ayant la même couleur parmi six couleurs différentes, en les juxtaposant dans un assemblage connexe, les petits cubes étant répartis au début du jeu de manière quelconque sur les faces du gros cube. 43 milliards de milliards de configurations pour le positionnement initial des petits cubes sont possibles si on tient compte des contraintes mécaniques nécessaires pour assurer leur positionnement sur le gros cube: en position centrale, en coin ou intermédiaire. La réussite du jeu de Erno Rubik est possible par l'enchaînement d'une série de rotations, « 20 étant le nombre de mouvements nécessaires pour ranger le cube dans les cas les plus difficiles…. les configurations exigeant 20 mouvements étant assez rares… environ 300 millions », suivant les termes des conclusions de Jean-Paul Delahaye, mathématicien, Professeur émérite au Laboratoire d'informatique fondamentale de l'Université de Lille qui a étudié, selon son expression, « ce numéro un de tous les casse-tête ». Dans le cas du pore équivalent d'un ensemble d'objets, comme il est envisagé ici par analogie, les interfaces élémentaires ne sont pas différenciées par leur couleur mais par leur orientation ce qui revient à envisager autant de faces dans la configuration terminale qu'il y a d'orientations possibles dans l'ensemble en comparaison de six pour le cube de Rubik dans le cas précité. Une fois effectué le regroupement des micro surfaces, la surface du pore équivalent doit apparaître sous la forme d'un polyèdre convexe, le plus souvent non régulier et dont le nombre de faces dépend à la fois de l'ensemble et du niveau de définition de la rectification des interfaces. Bien que non limité par définition par des contraintes dues à la contiguïté des éléments, l'ordonnancement des micros surfaces est un problème complexe qui fait appel à l'analyse combinatoire ainsi qu'à des algorithmes de calcul nécessitant des ordinateurs d'une grande capacité. Le raisonnement par analogie d'avec le jeu de Rubik permet d'imaginer la complexité des opérations qu'il faut réaliser pour construire le pore équivalent d'un ensemble d'éléments distribués dans l'espace. Heureusement d'autres méthodes, notamment celles développées sur la base des relations stéréologiques, permettent de construire dans la pratique le pore équivalent d'ensembles complexes, voir par exemple les figures 12 et 13 qui sont relatives à l'analyse d'un papier journal effectuée à l'E.S.R.F. de Grenoble par Christine Antoine et Rune Holmstad, chercheurs au Norvegian Pulp and Paper Research Institute à Trondheim (Norvège).

II-4. <u>Identification du pore équivalent par la stéréologie.</u>

La stéréologie est une méthode d'analyse des objets et de leurs ensembles qui met en œuvre les lois de probabilités géométriques. Les relations stéréologiques permettent d'évaluer dans un ensemble les propriétés de paramètres qui ont une dimension, n, à partir de la mesure de paramètres qui ont une dimension, n-1. Des propriétés linéiques peuvent ainsi être de proche en proche significatives de propriétés surfaciques et volumiques.
Par exemple, dans un ensemble bi phasique plan ou considéré comme tel, le nombre moyen d'interceptes c'est-à-dire de points d'intersection, d'avec les interfaces des éléments par unité de longueur de traversée dans une direction, $\gamma$, que l'on dénomme : nombre d'interceptes linéiques, $< P(\gamma) >$, est égal à la longueur projetée des interfaces des éléments de l'ensemble sur une droite perpendiculaire à, $\gamma$, rapporté à l'unité de surface.
Cette propriété établit dans un réseau plan une relation entre la distribution des projections orthogonales dans une direction, $\gamma$, des interfaces de densité de probabilité en orientation, $\Psi(\theta)$, et le nombre moyen des interceptes linéiques, d'une droite de direction, $\gamma$, tracée dans le réseau. Par inversion de cette relation il est possible de connaître la distribution de la densité de probabilité en orientation des interfaces, $\Psi(\theta)$, en fonction de la distribution des interceptes, $< P(\gamma) >$, mesurée dans le réseau.

Les fonctions analytiques qui représentent ces distributions, périodiques dans un intervalle de π radians, ne sont à priori pas connues aussi on les représente par des développements en série de Fourier dont les termes sont identifiés par l'intégration par parties de la relation stéréologique précitée. Lorsque le pore équivalent de l'ensemble est une ellipse ou un ellipsoïde, figures qui sont à privilégier dans le cas des ensembles désordonnés aléatoires comme il est indiqué et vérifié dans la pratique, voir les chapitres III et IV ci après, le pore équivalent est l'inverse géométrique de l'indicatrice des interceptes, $< P (\gamma)>$. La puissance de l'inversion permet de normer cette relation en la mettant à l'échelle de l'ensemble, à partir des caractéristiques morphologiques des éléments de la texture.

D'autres résultats importants pour l'étude des ensembles peuvent être obtenus par la stéréologie. Par exemple la valeur moyenne, toutes les directions étant confondues, des interceptes linéiques des interfaces dans un réseau plan, $<< P(\gamma) >>$, est reliée à la longueur, L, du contour des éléments par unité de surface suivant la formule : $<< P (\gamma) >> = 4 L/\pi$. Cette relation permet d'évaluer la longueur surfacique, L, qui est une caractéristique physique importante pour l'utilisation des matériaux fibreux conformés en feuille notamment dans le cas du papier et des textiles non tissés. La valeur de L est fonction du grammage c'est-à-dire de la masse surfacique de la feuille et de sa composition fibreuse caractérisée par la masse linéique des fibres.

Il est remarquable que la mesure de, $<< P (\gamma) >>$, réalisée dans un réseau de lignes plan, permet d'estimer la valeur de, π, en tant qu'espérance mathématique. L'expérience consiste à jeter au hasard et à répétition un objet, sur un réseau de lignes tracées dans un plan et de mesurer le nombre d'interceptes du contour de cet objet d'avec les lignes du réseau. Cette méthode dite du « jeté d'aiguille » a été proposée par le naturaliste G. Buffon en 1777 dans son « Supplément à l'histoire naturelle de ses Essais d'arithmétique morale ». Sa méthode a été mise en œuvre dans des travaux expérimentaux par les étudiants, élèves ingénieurs papetiers, à l'E.F.P.G, en 1964, en utilisant des feuilles de papier de très faible grammage : 2g/m², suffisamment minces pour être considérées comme constituant un réseau fibreux plan. Ces feuilles, voir la figure 6, ont été réalisées au laboratoire avec des fibres d'alfa fines et de forme régulière facilitant la détermination des interceptes d'avec des aiguilles jetées sur les photographies des feuilles agrandies 70 fois. Les résultats de ces mesures qui ont permis de déterminer la valeur de la constante, π, à la deuxième décimale près, confirment la validité des méthodes stéréologiques, tant du point de vue théorique que pratique, pour l'étude de propriétés du papier et plus généralement des matériaux à texture fibreuse.

Dans le cas d'un ensemble en trois dimensions, un théorème de stéréologie démontré notamment par Saltykov et Weibel, établit que le nombre moyen des interceptes linéiques des interfaces entre les phases d'une texture dans une direction, est égal à l'aire cumulée de la projection des interfaces sur un plan perpendiculaire à cette direction, par unité de volume de l'ensemble, sans tenir compte de la superposition des projections de ces surfaces[24].

En application de ce théorème on démontre que l'aire totale des interfaces entre les phases, rapportée à l'unité de volume de la texture, appelée surface spécifique, Sv, est égale au double du nombre moyen des interceptes linéiques d'avec les interfaces de la texture, toutes les directions étant confondues, mesurées dans un échantillonnage isotrope. Dans le cas d'un pore équivalent normé, c'est-à-dire dont l'aire est rapportée à l'unité de volume de la texture, la valeur du nombre moyen des interceptes linéiques des interfaces de la texture, toutes les directions étant confondues, est donc égale à celle de la demi aire de la surface du pore équivalent[25].

Les relations stéréologiques permettent ainsi d'identifier quantitativement le pore équivalent d'un ensemble d'éléments en analysant sa texture. Les mesures des interceptes sont réalisées dans les plans de coupes minces obtenues par tomographie avec un échantillonnage isotrope, les coupes étant considérées à la limite sans épaisseur. Les valeurs moyennes des mesures des interceptes linéiques, égales aux projections sur un plan orthogonal à chacune des directions des aires des interfaces de la texture rapportées à l'unité de volume, permettent d'identifier en retour par inversion la configuration géométrique optimale qui leurs correspond et qui représente le pore équivalent. Le choix d'un ellipsoïde facilite cette identification, les calculs d'optimisation ayant une convergence rapide en ce cas. La valeur moyennée des interceptes linéiques égale à la demi surface spécifique volumique de la texture est égale à la valeur de la demi aire de la surface de l'ellipsoïde normé. Dès lors, connaissant l'aire de l'ellipsoïde et les valeurs optimales de ses ellipticités déduites des mesures des interceptes linéiques, il est possible d'identifier par le calcul les valeurs des trois axes principaux du pore équivalent normé.

Dans le mythe dit de la caverne, Platon fait une analogie allégorique au sujet de la perception que nous avons des éléments qui nous environnent. Platon imagine que des prisonniers sont maintenus immobiles à l'entrée d'une caverne obscure, le dos tourné vers l'extérieur et devant laquelle se déplacent des objets éclairés par les flammes d'un feu situé à l'arrière. Dans ces conditions les prisonniers n'ont connaissance de ces objets qu'à partir de leurs ombres portées sur les parois de la caverne. C'est à quoi se ramène selon Platon la perception que nous avons du monde et des événements qui s'y déroulent.

Aujourd'hui les techniques d'imagerie parmi les plus performantes mises en œuvre dans différents domaines d'analyse pour l'étude des matériaux en cristallographie ou ceux dont la texture est amorphe ainsi que les tissus cellulaires dans le domaine biologique, sont basées sur l'interprétation de leurs ombres portées par un rayonnement intrusif ou diffracté par un faisceau d'ondes électromagnétiques ou de pression. Quatre siècles avant notre ère Platon a posé les bases de la stéréométrie en imaginant par la pensée la méthode qui est aujourd'hui mise en œuvre dans les équipements d'analyses qui fonctionnent en mode Scanner.

II-5. La dualité du pore équivalent et du pore moyen.

II-5-1. Définition du pore moyen.

Les mesures des traversées effectuées dans une texture permettent d'évaluer dans une direction, $\theta$, deux paramètres: d'une part le nombre moyen des interceptes linéiques d'avec les interfaces des pores et des particules, $<P(\theta)>$, et d'autre part la longueur moyenne de la corde interceptée dans cette direction entre les interfaces, $<g(\theta)>$. Ces paramètres sont inverses l'un par rapport à l'autre au facteur près de proportionnalité égal à : $2\varepsilon$, $\varepsilon$ étant la porosité de la texture.

L'indicatrice de la demi corde, $<g(\theta)>/2$, interceptée entre les interfaces de la texture, moyennée dans chaque direction, $\theta$, permet dans un système de coordonnées polaires de représenter ce qu'il est usuellement convenu d'appeler le pore moyen et son orientation dans de la texture. La forme moyenne des particules lui est complémentaire. Le pore moyen caractérise au sens probabiliste l'étendue de la phase poreuse qui jouxte la phase solide aux interfaces, moyennée dans chaque direction dans la texture. Ainsi contrairement à des pores occlus dans la texture le pore moyen n'est pas un espace qui définirait un contour ou une surface bornée physiquement identifiable et sans communications avec l'espace environnant, pas plus que ne l'est la forme moyenne ainsi définie dans le cas des particules.

Il est d'usage de définir la taille des pores comme étant égale à la longueur moyennée des cordes, $<<g(\theta)>>$, toutes les directions des traversées étant confondues. $<<g(\theta)>>$ est égale au quadruple d'une entité qu'il est convenu d'appeler le rayon hydraulique moyen, $m\hat{h}$, défini en tant que rapport de l'aire de la section droite normale à l'écoulement au périmètre mouillé des pores ou du rapport du volume rempli de fluide à la surface mouillée dans le cas d'un milieu poreux d'aire de section des pores uniforme saturée par le fluide. Cette définition qui est par essence analytique ne permet pas d'identifier, directement à l'observation, le rayon hydraulique moyen de la texture. Un théorème de stéréologie énoncé notamment par Tomkeieff démontre l'égalité du rayon hydraulique moyen, $m\hat{h}$, d'avec le quart de la taille des pores. On a : $<<g(\theta)>> = 4\ m\hat{h} = 4\ \varepsilon\ /\ Sv$, $\varepsilon$ étant la porosité et Sv la surface spécifique volumique du milieu poreux.

II-5-2. Le pore équivalent de forme ellipsoïdale.

Une propriété du pore moyen est sa similitude de forme d'avec le pore équivalent dans le cas ou celui ci est un ellipsoïde. Ce résultat résulte de l'équivalence entre d'une part la relation stéréologique précitée qui relie dans une direction donnée, $\theta$, la corde moyenne dans les pores d'un ensemble et l'aire des interfaces des particules projetée sur un plan perpendiculaire à cette direction et d'autre part la relation géométrique qui relie un diamètre conjugué d'un ellipsoïde dans une direction, θ, et l'aire projetée de la surface de l'ellipsoïde sur un plan perpendiculaire à ce diamètre[26].

Le choix d'un ellipsoïde en tant que configuration du pore équivalent n'est pas fortuit ni restrictif. En effet sa surface optimise les transferts d'énergie dans la texture d'un ensemble lorsque celui ci est contraint dans un champ de forces [27], voir les chapitres III et IV ci après. Ainsi la mesure des cordes moyennes dans les différentes directions entre les interfaces des phases permet-elle d'identifier directement et de manière simple le pore équivalent ellipsoïdal ce qui est le cas pour de nombreux ensembles d'objets. L'indicatrice de la corde moyenne, $<g(\theta)>$, dans les pores est semblable au pore équivalent, voir la figure 8, et par conséquent l'inverse géométrique de l'indicatrice des interceptes linéiques $<P(\theta)>$. La puissance de l'inversion est normée en fonction des paramètres morphologiques de la texture et des particules. Le pore équivalent ellipsoïdal et le pore moyen, entités qui sont définies de manière différente, sont des concepts duals ce qui justifie le qualificatif d'équivalent employé pour l'expression du premier des concepts précités.

La représentation du pore équivalent s'effectue dans un espace virtuel, son centre est localisé nulle part ou partout dans l'ensemble physique, seul l'orientation de la figure du pore équivalent est repérée par rapport au référentiel matériel de l'ensemble. Dans un ensemble homogène chaque partie est à l'image du tout dès lors qu'elle recouvre une étendue au minimum égale à celle du volume élémentaire représentatif de l'ensemble. Ainsi les représentations du pore équivalent de différentes portions d'un ensemble homogène sont autos similaires.

En conclusion de ce paragraphe je propose une analogie pour représenter le but et la méthodologie du concept du pore équivalent. Imaginons un orchestre au cours de son installation sur la scène, face au public. Venant de différents points de l'horizon les musiciens arrivent à la queue leu leu en formant une suite ordonnée, chacun gagnant sa place suivant la nature de son instrument en tenant compte des contingences de l'espace sur la scène. On a par exemple sur la gauche de l'orchestre les cordes des violons et des altos, à la suite plus à droite les violoncelles, à l'extrême droite les contrebasses, les instruments à vent prenant place en arrière plan, au milieu de l'orchestre et les percussions et les cymbales en position surélevée par rapport à la scène. Dans cette configuration ordonnée, l'ensemble instrumental peut résonner et réaliser à l'unisson le tutti qui permet au Maestro de conduire l'orchestre dans une exécution harmonieuse de sa partition sous les impulsions de sa baguette. C'est ainsi que l'ordonnancement des éléments dans un ensemble désordonné aléatoire, permet par la modélisation du pore équivalent d'obtenir une réponse globale et holiste lorsqu'il est sollicité dans un champ de forces.

# III

## Validation et applications du concept de pore équivalent.

### III-1. Les configurations à privilégier pour la représentation du pore équivalent.

D'après les remarques précédentes nous savons que des courbes et des surfaces centro symétriques, simplement connexes peuvent compte tenu des hypothèses d'homogénéité, convenir pour la représentation du pore équivalent de la texture des ensembles à structure désordonnée aléatoire. Dans le cas d'ensembles plans, ou considérés comme tels par projection, le cercle convient pour les milieux isotropes et l'ellipse pour les milieux anisotropes avec une symétrie bilatérale. De même dans le cas des ensembles tridimensionnels, la sphère et l'ellipsoïde triaxial sont à considérer en première hypothèse. Pour des ensembles à structure complexe la densité de la probabilité en orientation des éléments peut être représentée par une combinaison additive des rayons de courbure d'ellipses ou d'ellipsoïdes homocentriques, chacune de ces figures ayant ses propres dimensions d'axes.

Il n'est pas utile de chercher à affiner de la façon la plus détaillée qui soit la représentation du pore équivalent au regard des fluctuations observées dans les mesures quand celles-ci sont d'une amplitude modérée. Ceci pour deux raisons : d'une part on doit tenir compte de la précision des mesures qui limite le degré de signification des variations observées, d'autre part les fluctuations sont atténuées par l'homogénéisation du comportement des ensembles en présence de leurs sollicitations. Bien souvent une configuration médiane sous forme d'une ellipse ou d'un ellipsoïde est satisfaisante pour la représentation du pore équivalent des systèmes complexes ce qui limite notablement le nombre de paramètres à identifier. Cependant une analyse fine peut révéler des caractéristiques structurales intrinsèques aux matériaux qu'il est utile de chercher à identifier car elles sont liées à leurs procédés d'élaboration. Dans son ouvrage : « Introduction géométrique à quelques théories physiques », Emile Borel fait la part des choses entre ce qui est des effets des théories de la mécanique et les effets de la physique statistique. Il conclut que : « le choix est à faire selon les raisons que l'on a d'admettre que les fluctuations observées sont une conséquence des hypothèses physiques ou dérivent seulement des hypothèses statistiques ».

### III-2. La représentation du pore équivalent par des fonctions elliptiques.

Il a été constaté maintes fois que l'ellipse convenait pour la représentation du pore équivalent de la feuille de papier lorsque sa texture est analysée dans son plan de même que l'ellipsoïde lorsqu'elle est analysée dans ses trois dimensions. Ces observations ont été faites par exemple dans les cas du papier journal, des papiers d'impression, des papiers filtre, des papiers d'emballage. Des configurations elliptiques conviennent également dans le cas des matériaux textiles non tissés que ce soient les voiles de faible grammage : 12 g/m² ou les feutres avec un fort grammage : 1000 g/m², voire davantage[28]. Tous ces matériaux ont une texture fibreuse qui résulte de processus complexes conditionnés par des variables indépendantes qui interviennent au cours du procédé de fabrication. La longueur, la largeur, la courbure des fibres ou des filaments, leur localisation et leur orientation dans la texture fibreuse sont des paramètres distribués suivant des lois probabilistes.

Il est généralement constaté que la densité de probabilité en orientation pondérée par la longueur des éléments fibreux, $\Psi(\theta)$, vérifie dans le plan une loi de distribution qui correspond au rayon de courbure d'un cercle ou au rayon de courbure d'une ellipse d'ellipticité a/b et dans l'espace en trois dimensions à celui d'une sphère ou d'un ellipsoïde triaxial d'ellipticités a/b et c/b. Comme il a été remarqué plus haut dans le cas de structures complexes une combinaison linéaire des rayons de courbure de composantes elliptiques suivant différentes proportions convient également, chacune des composantes étant définie par son périmètre ou son aire, ses ellipticités et l'orientation de ses axes. Ces résultats sont observés quelque soit la nature des fonctions qui définissent les caractéristiques originelles des matières premières fibreuses et leur répartition dans la texture[29].

Dans le cas d'un ensemble d'éléments rectifiés dans un plan la densité de probabilité en orientation des éléments pondérée par leur longueur, $\Psi(\theta)$, peut être représentée par *le rayon de courbure normé par rapport à son périmètre* d'une ellipse rapportée à ses axes principaux et paramétrée par la valeur de son ellipticité, a/b, suivant l'expression :

$$\Psi(\theta) = \frac{1}{\pi} \frac{(1-e^2)}{(1 - f(e)(1-e^2 \sin^2(\Theta)))^{3/2}} \qquad \text{équation (1).}$$

avec : $f(e) = \sum_{k=1}^{\infty} \frac{e^{2k}}{2k-1} \prod_{p=1}^{k} \left(\frac{2p-1}{2p}\right)^2$, $e = \sqrt{1 - \left(\frac{b}{a}\right)^2}$ et $\frac{a}{b} = \sqrt[3]{A_F} \geq 1$

$-\pi/2 \leq \Theta \leq +\pi/2$.

La valeur du paramètre, $A_F$, est égale au rapport des valeurs maximum et minimum des mesures de la densité en orientation, $\Psi(\theta)$, obtenues dans deux directions respectivement perpendiculaires. L'expression de, $\Psi(\theta)$, s'obtient en pratique par une optimisation suivant la méthode des moindres carrés, en fonction des mesures de la répartition en orientation des éléments fibreux.

Le rapport, $\Psi(\theta=0) / \Psi(\theta=+/-\pi/2)$, atteint sa valeur maximum lorsque la fonction, $\Psi(\theta)$, a son maximum, $\Psi(\theta=0)$, centré sur le grand axe de l'ellipse.
On vérifie par ailleurs la relation :

$a/b = [(6 + 4a_1 + a_2)/(6 - 4a_1 + a_2)]^{1/2}$,                    équation (2),

où $a_1$ et $a_2$ sont les coefficients d'ordre 1 et 2 du développement en série de Fourier de la fonction, $\Psi(\theta)$, en fonction de l'orientation, $\theta$, des éléments fibreux. Le terme $a_0$ du développement correspond au cas d'une texture isotrope dans le plan ; sa valeur normée est égale à $1/\pi$.
La densité de probabilité en orientation, $\Psi(\theta)$, est corrélée avec les propriétés des matériaux, par exemple le module d'élasticité du papier, sa contrainte à la rupture en traction plane à mors jointifs, ainsi que les paramètres de fabrication, par exemple le taux de cisaillement dans l'épaisseur de la veine liquide de pâte à papier lors de la formation de la feuille sur la machine. Le paramètre $A_F$ caractérise l'anisotropie de ces propriétés. Sa relation d'avec l'ellipticité, a/b, est un cas d'espèce pour chacune des propriétés envisagées et peut être différente de la relation explicitée dans l'équation (1).
Par exemple dans le cas des modules d'élasticité en traction plane de la feuille de papier le rapport d'anisotropie, $A_F = Esm/Est$, vérifie les relations :

$A_F = [(6 + 4a_1 + a_2)/(6 - 4a_1 + a_2)]$,

$a/b = (Esm/Est)^{1/2}$                    équation (3).

Esm, étant la valeur maximum du module d'élasticité, généralement obtenue dans le sens de fabrication de la machine à papier et, Est, la valeur minimum obtenue dans le sens travers, perpendiculaire au sens de fabrication.

Différentes méthodes d'évaluation de la densité de probabilité en orientation des fibres dans les papiers et les textiles non tissés ont été réalisées avec des échantillons de tailles différentes. Les échantillons étaient prélevés au hasard dans une bande de papier de plusieurs mètres ou dans des liasses de feuilles. Dans la méthode par observation visuelle ou assistée par un capteur d'images, méthode qualifiée de méthode directe, on mesure l'orientation et la longueur rectifiée de fibres colorées représentant une fraction en masse de : 2,5 / 1000 dispersée dans la suspension de pâte à papier avant la formation de la feuille. L'espacement des fibres colorées observées dans le plan de la feuille est en moyenne de quelques mm dans le cas d'un papier de grammage : $50g/m^2$, de fibres de résineux raffinées à 25°S.R, à l'hydrafiner.
Les mesures peuvent être réalisées en scannant directement la feuille de papier à sa surface ou sur des fractions obtenues par des clivages successifs de la feuille dans son épaisseur. Jusqu'à sept fractions peuvent ainsi être obtenues dans une feuille de papier d'épaisseur 70 microns.
D'autres méthodes ont été expérimentées sur des champs d'observation de dimension millimétrique : par microscopie optique ou électronique, dans le plan de la feuille et dans des coupes minces transversales perpendiculaires au plan de la feuille, par diffusion de la lumière d'un faisceau laser de longueur d'onde , $\lambda= 820nm$, de section sub-millimétrique : 200 µm, impactant des feuilles de papier et des feutres textiles, soit à travers le plan de la feuille , par exemple figure 11, soit dans l'épaisseur des feuilles sur la tranche d'un empilement de feuilles.
La diffraction d'un rayonnement laser à travers des répliques transparentes de la surface du papier ou de voiles textiles non tissés permet d'analyser l'orientation des fibres à leur surface. Les champs d'observation impactés par le faisceau laser sont circulaires et de diamètre quelques mm. Les figures 9 et 10 sont des exemples de ces méthodes développées par Mario Pereira et Rita Salvado dans leurs recherches sur la structure des papiers et des non-tissés, au laboratoire d'optique du Professeur Paulo Fiadeiro à l'Universidade da Beira Interior (U.B.I). à Covilha au Portugal.
L'orientation des fibres peut être caractérisée également par la diffraction de rayons X à travers des feuilles de papier relativement minces ainsi que par l'analyse tomographique des images obtenues par absorption et contraste de phase de rayons X de haute énergie. Les échantillons de papier en ce cas ont une taille submillimétrique. Les figures 12 et 13 sont relatives à des analyses d'un papier journal, réalisées par cette méthode à l'European Synchrotron Radiation Facility (E.S.R.F) de Grenoble, par Christine Antoine et Rune Holmstad, chercheurs à l'Institut Norvégien de recherche en pâtes et papiers, P.F.I. à Trondheim.

Une série de papiers fabriqués en 1979 sur la machine pilote du Centre Technique du Papier à Grenoble, avec différentes valeurs d'anisotropies en orientation des fibres, a permis de faire des comparaisons entre ces différentes méthodes. Différents laboratoires universitaires et industriels de par le monde ont testé cette série de papier avec leurs propres méthodes basées soit sur : des mesures directes visuelles de l'orientation de fibres colorées, le comptage de leurs interceptes linéaires, les images de diffusion et de diffraction de rayonnements laser et de rayons X. La concordance de l'ensemble des mesures confirme à la précision près que la répartition de la densité de probabilité en orientation des fibres pondérée en longueur est au mieux conforme à la répartition des rayons de courbure d'une ellipse caractérisée par son ellipticité et par la direction de son grand axe suivant la direction dominante d'orientation des fibres. Le choix de la méthode de mesure de l'orientation des fibres s'effectue en fonction des matériaux étudiés et des moyens qui sont mis en œuvre au laboratoire ou dans la chaîne de production industrielle pour le contrôle de cette propriété.

III-3. Applications du pore équivalent dans la technique papetière.

La répartition en orientation des fibres dans la feuille est une caractéristique importante pour de nombreux usages du papier lorsque la feuille est soumise à des contraintes externes et/ou internes, de traction, de compression, de torsion, qui produisent des déformations de la feuille dans le domaine élastique et au delà jusqu'à sa rupture. Ces comportements s'observent avec les papiers utilisés en feuille simple ou dans des structures complexes à base de multi jets réalisées par transformation des feuilles de papier ou de carton.

Différents modes de sollicitation en traction plane et jusqu'à la rupture de la feuille, suivant et hors de ses axes principaux, ont été étudiées dans les conditions normalisées par l'I.S.O., avec des papiers de différentes anisotropies en orientation des fibres. La largeur des éprouvettes pour ces essais était de 1,5 cm et leur longueur de 18 cm.

Des mesures ont également été réalisées en rapprochant les mors de traction jusqu'à leur contact. L'écartement des mors étant considéré comme nul, le module d'élasticité et la contrainte à la rupture du papier peuvent être calculés dans les différentes directions du plan de la feuille en fonction de la distribution en orientation des fibres, $\Psi(\theta)$, de la masse volumique de la feuille et de la contrainte à la rupture ortho-radiale des fibres, mesurée dans les conditions ou le séchage de la feuille est contrôlé soit avec un retrait maximum, partiel ou nul. A l'inverse il est possible de calculer à partir des mesures de traction et des variations dimensionnelles du papier qui résultent des conditions de son séchage, les paramètres morphologiques de la texture de la feuille et les propriétés de résistance à la rupture en traction des fibres.

Lorsque les mors de traction ne sont pas jointifs, au delà d'un espacement de quelques fractions de la longueur moyenne des fibres, la répartition des contraintes dans la feuille est modulée par un effet de concentration de la contrainte aux interfaces des fibres et des pores. Ce phénomène dépend de la direction de la traction et de la distribution en orientation des fibres dans la feuille. Comparée au cas où l'écartement des mors est nul, la rupture de la feuille est initialisée à une valeur de contrainte lorsqu'elle est estimée uniformément répartie aux bornes de l'éprouvette, inversement proportionnelle au maximum de la concentration de contrainte qui s'exerce sur les fibres, dans la feuille. Le facteur de concentration de la contrainte peut être calculé sur le contour du pore équivalent de forme elliptique en fonction de son ellipticité et de la direction de la traction exercée sur la feuille. Les fibres qui initialisent la rupture de la feuille sont celles dont la direction correspond à la contrainte maximale sur le contour du pore équivalent.

Différents papiers caractérisés par l'ellipticité de leur pore équivalent ont été étudiés pour différentes directions de traction, suivant et hors les axes principaux de la feuille, jusqu'à leur rupture et pour différentes conditions de séchage de la feuille, à retrait nul ou à retrait libre. Les mesures ont confirmé le mode de raisonnement ci-dessus pour évaluer la résistance à la rupture du papier. La distribution elliptique de la répartition des fibres dans la feuille est validée ainsi que l'analyse de la traction en modélisant la répartition des contraintes dans la feuille sur le contour de son pore équivalent de forme elliptique. Il est possible ainsi de prévoir les conditions de fabrication du papier qui permettent d'optimiser la structure de la feuille en fonction de ses transformations ultérieures et de ses différents usages.

Les micros espaces formés entre les fibres et les irrégularités locales de la masse volumique de la feuille donnent au papier un aspect nuageux lorsqu'il est observé par transparence, voir par exemple la figure 20. La configuration du pore équivalent évaluée par un échantillonnage au hasard, de fibres colorées espacées de place en place à des distances supérieures au centimètre dans la feuille, est la même en moyenne que la configuration du pore équivalent évaluée à l'intérieur des grains de fibres qui forment les flocs à l'échelle de quelques millimètres : voir par exemple la distribution en orientation des fibres sur la figure 21. Cette autosimilarité tend à montrer qu'un même principe organisationnel forme la feuille de papier à ses différentes échelles au cours de sa fabrication, une hypothèse qui est développée au chapitre IV ci après.

III-4. Caractérisation de la structure des alliages métallurgiques et des matériaux à texture alvéolaire.

Des matériaux dont la texture n'est pas nécessairement fibreuse peuvent être analysés suivant le concept du pore équivalent. C'est le cas par exemple des alliages métalliques dont la texture est amorphe. Une caractéristique de ces matériaux est la géométrie des joints de grains qui est évaluée par la stéréométrie sur des coupes tomographiques après leur polissage. Il est souvent constaté que la figure qui est la mieux adaptée pour représenter le diagramme polaire de la longueur des interceptes entre les interfaces des grains est une ellipse ou une configuration qu'il est possible de déconvoluer en une somme pondérée d'ellipses homocentriques.
Un autre cas est celui de la caractérisation des mousses de polymères souples ou rigides utilisées dans le conditionnement des matériaux et dans l'habitat compte tenu de leurs propriétés mécaniques et d'isolation, de même que la caractérisation de la texture des mousses métalliques utilisées en tant que support catalytique dans les réactions électrochimiques ou pour la filtration de suspensions d'éléments pigmentaires dans des fluides. La texture de ces matériaux s'analyse sur les images de coupes tomographiques observées en microscopie électronique. Leur texture est souvent anisotrope dans les trois dimensions. La figure la mieux adaptée pour la représentation du diagramme polaire de la longueur moyenne des interceptes entre les interfaces dans ces coupes, que les pores soient fermés ou ouverts, est une ellipse dont les directions des axes et l'ellipticité dépendent de la direction des plans de coupes. A partir des différentes coupes il est possible d'obtenir par le calcul une forme moyenne des pores de la mousse et les caractéristiques du pore équivalent ellipsoïdal ainsi que son orientation par rapport aux axes principaux de la texture.

III-5. Caractérisation des irrégularités de la surface des matériaux .

Pour certains matériaux on cherche à caractériser la topographie de leur surface, un paramètre généralement dénommée : rugosité. Cette propriété est importante pour l'enduction des papiers, la pénétration des encres dans les supports au cours de leur impression, la lubrification et l'écoulement en parois des fluides, la glisse des matériaux sur un élément fixe par exemple sur la neige dans le cas des skis, la friction au contact du touché tactile de la peau, l'adhésion sur différents types de supports, les propriétés optiques de brillant, de lustre et de rétro diffusion de la lumière par les surfaces.
La rugosité peut se caractériser à partir de relevés topographiques effectués par des moyens mécaniques tactiles ou optiques sans contact. Ces relevés profilométriques correspondent à des coupes planes transversales du volume constitué par les creux et les bosses qui sont distribués à la surface, le plus souvent de manière aléatoire. L'analyse de ce volume par la stéréologie, interprétée suivant le concept du pore équivalent permet de caractériser la rugosité. La *surface conforme équivalente* qui est la terminologie utilisée en ce cas permet de quantifier de manière globale l'espace qui réalise la transition entre le milieu extérieur et le matériau.

Des éléments caractéristiques de cette interface sont: la porosité, la surface spécifique, la taille et la forme moyenne des creux et des bosses, la distribution des pentes des perturbations géométriques, l'anisotropie de la répartition en orientation des micros interfaces. Ces paramètres conditionnent le degré de lissé et le lustre de la surface de la feuille de papier après son calandrage, l'adhésion des films de polymères, l'état de surface des pièces métalliques en vue de leur assemblage, la friction et la lubrification à la surface des matériaux, le degré de brillant des revêtements ou leur matité par exemple pour l'aspect des carrosseries des véhicules et la rugosité des tôles métalliques après leur sablage ou leur grenaillage.

La réflexion d'un rayonnement électromagnétique sur un ensemble d'éléments qui présente des irrégularités de sa surface peut être modélisée en étudiant la réflexion d'un faisceau de rayonnement incident sur une surface lisse gauche et dont la courbure modélise la distribution de la densité de probabilité en orientation des micro facettes des éléments à la surface de l'ensemble. Un ellipsoïde convient théoriquement pour cette modélisation et permet de quantifier les irrégularités de la surface en fonction des mesures de réflexion du rayonnement. Des observations effectuées par voie aérienne de la surface de la canopée d'espaces forestiers ont été interprétées en fonction de modèles ellipsoïdaux ce qui a permis de différencier les espèces végétales.

III-6. Le concept de pore équivalent comparé aux analyses stéréométriques en pétrographie.

Un domaine d'étude proche de celui de l'analyse de la texture du papier est la pétrographie en géologie ou l'on cherche à caractériser la texture des roches et des sols à l'échelle microscopique dans le but de reconstituer les mouvements des grandes structures géologiques qui ont influencé leur morphogenèse. Les caractéristiques des roches et des sols sont la forme et la taille des poly cristaux minéraux ainsi que la configuration de leur réseau inter granulaire. Ces éléments sont identifiés par l'analyse stéréologique sur des coupes d'échantillons polies et éventuellement consolidées par des injections de résine polymérisée dans le cas des roches poreuses : les grès et les roches magasins pétrolifères.

La forme sphéroïdale des éléments granulaires qui sont dispersés ou agrégés dans les ensembles minéraux, privilégie les mesures des interceptes linéiques ainsi que les mesures planaires qui sont reliées par des équations stéréologiques à des valeurs fonctionnelles des particules en tant que valeurs globales et moyennées de leur volume, de l'aire de leur surface et de leur hauteur moyenne projetée, appelée épaisseur moyenne. Dès lors le choix d'une forme canonique telle celle d'un polyèdre ou d'un ellipsoïde en mesure de caractériser la forme moyenne des particules permet de définir les paramètres de forme et de taille des particules et leur orientation moyenne dans la texture de la roche. L'ellipsoïde triaxial est souvent choisi comme forme canonique pour les particules car sa configuration permet à partir de ses ellipticités de modéliser des formes variées évoluant à la limite entre celles d'un cylindre ou d'un disque avec pour configurations intermédiaires les ellipsoïdes oblongs allongés, la sphère et les ellipsoïdes aplatis. La forme ellipsoïdale des inclusions dans un matériaux polyphasé permet par ailleurs le calcul de ses propriétés de champ ce qui renforce l'intérêt de l'homogénéisation des phases particulaires par un ellipsoïde de forme moyenne. De nombreuses études pétrographiques ont confirmé l'intérêt de l'ellipsoïde en tant que forme canonique des particules.

Le concept du pore équivalent a été développé à l'origine pour l'étude des matériaux à base de fibres et pour lesquels il est plus facile de discrétiser l'orientation de leurs éléments que leur forme granulaire comme c'est le cas pour un échantillon minéralogique. L'équivalence de la densité de probabilité en orientation des éléments, pondérée par leur longueur ou l'aire de leur surface, d'avec le rayon de courbure d'une figure géométrique, qui est la base du concept du pore équivalent, permet ainsi de caractériser la texture du matériau et ses propriétés d'un point de vue global.

Dans l'analyse pétrographique les paramètres discriminants sont : le volume, l'aire spécifique, l'épaisseur moyennée de la phase particulaire. Ces paramètres permettent de définir la forme des éléments granulaires mais non celle de la texture de la roche considérée dans son ensemble.

Les deux méthodes d'analyse : pétrographie et pore équivalent, sont adaptées chacune à leur domaine d'application. Elles peuvent converger dans certains cas car elles ont en commun l'utilisation de relations stéréologiques adaptées à l'étude des ensembles d'objets.

# IV

## L'écoulement des fluides dans les milieux poreux

Le concept du pore équivalent peut être appliqué à l'étude des écoulements des fluides dans les milieux poreux dont la texture est désordonnée aléatoire. Dans ces milieux les caractéristiques géométriques des interfaces sont définies de place en place de manière stochastique. Il en est de même pour les déformations élémentaires de la phase fluide dans son déplacement à travers le milieu poreux et il n'est pas possible de prévoir que telle particule du fluide occupera telle place plutôt qu'une autre dans la texture. Ainsi les particules sont indiscernables les unes par rapport aux autres d'un point de vue analytique pour l'étude de l'écoulement.

Suivant le principe de la moindre action[30], de Moreau de Maupertuis, les déplacements du fluide d'un point à un autre dans le milieu poreux, s'effectuent de manière telle que l'action qui est l'accumulation du produit du déplacement des particules par leur quantité de mouvement au niveau élémentaire, soit minimale, ou ce qui revient au même, que la variation d'énergie cinétique des particules soit minimale pendant la durée de leur déplacement. Les déplacements de la phase fluide au cours de l'écoulement dans le milieu poreux sont irréversibles étant donnée leur caractère stochastique au niveau microscopique et s'effectuent donc avec une augmentation du désordre des particules dans l'ensemble : il y a augmentation de l'entropie du fluide[31].

Ces considérations générales, sont celles des écoulements des fluides dans la texture fibreuse des papiers, des cartons, des textiles non tissés et des feutres pendant leur utilisation ainsi que au cours de la mise en forme de ces matériaux sous l'effet de contraintes de différentes natures : mécanique, hydromécanique, gravitationnelle. Des analogies existent au point de vue formel entre les lois d'écoulement des fluides dans les milieux poreux, en régime stationnaire relativement lent, et les lois de transfert d'énergie : mécanique, électrique, électromagnétique et de propagation de la chaleur, d'où le double intérêt de l'étude des lois d'écoulement des fluides dans les milieux poreux[32].

IV-1. L'écoulement au niveau macroscopique.

L'étendue du domaine analysé dans le milieu poreux peut être délimité par celui de son volume élémentaire représentatif dont la surface du pore équivalent est une représentation statistiquement moyennée et conforme de celle des interfaces qui canalisent le fluide à travers la texture. Au niveau macroscopique l'étude du transfert du fluide dans le milieu poreux revient à analyser l'écoulement de filets de fluide relativement à la surface de son pore équivalent.

De manière générale le mouvement dans un petit intervalle de temps d'un élément de volume du fluide en écoulement se décompose au premier ordre en une translation et une rotation qui correspondent à un *déplacement* en bloc de l'élément de volume ainsi qu'en une *déformation* qui correspond à la décomposition de l'élément de volume en éléments plus petits tels que chacun d'eux se déplace relativement par rapport aux autres sous l'effet des contraintes imposées au fluide[33]. Dans le cas d'un fluide incompressible la divergence de la vitesse est nulle et la déformation s'effectue sans changement du volume de l'élément.

Etant données l'homogénéité de la texture poreuse et l'indiscernabilité des particules fluides due à leurs mouvements stochastiques on doit conclure que les déplacements et les déformations des éléments de volume sont, une fois moyennés, paramétrés de manière identique en chaque point de la surface du pore equivalent[34].

Ainsi pour chaque élément de volume du fluide sur la surface du pore équivalent la direction de l'axe de sa translation qui est également la direction de l'axe de sa rotation, le rayon de giration et l'amplitude de sa rotation, les taux de sa déformation sont dans une même durée simultanément les mêmes en tous les points de la surface du pore équivalent. L'invariance de ces paramètres caractérise leur appartenance à un groupe[35], spécifique du mouvement du fluide. La question est : quelle configuration du pore équivalent est compatible avec un groupe spécifique du mouvement des éléments du fluide sur sa surface?

IV-2. L'ellipsoïde, forme canonique optimale du pore équivalent des milieux poreux à texture désordonnée aléatoire .

Dans l'écoulement permanent et conservatif d'un volume de fluide incompressible, à débit constant la vitesse moyenne du débit varie à l'inverse de l'aire de la section droite qui conduit le fluide. Pour un périmètre donné, mouillé par le fluide, une configuration géométrique minimisant la vitesse de l'écoulement correspond à des aires conductrices délimitées par des couronnes circulaires cerclant la surface du pore équivalent[36]. D'autre part l'optimum d'efficacité pour la vitesse, permettant de minimiser l'action dans le déplacement du fluide, est réalisé lorsque la vitesse a une direction perpendiculaire aux sections de la veine du fluide en écoulement. Ainsi pour réaliser un tube de courant de forme optimale pour le fluide, la surface du pore équivalent doit pouvoir *être cerclée* et *tapissée* par un réseau de lignes de courant perpendiculaires à des cercles[37].

La surface d'un ellipsoïde peut être cerclée par les intersections de sa surface d'avec des plans parallèles et de direction perpendiculaire aux normales en ses points ombilics. Les centres des cercles sont alignés sur le diamètre qui joint deux points ombilics situés en position antipodale sur l'ellipsoïde. Il existe deux familles de ces sections circulaires situées symétriquement par rapport au plan principal de moyenne section de l'ellipsoïde. La sphère qui est un ellipsoïde dégénéré a une infinité de ces familles de plans de section circulaire[38].

Etant données ses propriétés géométriques l'ellipsoïde parait donc être un candidat potentiel à privilégier pour la surface du pore équivalent d'un milieu poreux. Effectivement l'intersection d'un ellipsoïde de forme générale tri axes et d'un groupe de cylindres droits de base elliptique avec une génératrice commune, axe neutre de symétrie du groupe, permet de modéliser au mieux l'écoulement dans les conditions d'une dépense d'énergie minimale. Une configuration qui satisfait cette condition est celle où chaque cylindre du groupe est tangent à la surface de l'ellipsoïde en son intérieur le long du grand cercle d'intersection de la surface de l'ellipsoïde et de son plan diamétral, voir les figures 14 et 15. Dans ces conditions la génératrice commune des cylindres, axe neutre du groupe, est le diamètre qui joint les points ombilics antipodaux de l'ellipsoïde. Cette génératrice commune constitue un axe de symétrie pour le groupe, chaque cylindre étant apparié avec un cylindre symétrique du premier par rapport à cet axe. Pour un ellipsoïde donné il existe deux groupes de cylindres qui réalisent ces conditions, chacun d'eux ayant comme axe neutre le diamètre de l'ellipsoïde qui joint les points ombilics antipodaux: voir la figure 16. Un seul de ces groupes est décrit ci après pour l'analyse de ses propriétés.

Compte tenu de la configuration géométrique réalisée par l'union de la surface de l'ellipsoïde et de chacun des cylindres elliptiques, en chaque point de leur intersection, les plans de section circulaire de ces surfaces ont une même direction perpendiculaire à la normale aux points ombilics. Ces plans confondus rendent conjointement optimal l'écoulement d'un fluide à travers ces sections. Ainsi la direction de *la normale* aux points ombilics de l'ellipsoïde est *en tous les points de la surface du pore équivalent* : la direction du déplacement avec une moindre action en *translation* des éléments de volume du fluide ainsi que l'axe de leur *rotation* avec un rayon de giration constant égal au rayon de la section circulaire oblique des cylindres elliptiques. La rotation couplée avec la translation réalise un mouvement hélicoïdal pour les éléments de volume, avec une vitesse de rotation définie en fonction d'une variable temporelle. Cette variable évolue suivant un axe dont la direction est celle de la normale aux points ombilics, dans un intervalle délimité par le diamètre de l'ellipsoïde entre ses points ombilics. Pendant son déplacement l'élément du fluide est contraint sur la surface du pore équivalent par une *déformation* qui équilibre les éléments du volume par rapport au plan de section droite des cylindres elliptiques.

Les arcs des courbes d'intersection de l'ellipsoïde et des cylindres elliptiques ont les propriétés de symétrie de chacune de ces surfaces. A chaque élément d'arc correspond un élément d'arc antisymétrique par rapport au centre de l'ellipsoïde et situé sur un cylindre symétrique du premier par rapport à l'axe neutre du groupe, diamètre des points ombilics. L'ensemble des courbes d'intersection lorsqu'elles sont étendues de part et d'autre de leur tracé constitue l'enveloppe modélisée et statistiquement moyennée des interfaces mouillées par le fluide d'un point de vue global. Cette enveloppe ellipsoïdale est définie en fonction d'un critère de dépense d'énergie minimale pour l'écoulement du fluide sur sa surface. Les points ombilics qui sont des points communs à l'ensemble des courbes d'intersection constituent des nœuds pour le réseau des trajectoires modélisées du fluide. La probabilité de présence des particules est de : 100 % en ces points et leur vitesse également probable dans toutes les directions dans le plan tangent à la surface du pore équivalent[39].

Suivant cette modélisation de l'écoulement, le fluide effectue d'un point de vue global dans la texture du milieu poreux un déplacement hélicoïdal de *pénétration* accompagné d'une *déformation*. Ce mouvement est par exemple celui qui est impulsé à un mélange de particules lorsqu'elles sont fractionnées par tamisage en fonction de leur taille. L'opération est réalisée par l'écoulement du mélange à travers des tamis constitués de toiles calibrées, disposées dans des plans parallèles successifs. L'ensemble des tamis est animé d'un mouvement de rotation dans leur plan ainsi que des impulsions transversales qui correspondent à des secousses dans la direction perpendiculaire au plan de la rotation.

Un autre exemple est celui de la formation de la feuille de papier sur les machines dites « à table plate ». Sur ces machines la suspension des fibres est mise en dépression sur une toile de filtration horizontale et simultanément soumise à un cisaillement dans l'épaisseur de la suspension et parfois à un branlement transversal de la toile filtrante dans la direction perpendiculaire au sens de la marche de la toile filtrante. L'effet combiné de ces mouvements facilite l'égouttage de l'eau à travers la feuille tout en contrôlant l'anisotropie de la texture fibreuse [40].

Quatre siècles avant notre ère Démocrite estimait « qu'il existe un seul genre de mouvement : le mouvement par secousse, qui s'exerce en tous sens », un concept repris par Epicure (341-271 av. J.-C.) qui écrivait « qu'il faut pour que le monde et les êtres soient engendrés, que se produise aussi spontanément une légère déviation latérale » [41].

IV-3. Les trajectoires ellipsoïdo-cylindriques: formalisme et propriétés.

L'intersection de l'ellipsoïde et de chacun des cylindres elliptiques du groupe, est une courbe cyclique, ellipsoïdo-cylindrique [42], une boucle en forme de huit avec un point double situé au point de tangence des deux surfaces, voir les figures 14 et 15. Cette courbe peut être représentée de manière unicursale en fonction d'une variable, u, qui est un angle dont le sommet est au centre de l'ellipsoïde et qui décrit un intervalle de $2|\pi|$ radians dans le plan diamétral d'intersection circulaire de l'ellipsoïde et de chacun des cylindres du groupe. Dans un repère d'axes orthonormés, Ox, Oy, Oz, les coordonnées d'un point de la courbe sont :

$$x = - a\,(\cos u \sin(u+w) \sin v - \sin u \cos v)$$
$$y = b \cos u \cos(u+w)$$
$$z = c\,(\cos u \sin(u+w) \cos v + \sin u \sin v), \quad \text{équations (4)}$$

avec pour paramètres :
   $a > b > c$, les demi axes principaux de l'ellipsoïde centré sur l'origine des axes et de directions celles des axes du repère,
   $\cos v = ((a^2 - b^2)/(a^2 - c^2))^{1/2}$ ; $\sin v = ((b^2 - c^2)/(a^2 - c^2))^{1/2}$,

w, un angle dont le sommet est au centre de l'ellipsoïde et situé dans le plan diamétral d'intersection circulaire de l'ellipsoïde, avec une valeur choisie dans l'intervalle de 0 à 2 π radians. Le paramètre, w, positionne la courbe sur l'ellipsoïde, le point double de la courbe étant choisi comme origine de la variable, u, sur le grand cercle d'intersection de l'ellipsoïde. L'origine des variations du paramètre, w, est le point de coordonnées : x = 0, y = b, z = 0, intersection de l'axe Oy avec l'ellipsoïde, voir la figure 15.

En incrémentant l'angle, w, d'une quantité 2|π /n| dans l'intervalle 0 à 2π, la valeur de, n, étant choisie aussi grande que l'on veut, l'ensemble des courbes recouvre la surface de l'ellipsoïde, voir par exemple la figure 17. L'ellipsoïde tri axes peut ainsi être construit d'une manière générale par une tapisserie de la courbe définie par les équations (4).

L'échange des valeurs de : u, et w, respectivement par les valeurs : u + π, et w + π, conserve les valeurs de chacune des coordonnées, x, y, z, tout en changeant leur signe. Les courbes appariées par cette transposition sont antisymétriques par rapport à l'origine des axes, une propriété qui résulte de la centrosymétrie de l'ellipsoïde comme il a été remarqué plus haut.

Ainsi le tracé d'une courbe à partir d'un point ombilic permet de revenir à son point de départ en suivant la branche de la courbe antisymétrique, à partir du point ombilic antipodal et en inversant le sens de rotation de la variable, u. Ce tracé effectué par deux demi boucles ellipsoïdo-cylindriques, appariées, réalise une boucle fermée sur la surface de l'ellipsoïde sans qu'il y ait de point double. Un exemple d'application de ce tracé est présenté au chapitre V, pour le décryptement de la figure du tai-chi, symbole de l'équipartition de l'énergie vitale dans la philosophie bouddhiste Taoïste.

Le périmètre de la courbe ellipsoïdo-cylindrique ne peut s'évaluer analytiquement compte tenu de l'intégration de la fonction elliptique qui figure dans son expression. La longueur de la courbe s'évalue numériquement à partir des intégrales des termes de son développement en série. La précision de cette évaluation dépend des variations de la courbure du tracé en fonction de son positionnement sur l'ellipsoïde. Dans un intervalle de variations de 2π radians du paramètre, w, et avec un pas permettant de réaliser une tapisserie fine de la surface de l'ellipsoïde, dans le cas de valeurs des ellipticités : a/b = b/c = 2,0, on constate que les variations de longueur de la courbe ne dépassent pas : ± 0,8%, de part et d'autre de sa valeur moyenne en fonction des variations du paramètre, w. Les écarts observés ont une période de π qui correspond aux propriétés de symétrie de ces courbes sur l'ellipsoïde.

Ces variations de la longueur de la courbe ellipsoïdo-cylindrique sont au demeurant faibles au regard des anisotropies élevées de l'ellipsoïde précité et en comparaison desquelles les anisotropies des structures matérielles rencontrées dans la pratique sont généralement inférieures Dans ces conditions on peut estimer que la longueur de la courbe ellipsoïdo-cylindrique, représentative des équations (4), est *quasi stationnaire* quelque soit son positionnement sur la surface de l'ellipsoïde[43]. L'ensemble de ces courbes forme ainsi un faisceau de lignes *homotopes* qui tapissent la surface de l'ellipsoïde, d'un point ombilic à l'autre en position antipodale.

La rotation de la variable, u, peut s'effectuer avec une probabilité égale dans un sens ou dans l'autre. Ainsi il est possible de regrouper les courbes aux points ombilics en doublets appariés dans lesquels la rotation peut être inversée d'une courbe par rapport à l'autre ce qui annule d'un point de vue global la résultante des moments cinétiques de points matériels parcourant les courbes en synchronisme.

IV-4. L'irrotationalité de l'écoulement au niveau macroscopique.

Les conséquences de la stationnarité de la longueur des courbes ellipsoïdo-cylindriques et de leurs propriétés de symétrie sont importantes à la fois du point de vue théorique et pratique. Dans le cas de l'écoulement permanent et conservatif d'un fluide incompressible ces courbes peuvent être considérées comme modélisant les lignes de courant probables du fluide, reconstituées d'un point de vue virtuel au niveau macroscopique de la texture. Ces courbes tapissent la totalité des interfaces des pores mouillées par le fluide et sont discrétisées en fonction de la moindre action pour l'écoulement [44].

Les points communs aux courbes sont les points ombilics antipodaux de l'ellipsoïde et représentent l'entrée et la sortie du fluide dans la texture poreuse modélisée. La durée du trajet du fluide est fonction du diamètre de l'ellipsoïde entre ces points. La *longueur du trajet et la durée* étant les mêmes pour les différentes lignes de courant, la vitesse moyenne l'est également, ce qui réalise l'équilibre dynamique du fluide en écoulement. L'action suivant les différentes lignes de courant étant la même sa valeur est stationnaire, minimale conformément aux hypothèses.

Le mouvement du fluide s'effectuant avec des rotations inversées dans chacune des lignes de courant appariées en doublet, la rotation de la vitesse moyenne du fluide est nulle d'un point de vue global. L'écoulement étant permanent et conservatif et le fluide incompressible, la divergence de la vitesse moyenne est nulle. Ainsi à rotationel nul la vitesse moyenne du fluide est proportionnelle négativement au gradient du potentiel qui est la pression motrice exercée sur le fluide.

L'ensemble des courbes ellipsoïdo-cylindriques constitue ainsi sur la surface du pore équivalent un faisceau divergeant et convergeant radialement aux points ombilics qui représentent dans cette modélisation les points source du fluide à son entrée et de drainage à sa sortie. Le faisceau des trajectoires, qui sont les lignes de courant du fluide, coupe les cercles centrés sur le diamètre des ombilics de l'ellipsoïde, chacun des cercles étant dans un plan équipotentiel perpendiculaire à la normale aux points ombilics, voir la figure 17. La vitesse de déplacement du fluide suivant les trajectoires se décompose en une composante normale aux plans équipotentiels et une composante de rotation dans ces plans ainsi qu'à des composantes dues à la déformation de l'élément de volume. La direction du mouvement de pénétration du fluide dans la texture est celle de la normale aux points ombilics. La direction de la résultante des mouvements est la direction du diamètre des points ombilics, images des points d'entrée et de sortie du fluide dans le milieu poreux modélisé.

Cette analyse est en accord avec les propriétés des écoulements de fluides à travers une texture poreuse et qui satisfont à la loi de Darcy [45]. Les courbes ellipsoïdo-cylindriques qui modélisent les lignes de courant du fluide dans la texture sont homotopes au mieux sur la surface du pore équivalent. Ainsi l'écoulement peut être considéré comme laminaire, avec une faible valeur du nombre de Reynolds compte tenu de la petite valeur du rayon hydraulique moyen, due à la grande extension de la surface spécifique du milieu poreux.

Pour une viscosité donnée la vitesse moyenne de débit du fluide est proportionnelle négativement au gradient de la pression motrice. Le coefficient de proportionnalité est le facteur de perméabilité, K, dont la dimension est celle du carré d'une longueur [46].

Ce facteur, K, dépend de la porosité de la texture, du carré de la taille des pores égale au rayon hydraulique moyen, de la forme des pores caractérisée par un coefficient, ko = 2χ, où, χ, est le facteur de circularité des pores, ainsi que du facteur de tortuosité, t, posé égal au rapport du trajet effectif du fluide à travers la texture poreuse à son trajet apparent évalué entre l'entrée et la sortie du milieu poreux. Différentes évaluations du facteur de tortuosité ont été proposées. Carman et Kozeny, par exemple, regroupent les deux termes : t, et, ko, en un paramètre unique : $k = ko \cdot t^2$, dont la valeur expérimentale est en moyenne égale à : 4,5 ± 1.

Dans la modélisation de l'écoulement suivant le concept du pore équivalent, le facteur de tortuosité, t, comprend deux termes : l'un intrinsèque à la texture poreuse est uniquement fonction des caractéristiques géométriques du pore équivalent défini par ses ellipticités, l'autre également fonction de la configuration du pore équivalent fait intervenir son orientation par rapport aux axes matériels de la texture. Ce deuxième terme est à l'origine des variations du facteur de tortuosité, t, et donc de la perméabilité, K, suivant la direction de l'évaluation de la perte de charge entre l'entrée et la sortie du fluide dans le milieu poreux.

La valeur théorique du paramètre, $k = ko \cdot t^2$, calculée suivant la modélisation du pore équivalent peut être considérée en moyenne égale à : 5,5, la circularité, χ, des pores étant choisie égale à : 1. Cette valeur du paramètre, k, calculée pour différentes anisotropies et pour différentes directions de l'évaluation de la perte de charge, estimées représentatives des cas probables rencontrés dans la pratique, est en bon accord avec la valeur expérimentale du paramètre, k, de Kozeny-Carman [47], qui est en moyenne : 4,5 ± 1. La modélisation de l'écoulement d'un fluide dans les milieux poreux désordonnés aléatoires, effectuée suivant le concept du pore équivalent, est ainsi validée par les mesures expérimentales de la perméabilité aux fluides.

L'intérêt de cette modélisation est de relier les phénomènes de transfert des fluides dans les milieux poreux, aux caractéristiques de leur texture, un aspect qui est important à la fois du point de vue théorique et pratique. La différenciation des composantes du mouvement du fluide suivant son déplacement par pénétration et rotation et sa déformation par compression et dilatation dans la texture poreuse est utile pour la prévision du drainage des liquides dans les sols et l'extraction des fluides stockés dans les roches magasins. D'autres cas où cette différenciation est importante sont par exemple : l'optimisation des propriétés barrières des matériaux poreux vis à vis des fluides liquides ou gazeux ainsi que les propriétés d'impression des papiers et des cartons dans le but de limiter le transpercement des encres au verso de la feuille et leur bavure à la surface au recto. Les figures 18 et 19 sont des exemples de cette modélisation pour un bon et un mauvais papier d'impression. Le bon papier se différencie du mauvais papier principalement par le taux de cisaillement transversal du fluide dans la texture qui est moindre et par une pénétration dans la feuille légèrement supérieure.

Une recherche effectuée dans le cadre de la Communauté Européenne a été réalisée en concertation par les constructeurs de machine à papier, les fabricants de feutres, les papetiers et des laboratoires de recherches Européens, dans le but d'améliorer le processus d'élimination de l'eau de la feuille sur la machine à papier. L'élimination de l'eau, après la formation de la feuille sur la table de fabrication, s'effectue dans des presses cylindriques pleines ou aspirantes en contact avec des feutres absorbants qui transportent la feuille dans la zone de pincement entre les cylindres. Cette opération a une influence sur la texture de la feuille tant à sa surface qu'à son intérieur. La nature du revêtement des presses, la forme de l'impulsion de pression, la nature des fibres qui entrent dans la composition des feutres, la texture superficielle et interne de ces milieux poreux, doivent être adaptées à la sorte de papier fabriqué, d'où l'intérêt de modéliser l'opération du pressage afin de prévoir l'évolution de la perméabilité de la feuille et des feutres dans la zone de pincement dans le but d'économiser l'énergie au pressage. Un aspect important de cette opération est de limiter le remouillage partiel de la feuille pendant la décompression des différentes structures poreuses à la sortie de la zone de pincement. La caractérisation de la structure de la feuille et de celle des feutres ainsi que leur évolution en compression et en décompression ont été effectuées dans cette étude suivant le concept du pore équivalent pour modéliser leur texture. Ces recherches ont permis une meilleure compréhension des phénomènes d'élimination de l'eau de la feuille par le pressage. Des améliorations du procédé ont été réalisées entraînant une diminution notable de la consommation d'énergie pour la fabrication du papier.

V

**L'équipartition bipolaire des principes du Yin et du Yang**

Lorsque l'ensemble des éléments analysé est isotrope le pore équivalent de l'ensemble est une sphère. La courbe représentative des équations (4) est sphéro-cylindrique et la directrice des cylindres du groupe est un cercle de diamètre égal à la moitié du diamètre de la sphère. Cette courbe a été formulée par Eudoxe de Cnide dans ses travaux en astronomie (vers 360 avant J.-C). Il l'aurait dénommée hippopède, vocable qui en grec ancien désigne l'entrave que l'on met aux pieds d'un cheval pour l'empêcher de s'écarter. Cette courbe découpe sur la sphère une double fenêtre de Viviani, voir la figure 22.

L'hippopède et ses projections ont des propriétés singulières remarquables. Sa projection orthogonale sur *un plan diamétral vertical* de la sphère, perpendiculaire au diamètre qui passe par le point double de tangence du cylindre et de la sphère, est une lemniscate de Gérono. Cette courbe qui forme une boucle fermée, avec un point double centre de symétrie de la figure, a un tracé proche du chiffre 8 qui représente le symbole de l'infini.

La courbe réalisée par un doublet de deux demies hippopèdes développées sur des cylindres appariés en position symétrique par rapport à l'axe du groupe, projetée sur *un plan diamétral horizontal* de la sphère, perpendiculaire à l'axe du groupe, est semblable à la figure du tai-chi, voir la figure 23. Cette figure est le symbole dans la philosophie bouddhiste Taoïste, de la dynamique des interactions de l'énergie vitale, le Qi, qui se concrétise dans deux principes de nature opposée : le Yin lunaire, moite sombre et féminin et le Yang solaire, céleste chaud lumineux et masculin,

La similitude de ces figures résulte de la configuration de l'hippopède dans l'espace. La surface gauche dont le contour s'appuie sur les deux demies hippopèdes appariées en position de symétrie par rapport à l'axe neutre du groupe est une surface réglée, minimale, en forme de paraboloïde hyperbolique qui partage le volume de la sphère en deux parties égales. Cette équipartition projetée dans le plan diamétral horizontal correspond à l'équipartition du Yin et du Yang, réalisée sur la figure du tai-chi en blanc et en noir de part et d'autre d'un tracé délimité par deux demis cercles en position antisymétrique par rapport au centre de cette figure.

Deux points singuliers sont identifiés sur la figure du tai-chi. Ces points sont alignés avec le centre et les points où la courbe des demis cercles est tangente au grand cercle du contour du tai-chi. Ces points situés sur un diamètre de ce grand cercle, à la distance d'un demi rayon du centre, sont les traces des axes des cylindres sur lesquels sont développées dans l'espace les courbes des demies hippopèdes qui vont d'une extrémité à l'autre de l'axe du groupe des cylindres, avec des rotations inversées l'une par rapport à l'autre et un échange de leur courbe à chaque extrémité de l'axe neutre du groupe où leurs tangentes sont isoclines. Dans la figure du tai-chi ces deux points sont identifiés par une teinte complémentaire de celle de la surface sur laquelle ils se détachent, en blanc ou en noir. Cette disposition rappelle la nature opposée des éléments situés de l'autre coté de la surface qui effectue l'équipartition du volume dans l'espace et qui seraient observés de l'autre coté du plan sur lequel s'effectue la projection de la figure du tai-chi.

Un mobile qui parcourt la courbe, dans un sens ou dans l'autre, en suivant l'enchaînement des branches des demies hippopèdes, part du centre de la figure du tai-chi qui correspond dans l'espace au sommet de la sphère sur l'axe neutre du groupe des cylindres, pour revenir à son point de départ en ayant réalisé une *boucle fermée* constituée par une *descension* jusqu'à l'antipode de son point de départ puis une *ascension*, inverse de la descension suivant l'axe du groupe. Le demi parcours réalisé dans l'hémisphère supérieur de cette boucle correspond en projection à la courbe constituée par les demis cercles qui séparent les éléments du Yin et du Yang dans le plan de la figure du tai-chi.

Par des variations, en augmentation ou en diminution, du paramètre, w, dans les équations (4), le tracé des hippopèdes effectue un *déplacement transversal* sur la surface de la sphère ce qui correspond à une rotation dans un sens ou dans l'autre de la courbe du tai-chi par rapport à son centre. Ainsi les propriétés spécifiques des principes du Yin et du Yang peuvent s'échanger de place en place dans l'espace, sans discontinuités et en douceur. Lorsque l'ensemble des éléments est anisotrope l'équipartition est réalisée non plus en s'appuyant sur le tracé de l'hippopède sphéro-cylindrique mais sur des courbes ellipsoïdo-cylindriques définies de manière générale par les équations (4), voir par exemple la figure 24.

L'équipartition réalisée dans l'espace et projetée dans le plan, peut être qualifiée de bipolaire car elle est réalisée par deux configurations géométriques qui sont développées sur deux cylindres appariés en position symétrique par rapport à un axe neutre avec des rotations inversées qui les différencient suivant des principes opposés tels que le Yin négatif et le Yang positif [48].

Ce décryptement de la figure du tai-chi est en accord avec les représentations de Bouddha en majesté et notamment de sa gestuelle, voir par exemple la figure 26. Les postures rituelles du Qi-Gong et du T'aï Chi Chuan, sont à l'image de la variété des formes et de la fluidité avérée des courbes sphéro et ellipsoïdo-cylindriques des équations (4), voir par exemple l'illustration faite figure 25.

# VI

## Conclusions

Une propriété discriminante des ensembles d'objets distribués dans l'espace de manière désordonnée aléatoire est la répartition en orientation de leurs interfaces. Cette répartition s'obtient généralement par l'analyse d'images échantillonnées dans les ensembles. La densité de la probabilité en orientation des interfaces de leur texture pondérée par leur longueur ou leur aire peut s'interpréter en tant que rayon de courbure d'un contour ou d'une surface qui caractérise d'un point de vue statistique la géométrie de la texture dans l'espace en deux ou en trois dimensions. Cette figure que j'ai appelée le pore équivalent dont la forme est le plus souvent elliptique ou ellipsoïdale est semblable à celle du pore moyen défini par la corde moyenne évaluée dans les différentes directions entre les interfaces de la texture.

Suivant ce concept il est possible de modéliser la configuration spatiale des ensembles désordonnés aléatoires, pour étudier leur comportement lorsqu'ils sont sollicités dans des champs de forces, tout en simplifiant l'analyse. Ainsi un phénomène qui se produit dans un ensemble plan peut s'analyser sur le contour linéaire de son pore équivalent et un phénomène qui se produit dans le volume d'un ensemble peut s'analyser sur la surface gauche de son pore équivalent. Ce concept d'analyse a été validé en pratique pour l'étude de matériaux dont la texture est stochastique tels que les papiers et les cartons, les feutres, les voiles textiles non-tissés, les alliages métalliques avec des joints de grains, les mousses de polymères alvéolaires, les roches et les sols, les matériaux ayant différents degrés de rugosité et de lustre à leur surface.

L'ellipse et l'ellipsoïde et des compositions multimodales additives de ces figures, sont les configurations les mieux adaptées pour représenter la forme du pore équivalent ; elles permettent de caractériser la répartition en orientation des interfaces des ensembles de manière satisfaisante. Qu'une loi déterministe qui définit la courbure d'une configuration géométrique elliptique, s'impose pour représenter la distribution probabiliste de l'orientation des interfaces dans les ensembles désordonnés aléatoires est un fait remarquable qui nous interroge. Il est remarqué que la distribution des rayons de courbure d'une ellipse en fonction de leur orientation est proche, jusqu'à des valeurs d'anisotropie moyennement élevées, d'une Gaussienne tronquée, circulaire, ce qui caractérise une répartition des interfaces des ensembles avec une entropie maximale.

L'écoulement des fluides dans les milieux poreux a été modélisé en définissant les trajectoires probabilistes des particules du fluide conformément aux principes de la moindre action de Moreau de Maupertuis et d'entropie statistique maximale selon Boltzmann. Aux faibles valeurs du nombre de Reynolds la laminarité et l'irrotationnalité de l'écoulement ont été établies d'un point de vue global à l'échelle macroscopique, résultats qui sont en accord avec ceux obtenus par d'autres modes de raisonnement.

Le milieu poreux étant homogène et les particules du fluide indiscernables les unes par rapport aux autres compte tenu de leurs échanges stochastiques continuels d'un élément de volume dans un autre on doit conclure que la quantification du mouvement des particules du fluide est identique en chaque point de la surface du pore équivalent qui est statistiquement conforme à l'espace tangentielle des interfaces dans la texture. Le mouvement des particules du fluide peut ainsi être quantifié de manière générale sur la surface d'un groupe de cylindres elliptiques dont les intersections d'avec la surface du pore équivalent définissent les trajectoires probabilistes des particules d'un point de vue virtuel et global. Dans une configuration particulière de l'union du pore équivalent ellipsoïdal et du groupe de cylindres, ces trajectoires ont une longueur sensiblement stationnaire pour aller d'un point à un autre en position antipodale et correspondent à la même action, minimale, qui est mise en œuvre par le fluide dans son mouvement. Les trajectoires tapissent l'ellipsoïde par un faisceau de lacets, en boucles ouvertes ou fermées suivant les appariements qui sont possibles en leurs points de regroupement et de tangence isocline.

Dans le cas ou l'ensemble des éléments est isotrope les trajectoires probabilistes et modélisées des particules du fluide en mouvement dans la texture sont des hippopèdes. Cette courbe sphéro-cylindrique permet de construire par sa projection plane la figure du tai-chi qui symbolise dans la philosophie bouddhique Taoïste l'équipartition de l'énergie vitale suivant deux principes opposés : le Yin et le Yang.

Dans l'introduction de son ouvrage sur le calcul des probabilités, Henri Poincaré a écrit: « Les anciens distinguaient les phénomènes qui semblaient obéir à des lois harmonieuses, établies une fois pour toutes, et ceux qu'ils attribuaient au hasard, c'étaient ceux qu'on ne pouvait prévoir parce qu'ils étaient rebelles à toute loi. Dans chaque domaine, les lois précises ne décidaient pas de tout, elles traçaient seulement les limites entre lesquelles il était permis au hasard de se mouvoir ».

Dans cette étude des ensembles désordonnés aléatoires il est envisagé un lien entre ces deux domaines. Le hasard est la variable d'ajustement qui est nécessaire et suffisante pour rendre compatible le comportement de ces ensembles à la fois suivant des lois : déterministes d'une part, conformément au principe de la moindre action et probabilistes d'autre part, avec une entropie maximale dans les conditions de leurs sollicitations.

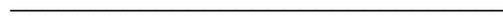

**Figures 1 à 26**

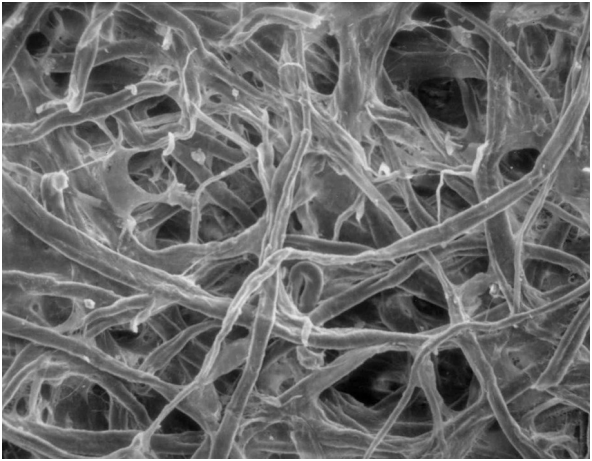

Fig. 1: *Papier buvard, vue en plan, G=80X.*

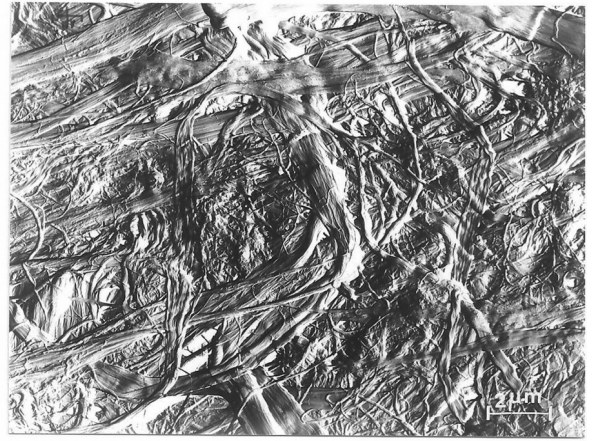

Fig. 2 : *Papier pour condensateur, vue en plan.*

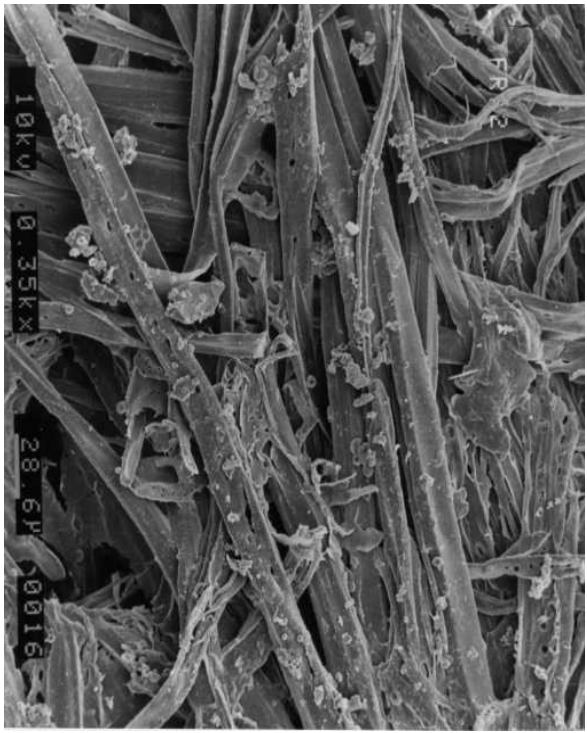

Fig.3 : *Papier du nid de guêpes, vue en plan*

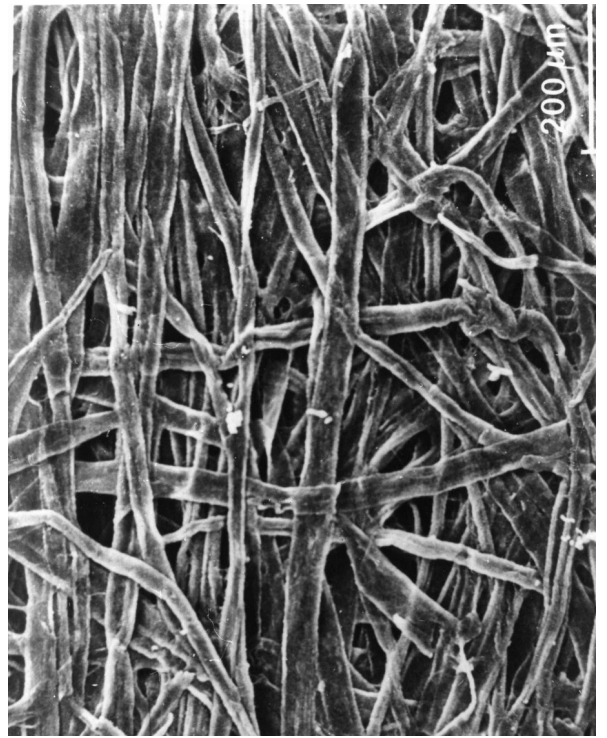

Fig.4 : *Papier d'emballage, vue en plan.*

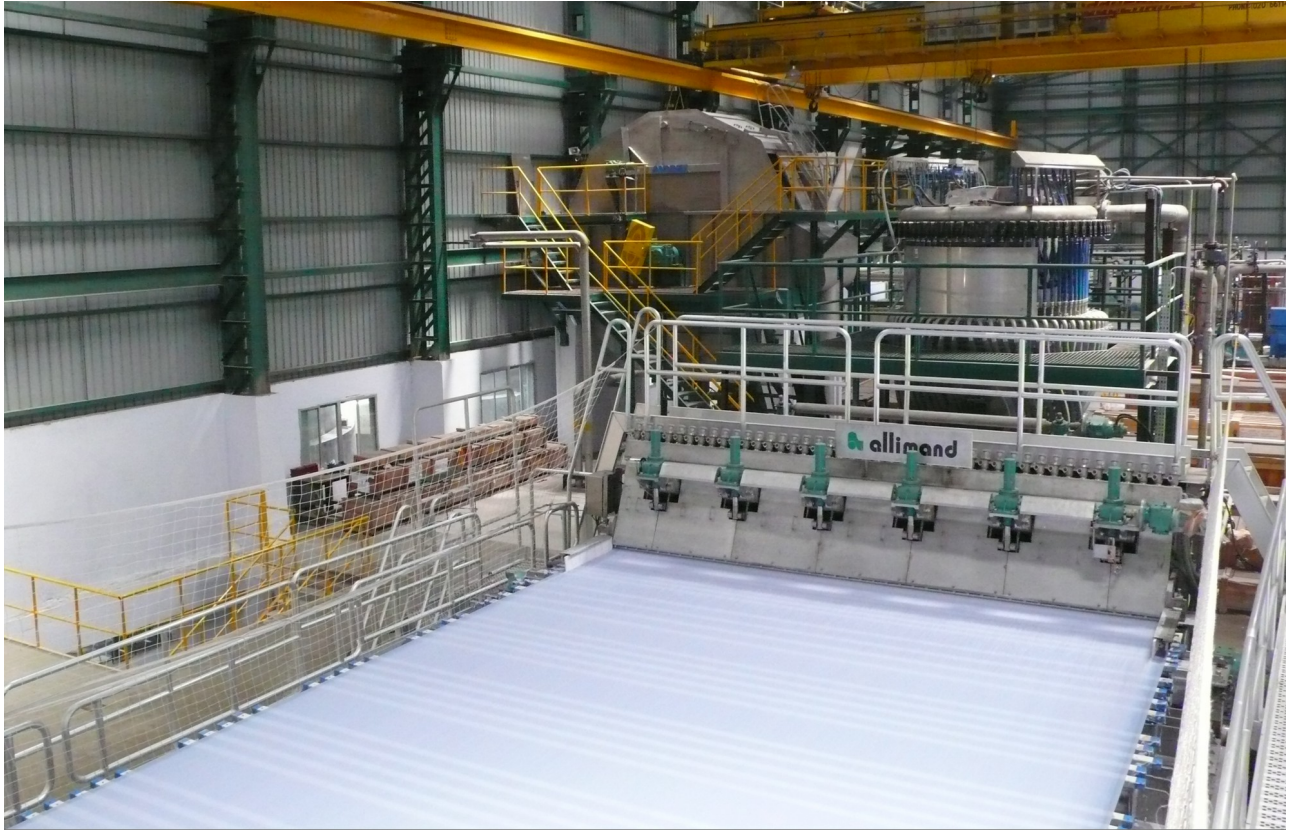

Figure 5. Formation de la feuille d'un papier d'impression à base de pâte de paille et de plantes annuelles, vitesse de production de la machine : 850 m/mn.
Machine à papier du *Groupe Trident* au Punjab en Inde. Concepteur et Constructeur de la machine : *Allimand* à Rives, France.

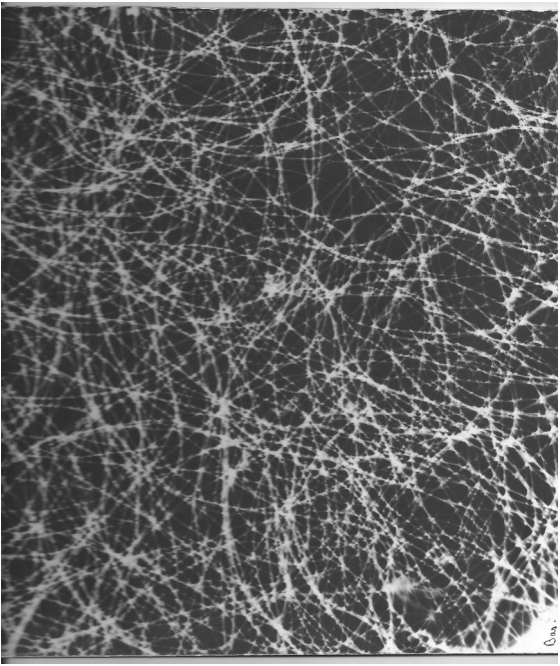

Fig. 6 : Feuille de papier extrêmement mince, de grammage: 2 g/m2, fibres d'alfa, G=24 X.

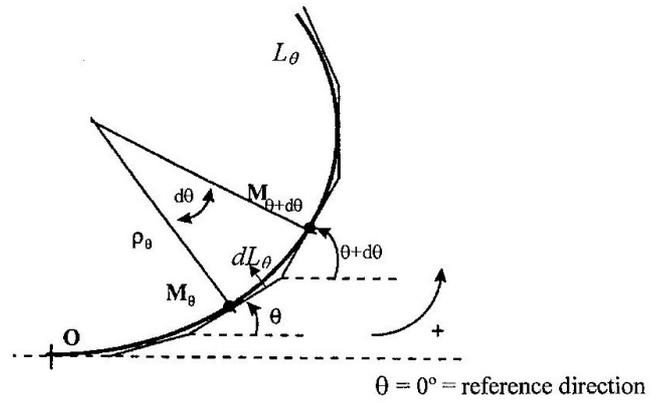

Fig. 7 : Le pore équivalent d'un réseau plan: enveloppe les interfaces $dL_\theta$ rectifiés et hiérarchisés en fonction de leur direction et mis bout à bout à la queue leu leu.

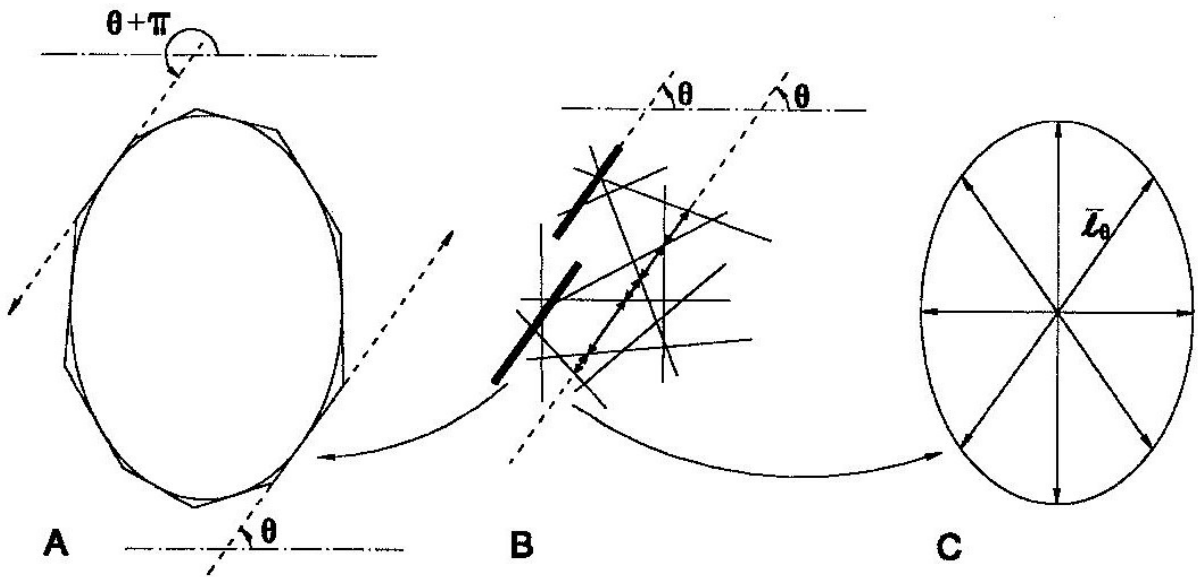

Fig 8: La dualité du pore équivalent elliptique, A, et du pore moyen, C, d'un réseau rectifié dans le plan, B.

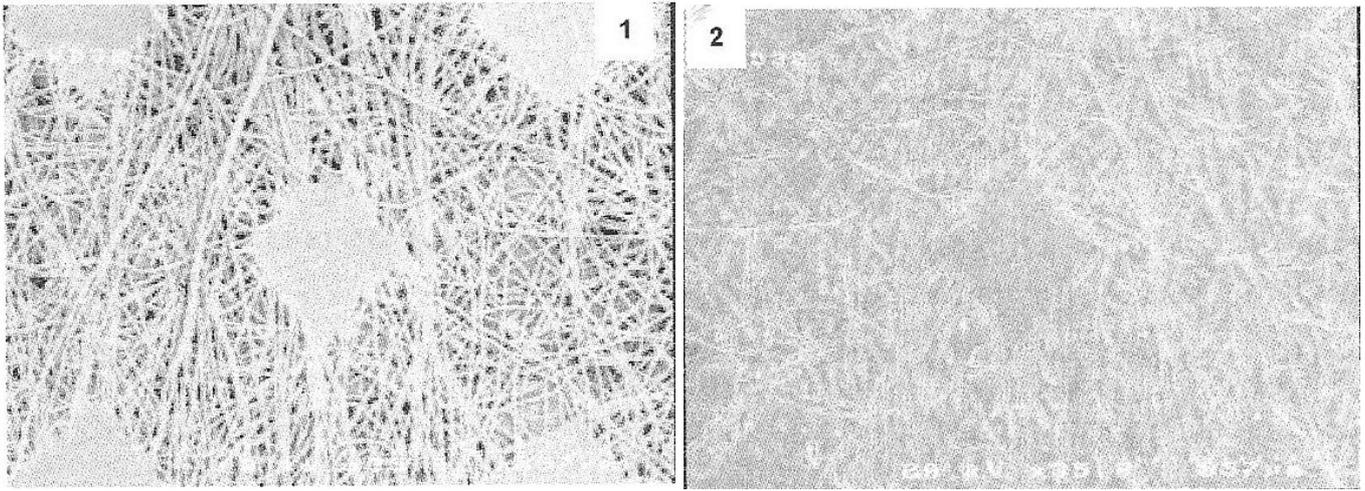

*Fig. 9 : **1,** Image en microscopie électronique de la surface d'un non tissé spunbond, G= 35 X.*
***2,** Réplique de la surface du non tissé fig.9-1 sur un film transparent de polymère thermoplastique.*

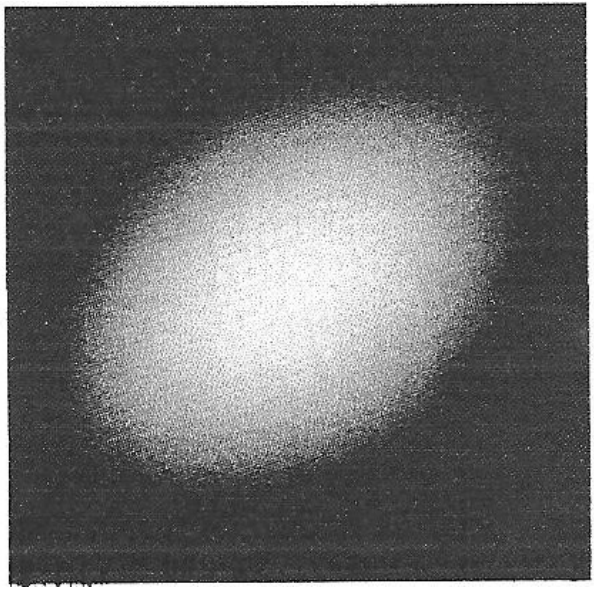

*Fig. 10 : Image de diffraction par transmission d'un faisceau laser de section 10 mm, impactant la réplique transparente de la surface d'un non tissé spunbound (figure 9-2). L'élongation de l'image est la plus marquée perpendiculairement à la direction du maximum d'orientation des fibres.*

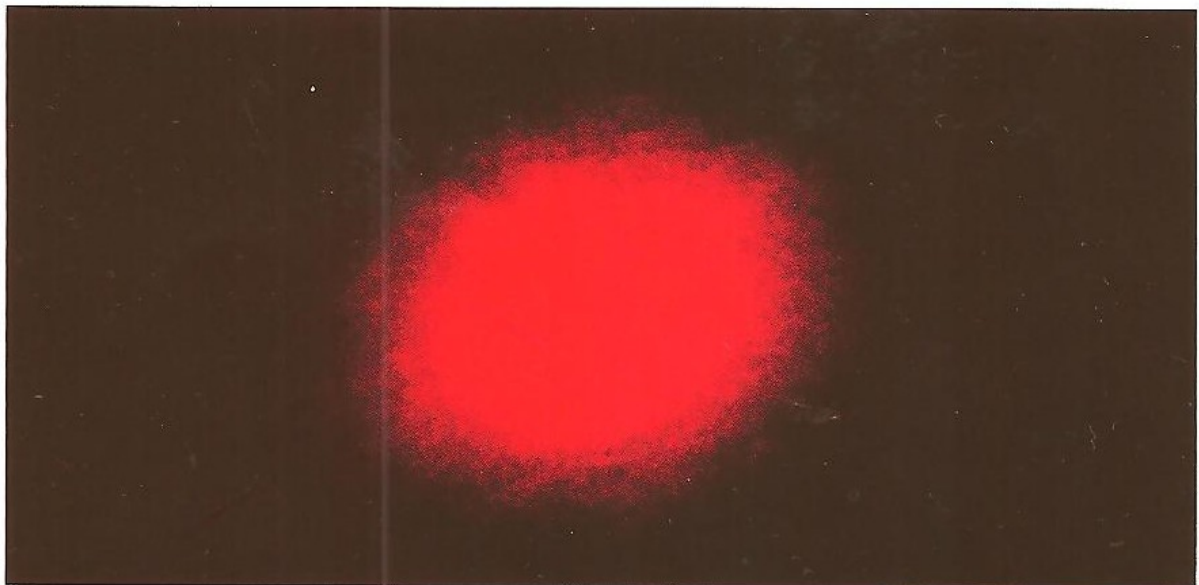

*Fig.11 : Image de diffusion par transmission d'un faisceau laser de section 0,2 mm, impactant une feuille de papier d'emballage, mesure en continu sur machine à papier par le procédé Lippke.*

*L'élongation de l'image est la plus marquée dans la direction du maximum d'orientation des fibres.*

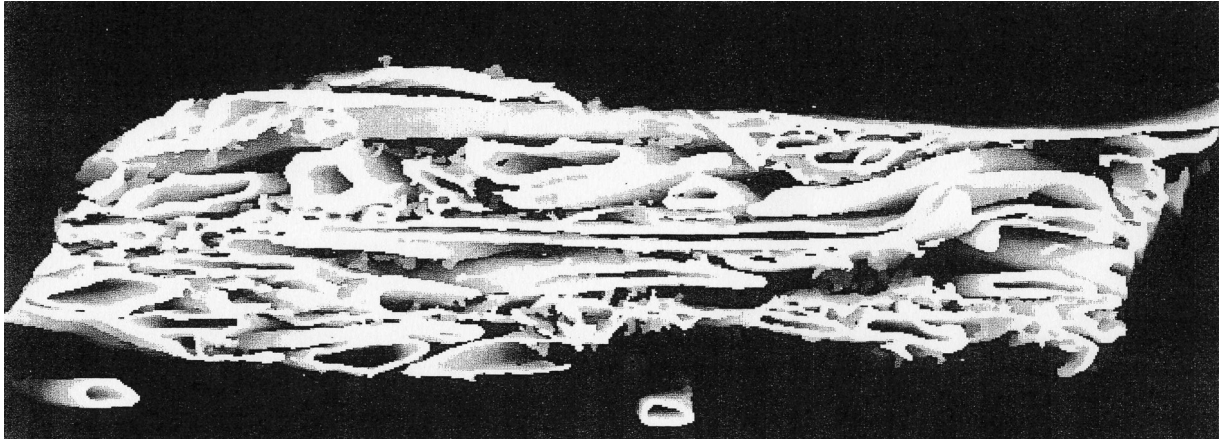

*Fig.12 : Vue d'une coupe transversale d'un papier journal de 45 g/m2, micro tomographie aux rayons X, à contraste de phase réalisée à l'ESRF de Grenoble, G=350 X.*

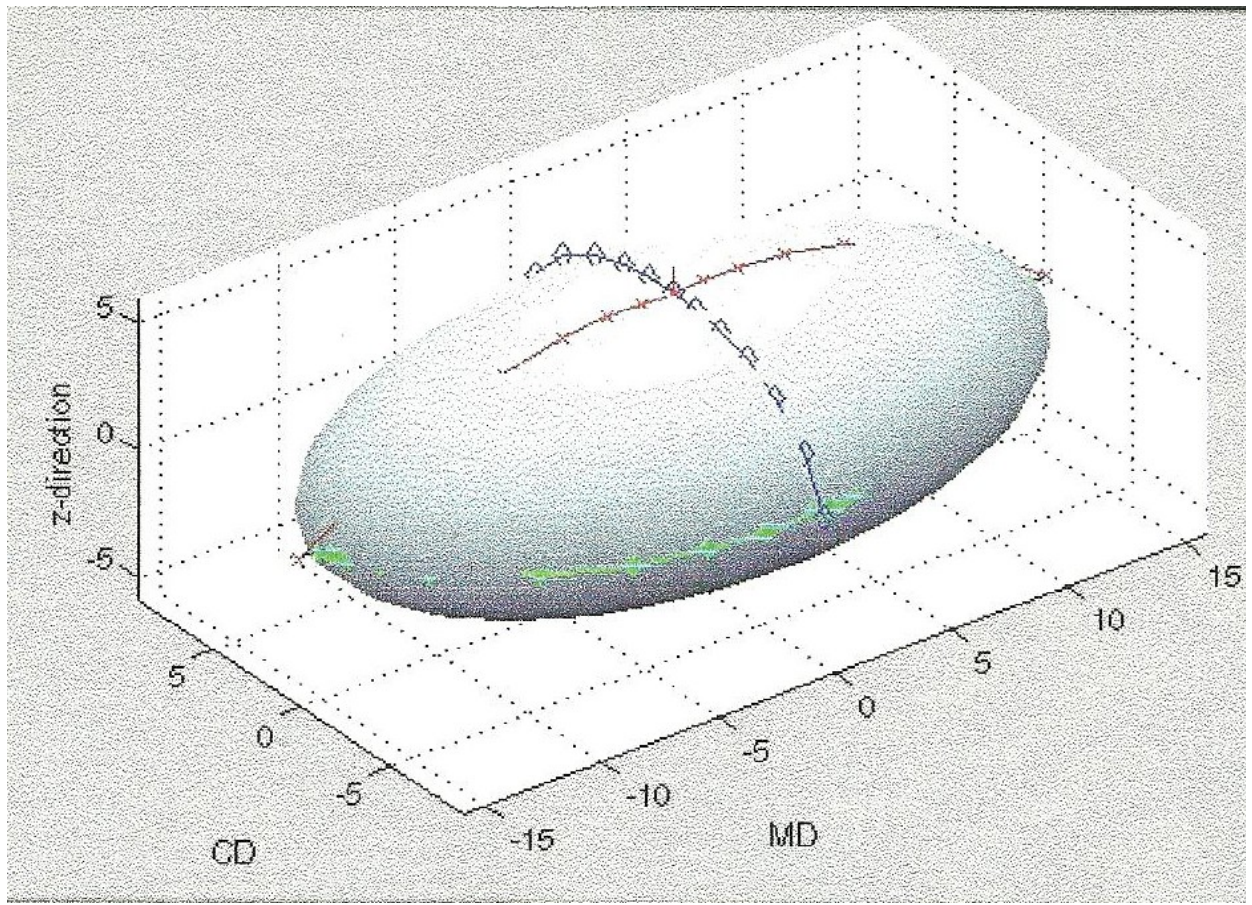

*Fig.13 : Le pore équivalent ellipsoïdal d'un papier journal de 45 g/m2 ; méthode des interceptes realisée en trois dimensions à l'E.S.R.F. de Grenoble sur des coupes micro-tomographiques telles que celle de la figure 12, échelle des axes en µm .*

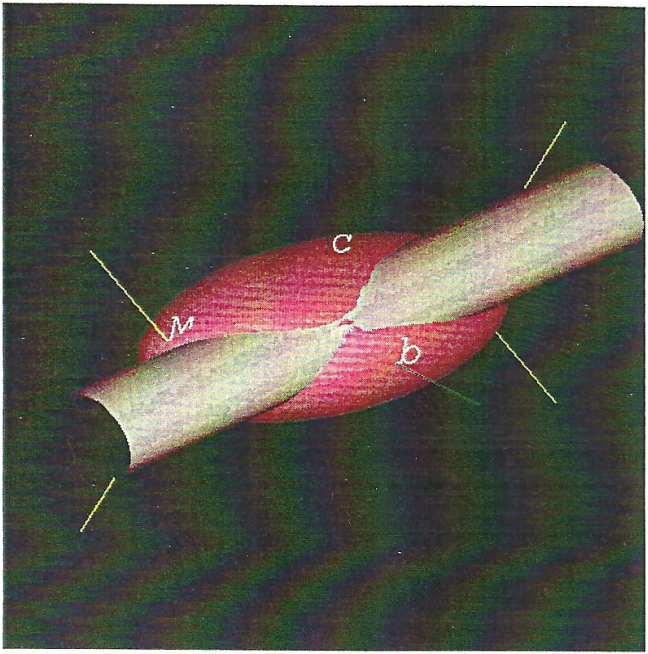

Fig. 14 : L'intersection d'un ellipsoïde tri axes et d'un cylindre elliptique tangent, avec une génératrice du cylindre diamètre des ombilics.

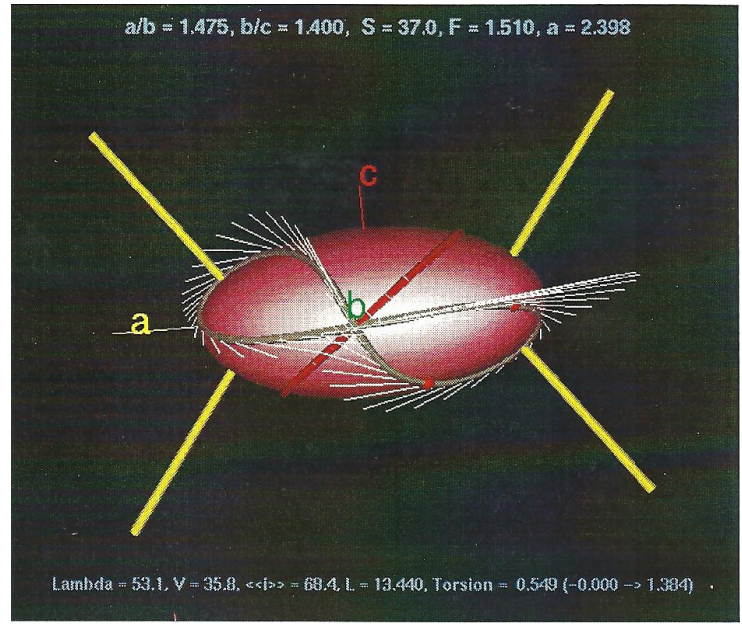

Fig. 15 : Courbe ellipsoïdo-cylindrique(en gris) avec la représentation des normales aux ombilics (en jaune) et de l'intersection circulaire de l'ellipsoïde dans un plan diamétral(en rouge) .

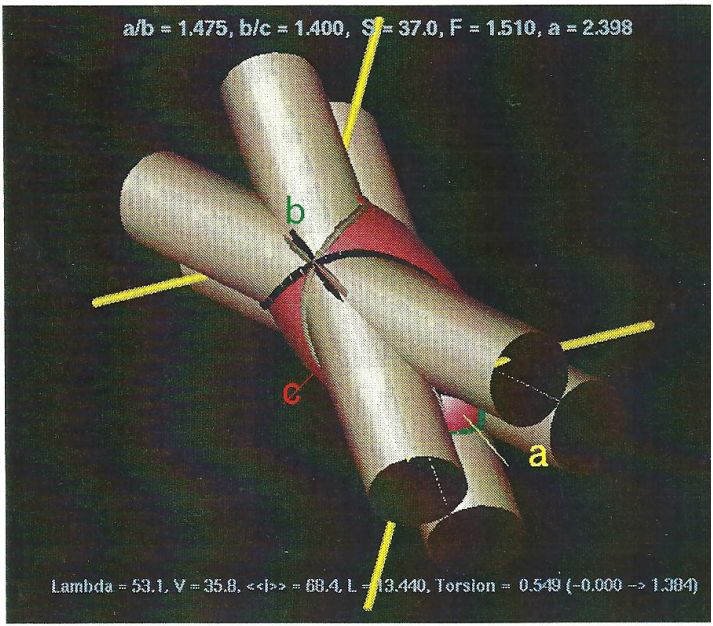

Fig.16 : Cylindres elliptiques appariés dans chacun des groupes par symétrie par rapport à l'axe neutre du groupe, diamètre des ombilics,

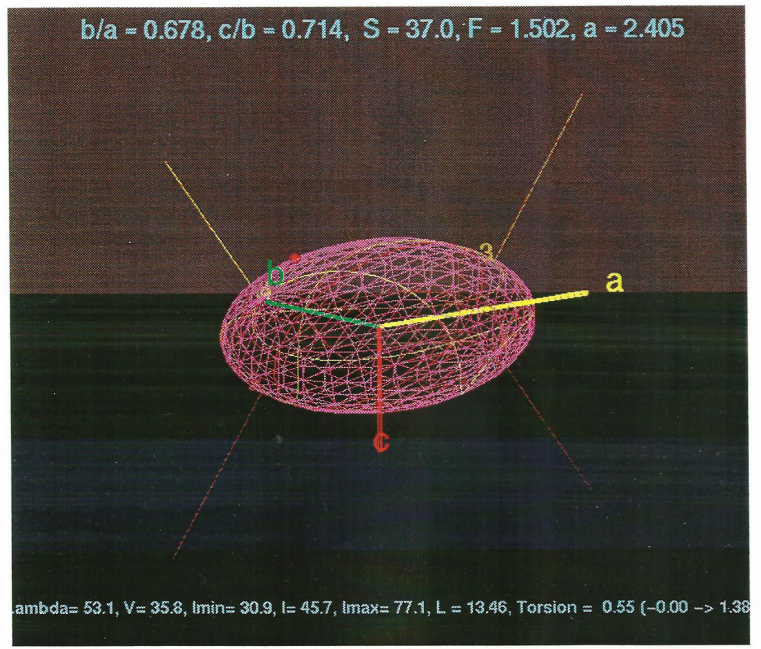

Fig.17: Réalisation d'un ellipsoïde par une tapisserie de courbes ellipsoïdo- cylindriques, (équations (4)).

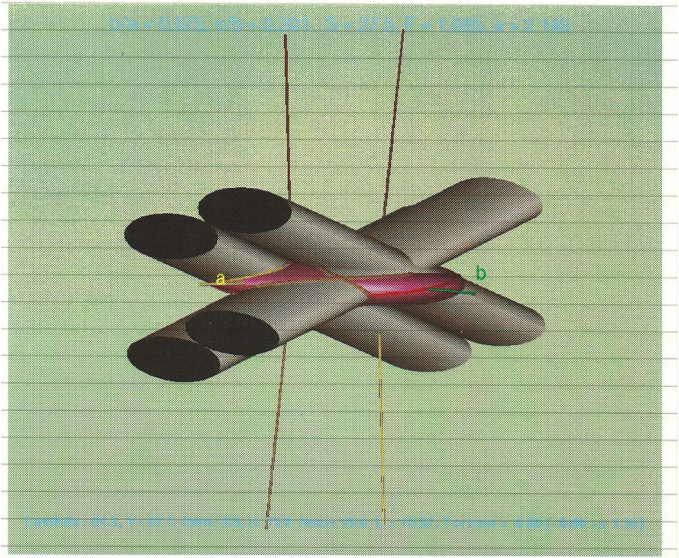

*Fig.18: Un papier bon pour l'impression. La pénétration de l'encre dans la feuille est satisfaisante, la tortuosité $t_z=3,4$.*

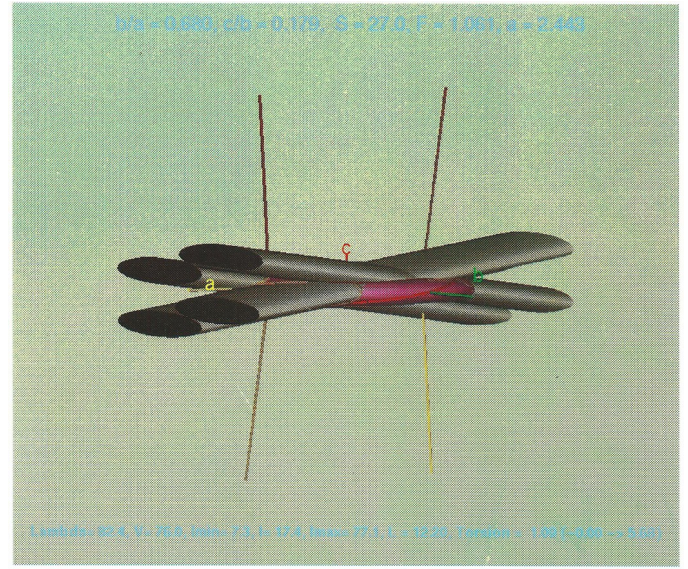

*Fig. 19 : Un papier mauvais pour l'impression. L'encre a tendance à baver en s'étalant à la surface de la feuille, la tortuosité $t_z=3,9$.*

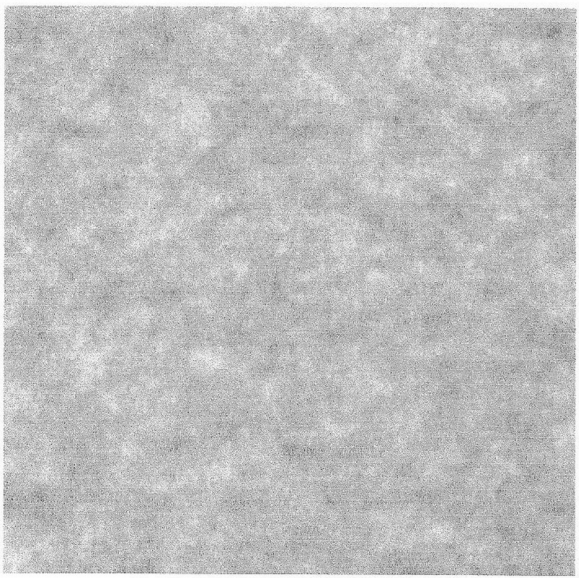

*Fig.20 : Feuille de papier d'impression vue en transmission de la lumière, montrant la répartition en flocs des fibres dans la feuille, G=1.*

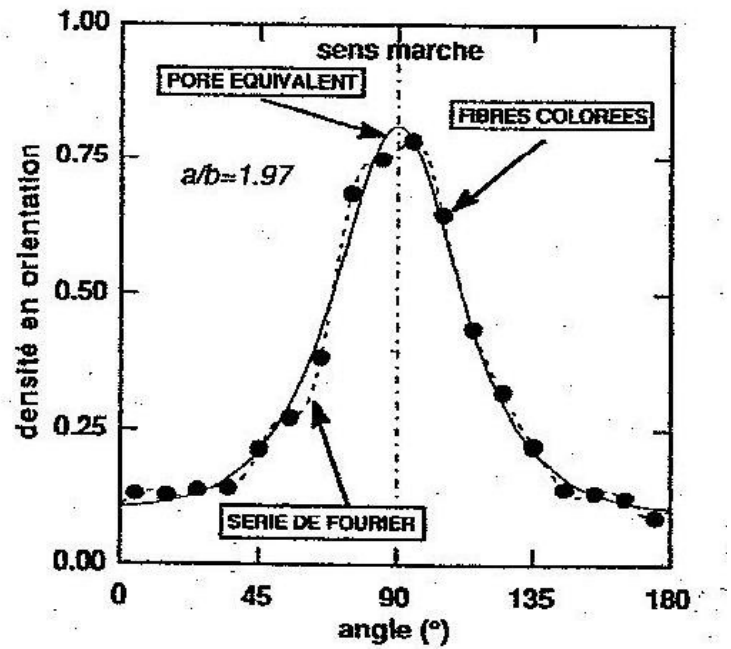

*Fig.21 : Distribution en orientation des fibres colorées dans la feuille.*

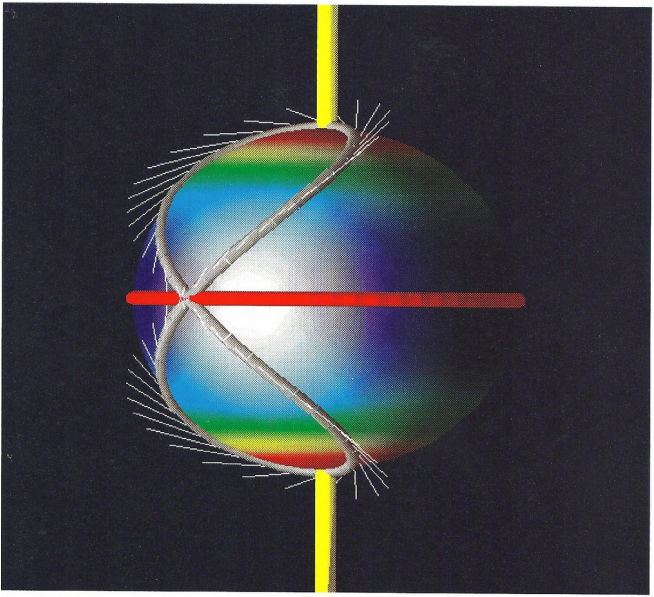

*Fig. 22 : L'hippopède, courbe sphéro-cylindrique, d'Eudoxe de Cnide*

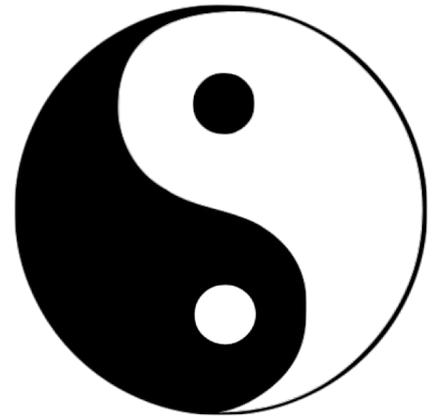

*Fig.23 : La figure du tai-chi, symbole Taoïste de l'équipartition des principes du Yin et du Yang.*

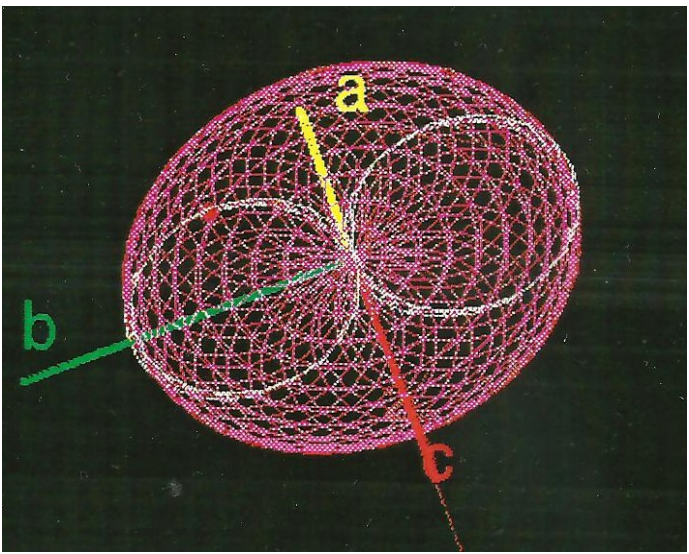

*Fig24 : Equipartition bipolaire anisotrope, vue en bout de l'axe neutre, diamètre des ombilics de l'ellipsoïde ; paramètres des équations (4) : b/a=0.678, c/b=0.714, S=37.0, a=2.40.*

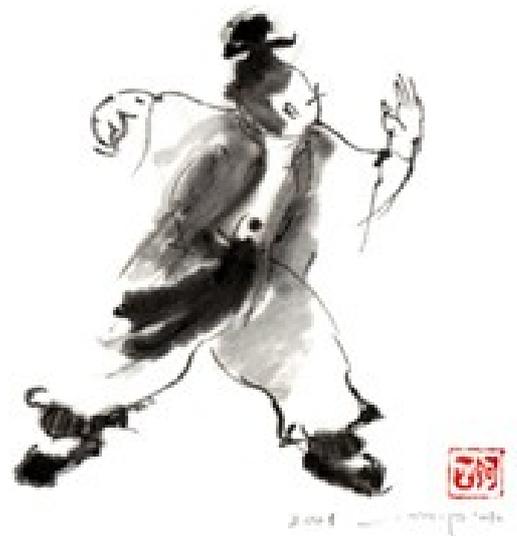

*Fig. 25 : Vue artistique d'une posture du Qi -Gong.*

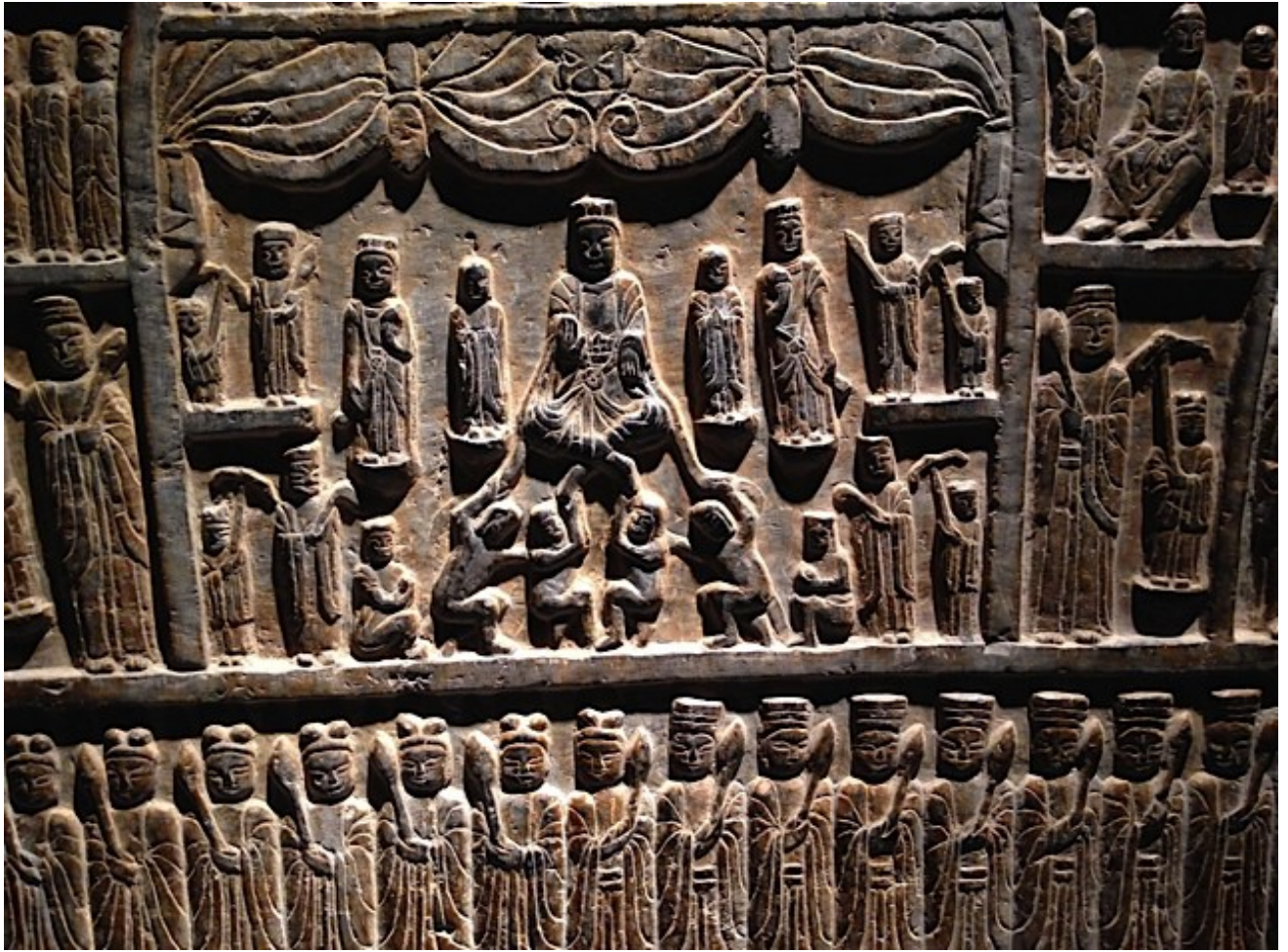

*Fig.26 : La fresque des mille bouddhas, Zhou du Nord - Shangaï muséum.
Remarquer : l'orientation verticale et inversée des paumes de la main de Bouddha et des doigts de sa main gauche, la courbure centro-symétrique des plis de sa tunique agrafée au centre, le mouvement des bras des supporteurs accroupis imprimant des rotations inversées, la différenciation des attributs des moines dans les parties droite et gauche de la fresque.*

**Appendices**

**Glossaire.**

-Moteur, ce mot employé en tant qu'adjectif signifie : qui produit ou transmet le mouvement, cf. : Dictionnaire Larousse.
-Déterminisme, propriété d'un système physique qui manifeste une causalité forte : à un cause donnée correspond un effet unique. C'est le cas pour les systèmes dont l'état évolue selon une loi déterminée en mécanique classique.
-Stochastique, adjectif qui signifie qui est de nature aléatoire.
-Holisme, « nom qui vient du mot grec, holos, entier : doctrine épistémologique selon laquelle face à l'expérience, chaque énoncé scientifique est tributaire du domaine tout entier dans lequel il apparaît ; l'adjectif : holiste ou holistique, signifiant relatif à l'holisme ». cf. : Dictionnaire Larousse.
-Equivalent signifie que toute expérience physique menée dans le cadre du premier modèle fournira des résultats non distinguables de ceux obtenus en travaillent dans le cadre de la deuxième théorie.
-Invariance conforme, c'est l'invariance par rapport à des transformations géométriques qui conservent les angles, sans nécessairement conserver les dimensions ; cf. : Jean-Bernard Zuber, « L'invariance conforme et la physique à deux dimensions », La recherche, N° 251, 1993, volume 24, pp.143-151.
-Les dualités sont à distinguer des symétries d'une théorie car ces dernières sont par définition les transformations sous lesquelles une théorie donnée est strictement invariante.
-Homotope, se dit de deux trajets s'ils se transforment l'un dans l'autre par une déformation continue sans traverser de lignes singulières, cf. : G.Lochak, « Géométrisation de la physique », coll. Champs ,364 , 1999, p. 117.
- L'action est le produit de la masse du mobile par sa vitesse et par la longueur du déplacement. C'est aussi le produit de l'énergie cinétique du mobile multipliée par l'intervalle de temps du déplacement.
- L'énergie vitale est l'énergie motrice qui anime toute chose. Dans la médecine traditionnelle chinoise le flux de l'énergie vitale est désigné par le mot, Qi.

**Notes complémentaires.**

**1.** L'ordre: « C'est tout ce qui est répétition, constance, invariance, tout ce qui peut être mis sous l'égide d'une relation hautement probable, cadré sous la dépendance d'une loi » .
Le désordre **:** « C'est tout ce qui est irrégularité, déviations par rapport à une structure donnée, aléa, imprévisibilité », propos de Edgar Morin dans son ouvrage: « Introduction à la pensée complexe », Ed. Du Seuil, 2005, Vivre et traiter avec le désordre, p. 118.
La structure caractérise de manière conceptuelle l'ordonnancement d'un matériau. Par exemple dans un matériau poreux les paramètres de la structure sont la distribution spatiale de l'orientation des interfaces entre les phases, le rayon hydraulique moyen défini par le rapport de l'aire au périmètre des pores ou de leur volume rapporté à leur surface.
« La texture signifie proprement l'arrangement et la liaison de différents corps ou filets minces, mêlés et entrelacés comme dans les toiles d'araignée, dans les draps, étoffes, tapisseries, etc. », définition donnée par Diderot et d'Alembert, Encyclopédie, (1751-1772) et rappelée par P-M. de Biasi et K. Douplitzky dans leur ouvrage : « La saga du papier ». A la différence de la structure, la texture est définie par des caractéristiques objectives, directement perceptibles par nos sens, par exemple la forme des particules dans la texture fibreuse ou granulaire d'un matériau.
« La découverte géniale du mathématicien Evariste Galois consista à faire passer la structure avant l'objet et à définir celui-ci à partir de celle là », un commentaire de Georges Lochak, dans son ouvrage : « La géométrisation de la physique », collection Champs, Flammarion, 1999, pp. 179-181.

**2.** Quelques définitions du hasard.
Le hasard, « Ce nom vient aussi du jeu de dés car dans l'Espagne mahométane, les pointes sur les faces des dés étaient des fleurs d'oranger (azahar) », remarque faite par : Fernando Corbalan et Gerardo Sanz, dans « Le monde est Mathématique », La conquête du Hasard, Edition : Le Monde, 2013, p. 60.
Le hasard c'est la conjonction d'événements dont l'un au moins est aléatoire. L'adjectif aléatoire caractérise un état hasardeux dont les variations sont redevables d'une loi de probabilité comme c'est le cas par exemple pour une variable aléatoire ; on dit également une variable stochastique.
Von Neumann en 1932 estime que le hasard doit être vu comme une sorte « d'incomplétude » essentielle, en référence au théorème de Gödel **,** cf. : Igor et Grichka Bogdanov , « La fin du hasard », Grasset 2013, p.168.
« Le hasard quand un événement se produit avec un manque apparent de causes et que l'on ne sait pas ce qu'il va se passer… Le hasard n'est que la mesure de notre ignorance », a dit Henri Poincaré dans: « Calcul des probabilités », les grands classiques Gauthier-Villars.
« Ce que nous appelons le hasard c'est peut être la logique de Dieu » a dit Bernanos. « C'est peut-être pour cela qu'un beau jour Einstein a lancé en souriant : le hasard, c'est Dieu lorsqu'il se promène incognito ! ».
« En mécanique quantique le hasard ne peut se définir que là ou il y a une observation car les fonctions d'onde sont parfaitement déterminées et seule leur réalisation une fois observée est aléatoire. Les fonctions d'onde portent en elles la marque des symétries de l'objet quantique. En microphysique, tout comme pour le dé à jouer, la symétrie impose ses contraintes au hasard. Ici, à la notion d'invariance au cours du temps se superpose celle de l'invariance face au hasard. L'harmonie spatiale entraîne l'harmonie numérique. Lorsque le hasard prend forme, les événements possibles sont fortement limités », selon G.Lochak, S.Diner, D.Fargue, dans « L'objet quantique » pp. 105-107, Ed. Champs Flammarion.
« Dans le plus grand des hasards se trouve la plus grande détermination…plus on est indifférent moins on est libre.., d'où l'existence d'une libre nécessité, je vois le bien mais je fais ce qui est mal », a dit Spinoza, philosophe rationaliste Hollandais (1632-1677) qui réfute l'idée du libre arbitre.

« Le hasard n'est qu'un aspect de la nécessité elle-même », Pierre-Marie Morel, dans : « Democrite, l'Atomisme ancien », AGORA, les classiques, p.13.

« Un peu de désordre = beaucoup de profit », selon Eric. Abrahamson et David. Fridman. En fin de compte on pourrait dire que le hasard fait bien les choses !

Fernando Corbalan et Gerardo Sanz, opus cité, p. 60, ont écrit : « Il existe d'autres phénomènes (que les phénomènes déterministes) pour lesquels les résultats sont différents et imprévisibles, même à partir d'une situation initiale identique : ce sont les phénomènes aléatoires par exemple le lancer d'un dé (le mot aléatoire vient du latin alea = jeu de dés). En le lançant toujours de la même façon, le résultat est à chaque fois imprévisible. Le résultat d'une expérience aléatoire dépend du hasard ».

Yakov G. Sinaï, prix Abel 2014, a étudié « L'aléatoire du non aléatoire » dans : « Chaos et déterminisme », première partie, Approches mathématiques, 4, pp. 68-87, ouvrage traduit du russe par S. Diner, Ed. Seuil, S 80, sous la direction de A.Dahan Dalmedico, J.-L. Chabert, K. Chemla, 1992.

**3-** La complexité « est un tissu (complexus : ce qui est tissé ensemble) de constituants hétérogènes inséparablement associés : elle pose le paradoxe de l'un et du multiple…..Mais alors la complexité se présente avec les traits inquiétants du fouillis , de l'inextricable, du désordre, de l'ambiguïté, de l'incertitude….La difficulté de la pensée complexe est qu'elle doit affronter le fouillis ( le jeu infini des inter rétroactions), la solidarité des phénomènes entre eux, le brouillard, l'incertitude, la contradiction. ». Citation extraite de l'ouvrage de Edgar Morin: « Introduction à la pensée complexe », 2005, Edition du Seuil, Essais 534, pp.21-22. .

La complexité génère le hasard : «Mais si ce joueur bat les cartes assez longtemps, il y aura un grand nombre de permutations successives ; et l'ordre final qui en résultera ne sera régi que par le hasard ; je veux dire que tous les ordres possibles seront également probables. C'est au grand nombre des permutations successives, c'est-à-dire à la complexité du phénomène, ce que ce résultat est du », a écrit Henri Poincaré dans son ouvrage : « Calcul des probabilités », pp. 8-10, opus cité.

**4-** Le nombre de fibres par mm³ dans une feuille de papier est très sensiblement le même que le nombre de neurones par mm³ de substance grise dans le cortex du cerveau humain. Dans le cas d'un papier les fibres sont en moyenne réparties dans des feuillets superposés dans l'épaisseur de la feuille, par exemple: 6 à 7 feuillets dans le cas d'un papier pour impression de 60 microns d'épaisseur. Des chercheurs brésiliens ont trouvé que les circonvolutions du cortex du cerveau dans la boite crânienne étaient décrites par la même loi mathématique que celle des plis d'une feuille de papier froissé.

**5-** Au vingtième siècle les technologues ont utilisé l'appellation de : « Grand univers » pour désigner « le format d'une feuille de papier de pâte fine, servant généralement pour les impressions lithographiques, et ayant environ 1 m sur 1,3 m, (ce qui correspond à environ un milliard de fibres), cf. Le Nouveau Larousse illustré, Dictionnaire universel encyclopédique, Paris, Librairie Larousse. L'instances internationale de normalisation, I.S.O, définit le format des feuilles de papier en référence aux termes de différentes séries : An, Bn, Cn, qui délimitent la surface de feuilles de papier rectangulaires, en fonction de leurs dimensions en longueur et en largeur.

La multiplicité des usages du papier a été déclinée de manière plaisante en 1965 dans la chanson : « Les petits papiers », parole et musique de Serge Gainsbourg, interprétée avec succès par la chanteuse française Régine.

**6-** 75 milliards de fibres réparties en 0,3 secondes dans une feuille de : 10 m x 10 m sur la machine à papier, c'est sensiblement le nombre de neurones dans le cerveau humain et la moitié du nombre d'étoiles dans notre galaxie de la voie lactée.
-

**7-** La taille d'un objet compact peut être évaluée par la longueur moyenne des sécantes interceptées par cet objet lorsqu'il est traversé de part en part par des droites, dans toutes les directions. Suivant cette définition Tomkeieff a établi en stéréologie que la taille d'un objet est égale à quatre fois l'inverse de la surface spécifique volumique de l'objet. Dans le cas d'un milieu poreux à porosité uniforme, la taille moyenne des pores évaluée en tant que sécante moyenne dans les pores, <<g (θ)>>, est égale à quatre fois la valeur de la porosité rapportée à la surface spécifique volumique de la texture poreuse.

La terminologie qui définit les domaines qualifiés de macro, méso, microscopiques dans un ensemble d'éléments ne se réfère pas à des dimensions spatiales fixées dans l'absolu mais elle caractérise des champs d'observation classés relativement les uns par rapport aux autres. Ce classement correspond à la possibilité d'effectuer l'observation d'éléments de taille différente. Ainsi dans une feuille de papier le domaine macroscopique permet d'identifier les flocs qui sont des agrégats de fibres de dimension moyenne de quelques millimètres et répartis de manière homogène dans la feuille. Pour ce faire la dimension du champ d'observation est proche du décimètre. Le domaine mésoscopique permet d'observer les fibres à l'intérieur des flocs. Il correspond à des champs d'observation dont la dimension est de l'ordre de quelques millimètres. Le domaine microscopique permet d'observer les parois des fibres ainsi que leurs constituants fibrillaires. Ce domaine correspond à des champs d'observation de quelques micromètres. Les fibrilles dont la longueur est de quelques micromètres et la largeur de quelques nanomètres sont constituées par des associations de macromolécules de cellulose. La macromolécule de cellulose est un enchaînement de molécules de glucose dont la taille est inférieure au nanomètre et constituée d'atomes de carbone, d'hydrogène et d'oxygène, Ref. : J.J. Hermans, « La cellulose », ouvrage traduit de l'allemand par Marcel Chêne, professeur à l'E.F.P- I.N.P de Grenoble, 1950.

Selon les définitions de G. Matheron dans son ouvrage « Eléments pour une théorie des milieux poreux », Ed. Dunod, pp.86-89, Genèse de la loi de Darcy : « Le niveau macroscopique, correspond à des éléments de volume suffisamment grands, vis-à-vis des dimensions granulométriques, pour que le milieu poreux puisse être regardé comme homogène… Un milieu poreux homogène se caractérise expérimentalement par l'apparition de propriétés macroscopiques constantes : porosité moyenne, perméabilité, etc.…, pourvu qu'elles soient mesurées à l'échelle de volumes grands vis-à-vis des dimensions granulométriques et ne se modifient pas dans l'espace. Cette homogénéité est de nature purement statistique. Au niveau granulométrique le milieu reste irréductiblement hétérogène ». Un ensemble homogène se reproduit de place en place, c'est ainsi que l'entend Henri Poincaré dans sa définition de la loi d'homogénéité, cf.: « La science et l'hypothèse », Ed. La Bohème, Les sillons littéraires, 1992, Chap.IV, l'espace et la géométrie, pp. 86-87.

**8-** Le grammage d'un papier est sa masse par unité de surface. La valeur normalisée, I.S.O., du grammage s'exprime en g/m2. Elle se détermine en effectuant un échantillonnage de feuilles au hasard dans un lot industriel. Le grammage est une caractéristique essentielle du matériau papier. Il est relié à ses propriétés d'usage et permet d'évaluer sa valeur marchande.

**9-** L'appellation « papier » est éponyme pour de nombreux matériaux formatés en feuille bien que non réalisés suivant le procédé papetier qui est majoritairement fabriqué à partir de la cellulose des fibres végétales. Par exemple il est d'un usage courant de parler du « papier d'aluminium » dans le cas de ce métal laminé en feuille très mince.

**10-** L'International Association of Scientific Paper Makers, (I.A.S.P.M), regroupe les chercheurs spécialisés dans les études du procédé de fabrication du papier et de ses propriétés d'usage. La Fiber Society regroupe les chercheurs dont les travaux sont dédiés à l'étude des matériaux fibreux et plus spécifiquement : les textiles, les non tissés, les feutres, les filins pour amarrage.

**11-** La courbure totale en un point d'une interface est proportionnelle à l'inverse du produit des rayons de courbure principaux de l'interface. La courbure totale, également dénommée courbure de Gauss, peut être positive : en forme de vallée ou de colline, négative : en forme d'une selle de cheval, ou nulle. La courbure moyenne est la moyenne arithmétique des courbures principales en un point.

**12-** L'idéogramme chinois du papier est dérivé de celui de la soie, voir l'ouvrage de Michel Soutif sinophile érudit, Président honoraire de l'Université scientifique et médicale de Grenoble et Professeur associé à l'Université de Shanghai : « L'Asie source de sciences et des techniques », collection Grenoble Sciences, 1995, p. 196.
Cette origine du papier est également celle qui est citée par Erik Orsenna, membre de l'Académie Française, dans son ouvrage :  « Sur la route du papier », Petit précis de mondialisation III, Le livre de poche 32917, Ed. Stock, 2012, p. 19.

**13-** Les origines du matériau papier dans sa forme de réalisation artisanale ont été étudiées par differents auteurs notamment par : Erik Orsenna, dans son ouvrage, opus cité.  Pierre-Marc de Biasi et Karine Douplitzky dans  « La saga du papier », ARTE Editions, Adam Boro, 1999,  chap. II, pp. 25-39, chap. IV, pp.59-61. Gerard Coste, « Le papier la belle histoire », CERIG, Avril 2004.

**14-** Une présentation de la comparaison du papier des guêpes cartonnières et du matériau papier tel qu'il est fabriqué industriellement de nos jours, est ainsi faite au Science Muséum de Toronto, Canada, dans les salles dédiées à la communication ainsi qu'au musée du papier de OJI  Paper Cie, à Tokyo, Japon.

**15-** Le papier est un matériau qu'il faut privilégier pour un développement qui soit durable sur la planète. La matière première du papier, la cellulose, est une macromolécule d'origine naturelle renouvelable, sa formule chimique est : $(C_6 H_{12} O_6)_n H_2 O$. Sous sa forme fibreuse cette matière première peut être réutilisée, en moyenne 6 à 7 fois successives, pour la fabrication de nouveaux papiers ou de cartons, après triage et remise en pâte des vieux papiers recyclés. Il existe des gammes de papiers qui sont à 100% produites avec du papier recyclé, cf. : Gilles Pelissier, Cosmétique Mag. N°106 mars 2010, « Le papier, vecteur d'une communication de Qualité ». De nombreux articles d'usage courant et des matériaux sont fabriqués à partir de fibres cellulosiques dans des conditions ou le bilan carbone est neutre justifiant la certification du FSC. Ce label du Forest Stewardship Council est propre à l'industrie du bois et celle du papier. Il est basé sur une charte de gestion responsable de la forêt qui implique sa régénération et bénéficie du soutien d'ONG, de Greenpeace et de WWF.
Des traitements chimiques de gazéification et de combustion, l'un comme l'autre source d'énergie, transforment en fin de leur utilisation les vieux papiers en dioxyde de carbone atmosphérique qui en combinaison avec l'eau et sous l'effet de la lumière absorbée par les pigments chlorophylliens génère la cellulose par un processus photosynthétique suivant la réaction : $6 CO_2 + 12 H_2 O = C_6 H_{12} O_6 + 6 O_2 + 6 H_2 O$. C'est ainsi que se renouvelle la matière première du papier tout en étant la source de production d'oxygène dans l'atmosphère. Il existe également une génération de la cellulose par des processus bactériens par exemple par la souche : Acetobacter xylinium.
La cellulose utilisée dans la fabrication du papier sous sa forme fibreuse, est générée dans un intervalle de 6 à 25 ans suivant les espèces végétales de bois, feuillus ou résineux. L'industrie de la pâte et du papier utilise souvent les bouts fins des arbres et les tombées du sciage, appelées dosses, sous produits de l'équarrissage des grumes qui sont destinées à la fabrication des planches pour l'industrie du meuble et des bois de charpente dans la construction de l'habitat. Des forêts de culture sont plantées et gérées suivant une logique de développement durable pour ces différentes utilisations.
Le processus de fabrication de la pâte à papier et du papier s'autoalimente en énergie par combustion de la lignine, une macromolécule organique amorphe intriquée avec les enchaînements macromoléculaires cellulosiques et qui rigidifie la structure fibreuse des végétaux. La lignine est éliminée de la paroi des fibres par  dissolution au moyen de processus chimiques, par exemple le procédé Kraft qui permet de récupérer les produits chimiques utilisés. Par combustion de la lignine la production de la pâte à papier ainsi que celle du papier est excédentaire au point de vue de son bilan énergétique. Ainsi dans la majorité des cas le procédé papetier est producteur et exportateur d'énergie.

La fermentation en éthanol des sucres obtenus par l'hydrolyse des hémicelluloses extraites du bois dans la fabrication de la pâte à papier, peut être la source de biocarburants dit de seconde génération. Les recherches effectuées à Grenoble I.N.P- PAGORA, par Dominique Lachenal Professeur, médaille d'Or de l'Académie d'Agriculture de France, Christine Chirat, Maître de conférence et l'équipe des chercheurs en bio raffinerie du laboratoire de Génie des Procédés Papetiers, (L.G.P.2), ont montré que 75 l d'éthanol peuvent être obtenus en moyenne par tonne de bois pinus pinaster avant traitement pour la fabrication de pâte à papier. L'émission de $CO_2$ par combustion de l'éthanol est réduite de 50% par rapport à la combustion des carburants d'origine fossile ce qui est un bien meilleur résultat qu'avec l'utilisation des bio carburants de première génération produits à partir du colza ou du mais pour lesquels la diminution d'émission de $CO_2$ n'atteint que 30 à 25 % en moyenne.

L'utilisation du papier et du carton est bénéfique à long terme pour l'activité humaine dans un développement durable, pour autant que la composition de l'atmosphère, les réserves d'eau et les caractéristiques spectrales du rayonnement solaire soient conservées. Imaginons ce que serait un monde sans papier, une conjecture illustrée de manière plaisante et instructive dans un film destiné au grand public, réalisé dans les années 1970 par l'IRFIP, un organisme professionnel de formation pour l'industrie papetière à l'E.F.P- PAGORA de Grenoble - I.N.P.

**16-** Nicolas Robert fut l'inventeur de la machine à fabriquer le papier en continu. Son brevet fut acheté par son patron Firmin Didot puis vendu par le beau frère de ce dernier à un industriel anglais qui l'exploita et le mis en œuvre avec succès sous l'appellation de : machine Fourdrinier.

**17-** Le calcul de la taille des pores, <<g (θ)>>, dans un milieu poreux s'effectue par l'application de l'équation stéréologique de Tomkeieff : <<g (θ)>> = $4 \varepsilon / S_v$ où $\varepsilon$ désigne la porosité et $S_v$ la surface spécifique volumique du milieu poreux. Il est possible d'évaluer la distribution en taille des pores dans une texture poreuse par différentes techniques de mesure. Ainsi des papiers ont été analysées à l'E.F.P.G – PAGORA- Grenoble - INP durant les années 1970 et à la suite, avec un prosimètre à mercure, prototype à absorption de rayons, γ, réalisé au C.E.A. de Grenoble. Les résultats de ces recherches ont été utilisés pour caractériser la texture et les propriétés physiques des papiers destinés à différents usages. Ces travaux ont été publiés par : R.Chiodi, R. Kedadi et J.Silvy, dans la revue de l'Association technique de l'industrie papetière (A.T.I.P). Les propriétés de papiers d'impression avec différentes compositions fibreuses et teneur en charges minérales ont été étudiées par la porosimétrie à l'Universidade de Coimbra et à l'Universidade da Beira Interior au Portugal : voir dans la bibliographie la publication des travaux de : M. Moura, A.Claro, A. Ramos, A. Costa, J.Silvy, M. Figueiredo.

**18-** Lorsque la feuille papier est comprimée en feuille très mince jusqu'à atteindre une épaisseur de quelques micromètres celle ci est fortement densifiée et sa capacité électrique est très élevée. L'énergie électrique peut ainsi être stockée dans des condensateurs au papier de très forte capacité qui sont utilisés notamment dans le domaine médical pour les défibrillateurs cardiaques et en aérotechnique pour la mise à feu des boosters des moteurs d'avions et des fusées. Les condensateurs au papier ont par ailleurs des propriétés auto cicatrisantes qui sont dues à la fusion du polymère cellulosique dans le plasma qui est généré en présence de décharges électriques. Associée à la texture fibreuse de la feuille cette propriété diminue les risques de claquage des condensateurs au papier, d'où leur grande fiabilité.

**19-** Le volume analysé par les micros tomographies aux rayons X peut être par exemple de 1,4mm x 1,4mm x (épaisseur de la feuille de papier), voire : 0,5mm x 0.2 mm x (épaisseur de la feuille de papier), suivant le degré de résolution de l'image qui peut atteindre, 0,2 μm, en fonction des méthodes de traitement du signal. Voir à ce sujet les travaux de recherche effectués à l'E.S.R.F. de Grenoble par Jean-Francis.Bloch, Maître de conférence à l'INP-Pagora et Sabine.Roland du Roscoat, Chercheuse à l'Université Joseph Fourier ainsi que Christine.Antoine et Rune.Holmstadt antérieurement Chercheurs au Norvégien Pulp and Paper Institute, (P.F.I), à Trondheim.
.

**20-** Dans le cas d'un ensemble plan l'indexation du rayon de courbure est faite en référence à la direction, Θ, de l'élément rectifié qui est celle de la tangente à la courbe du pore équivalent. Dans le cas d'un ensemble analysé dans un espace tridimensionnel, l'indexation du rayon de courbure est faite en référence à l'orientation de la normale à la surface de l'élément, repérée par ses cosinus directeurs dans un repère d'axes orthonormés.

Le rayon de courbure des interfaces des éléments rectifiés dans un ensemble plan désordonné aléatoire vérifie une équation intégro-diférentielle en fonction de la direction des éléments. Cette relation se vérifie dans le cas ou le rayon de courbure est celui d'une ellipse d'où le choix qui est fait d'utiliser cette figure, d'autant qu'elle convient maintes fois dans la pratique pour représenter le pore équivalent des ensembles d'objets, cf. : « Etude structurale de milieux fibreux », J. Silvy, thèse de doctorat, 1980, pp.92-98, équ.46- 51-56.

**21-** Le concept du pore équivalent a été exposé la première fois lors d'une session de perfectionnement de l'I.R.F.I.P, destinée aux ingénieurs papetiers, en juillet 1968 à l'E.F.P-PAGORA- Grenoble-I.N.P, puis lors de la Conférence Internationale de Physique Papetière, en 1971 à Mont Gabriel Lodge au Canada, P.Q. Plusieurs versions ont été faites suite à la présentation de ce concept en 1980 dans la thèse d'état de J.Silvy.

Une version intégrale a été publiée, en espagnol en 1986 par F.Astals, Professeur à l'Université Politecnica de Catalunya à Terrassa dans la Monografia N° 6 de Fisica del Papel (103 pages).

J. Grolier, A. Fernandez, M. Hucher, J. Riss ont présenté ce concept dans leur ouvrage: « Les propriétés physiques des roches, Théorie et modèles », Le concept du pore équivalent, (Silvy, 1980,1985), pp. 114-116, Masson Ed., ISBN : 2-225-82306-5,1990.

C. Schaffnit, J.Silvy, K.J.Dodson ont développé le concept du pore équivalent en anglais dans : Nordic Pulp and Paper Research Journal, N°, 3, 1992, pp121-125.

Une présentation du concept du pore équivalent est faite dans l'ouvrage : « Handbook of physical and mechanical testing of paper and paperboard », vol. 2, pp. 306-310, Marcel Dekker Inc., New York, édité en 1984 aux U.S.A.par Richard.E.Mark, Senior Research Associate à l'Empire State Paper Research Institute (E.S.P.R.I) de Syracuse

**22-** Deux concepts sont dits dual lorsqu'une expérience physique menée dans le cadre du premier concept fournit des résultats indistinguables de ceux obtenus en travaillant dans le cadre du deuxième concept. La dualité s'applique dans le cas du pore moyen et du pore équivalent lorsque celui-ci est de forme ellipsoïdale.

**23-** La figure de diffraction d'un faisceau laser qui recouvre le tracé du pore équivalent imprimé sur une feuille transparente de polymère est identique à celle obtenue par l'impact du faisceau laser recouvrant la réplique transparente de la surface de la texture. Cf. : Mario. José. Peirera, « Reconhecimento do padrâo optico da estructura da folha de papel », Thèse de doctorat, Universidade da Beira Interior, 2002, Chap. IV et V.

La projection du pore équivalent sur un plan dans une direction, est semblable à celle de la tache de lumière diffusée, observée à la limite de perception dans un plan, par un faisceau laser qui impacte le papier sur ses faces externes ou sur la tranche d'un empilement de feuilles.

Ces propriétés de diffraction et de diffusion multiple de la lumière sont mises en oeuvre pour le contrôle de la répartition des fibres à la surface et à l'intérieur des feuilles de papier ainsi que des textiles non tissés : voir la bibliographie à ce sujet ainsi que les figures, 9, 10 et 11.

**24-** Cette propriété est une généralisation à l'espace tridimensionnel de la relation établie en stéréologie dans le cas d'un réseau plan. Dans l'espace tridimensionnel la direction des traversées est définie par les cosinus directeurs de la droite support des traversées.

**25-** Ce résultat peut être obtenu également en appliquant le théorème de Gauss-Cauchy suivant lequel : l'aire moyenne de la projection de la surface d'un polyèdre convexe quelconque est égale au quart de l'aire de la surface de ce polyèdre. La forme moyenne se définit à partir des interfaces accessibles dans les pores, les aires en contact entre les particules n'étant pas prises en compte.

**26-** Cette équivalence entre une propriété géométrique de l'ellipsoïde et une propriété stéréologique d'un ensemble d'objets est un élément clé pour l'analyse des ensembles désordonnés aléatoires suivant le concept du pore équivalent. Cette propriété dérive d'un des théorèmes d'Apollonius de Perga qu'il a énoncé en -260 (environ) av.J.-C dans le cas de l'ellipse. Le théorème d'Apollonius se généralise dans le cas de l'ellipsoïde et permet de calculer le volume de cette figure en fonction de trois diamètres conjugués et de leurs angles respectifs deux à deux. En faisant apparaître dans l'expression du volume l'aire projetée de la surface de l'ellipsoïde sur un plan perpendiculaire à l'un des diamètres, j'ai rapproché le résultat obtenu de l'équation stéréologique qui relie les projections sur un plan des aires des interfaces d'une texture et la longueur moyenne entre les interceptes dans la direction perpendiculaire à ce plan.
Ce résultat est cité et ré argumenté parfois par differents auteurs notamment : J.Riss dans sa thèse : « Principes de stéréologie des formes en pétrographie quantitative », thèse d'état, Univ.Orléans, 1988, chap.IV, Autre forme équivalente : « le pore équivalent », pp.304-309. Des développements basés sur cette relation ont été faits par : N. Bodin, dans sa thèse: « Description de la topographie des surfaces à l'aide de structures homogénéisées équivalentes conformes », soutenue à l'Université de Franche-Comté, en 1999.

**27-**Cette hypothèse est à rapprocher d'une conjecture énoncée par Eshelby et démontrée par la suite par H. Kang, qui établit que la forme elliptique d'une inclusion ou d'un vide de petite taille dans un matériau considéré par ailleurs comme homogène est la forme optimale qui minimise l'énergie dépensée dans le cas de sollicitations uniformément réparties sur l'ensemble, celles-ci pouvant être de nature thermochimique, mécanique, magnétique ou électrique, voir la bibliographie citée sur ce sujet.

**28-** Des feutres et des mats géotextiles de fibres ont été analysés par diffraction ainsi que par diffusion d'un faisceau laser, à l'Université de Beira Interior au Portugal. Ces analyses ont été complémentées en les corrélant avec des essais physiques réalisés sur ces mêmes échantillons au laboratoire de mécanique textile du Professeur Drean de l'Ecole Nationale Supérieure des Industries Textiles de l'Université de Mulhouse, en France. Les matériaux non-tissés et les géotextiles ont été réalisés au Centre of technology Textile du Saint- Hyacinthe-Polytechnic Institute de Montreal, au Canada.
Pierre Chevalier, Ingénieur E.F.P. doctorant au laboratoire de génie des procédés papetiers, L.G.P.2, E.F.P.G.-PAGORA de Grenoble- I.N.P., a modélisé la texture des feutres de machines à papier suivant le concept du pore équivalent dans sa thèse: « La modélisation et les propriétés de perméabilité des feutres », réalisée en collaboration avec la Société Binet- feutres, à Annonay, France, en 1995. Sa recherche a été primée en 1996 par l'Association Européenne des fabricants de feutres.

**29-** Il a souvent été constaté que le rayon de courbure soit d'une ellipse soit d'un ellipsoïde, lorsqu'il s'exprime normé de manière relative c'est à dire rapporté au périmètre ou à l'aire de ces figures, était la mesure la mieux adaptée pour caractériser la densité de probabilité en orientation pondérée des particules dans une texture, en comparaison des distributions circulaires de Gauss tronquée (truncated Normal distribution), de Von Mises ou de Cauchy Lorentz ( wrapped cercle Cauchy distribution ) souvent utilisées pour l'analyse des matériaux. Voir les travaux de :
M-C. Peron, Y.Berthoumieu, J-P. Da Costa, C. Germain et al, « Modèles circulaires pour la caractérisation de textures orientées », Université de Talence et Gradignan, France.
J. Silvy,« Etude structurale de milieux fibreux », thèse d'état, I.N.P.G. et U.S.M.G, opus cité,1980, pp.97-157.
Ch. Baratte « L'adéquation des feutres de papeterie en fonction de la morphologie des fibres et des procédés de fabrication du papier, une étude réalisée par modélisation sur ordinateur avec ses vérifications expérimentales », p.113, thèse réalisée au L.G.P.2 de Grenoble-I.N.P., France.
Torbjorn Wahlstrom : "Prediction of fibre orientation and stifness distribution in paper, Fundamental research symposium on paper ", TAPPI, B.P.M.A, Cambridge, 2009.

De nombreux matériaux qu'ils soient naturels ou artificiels ont un pore équivalent ellipsoïdal. La géométrie de l'ellipsoïde permet de prendre en compte les symétries bilatérales qui s'établissent au cours de l'élaboration de la structure des matériaux ou des ensembles dans des champs de forces gravitationnelles, hydromécaniques et sous l'effet des tropismes environnementaux. D'Arcy Thompson dans son ouvrage « On grow and form », publié en 1917, a analysé la symétrie des structures naturelles animales, végétales et minérales. Hermann Weyl dans son ouvrage : « Symétrie et mathématique moderne », Princeton, 1952, Flammarion, 1964, montre l'importance des facteurs environnementaux qui sont déterminants pour l'établissement des symétries structurales dans le domaine du vivant. Cet auteur cite « les potentialités prospectives de développement des membres des amphibiens à partir de bourgeons telles que les a étudiées R.G.Harrison dans des expériences de transplantation de disques prélevés sur le corps, mettant en évidence que « *l'axe antéropostérieur* est déterminé à une époque où la transplantation peut encore inverser *l'axe dorso-ventral* et *l'axe médian latéral* » de leur corps ». Réf : H.Weyl, « Symétrie et mathématique moderne », Chap. sur la symétrie bilatérale, p.43, Edi.Champs Flammarion, 1964. Les études des botanistes sur la phyllotaxie de l'émergence des feuilles dans les végétaux ont montré qu'il existe parfois un ordre caché derrière l'irrégularité, certains modèles déterministes pouvant intégrer des fluctuations aléatoires. Consulter à ce sujet l'ouvrage de synthèse de Francis Hallé : « Aux origines des plantes », Tome 1, chap. 4, pp.155-163.
Boris Zhilinskii et Guillaume Dhont, de l'Université du Littoral côte d'Opale, France, ont analysé différents aspects de ces questions dans leur ouvrage : « Symétrie dans la nature », publié par les Presses Universitaires de Grenoble, Octobre 2011.
Les pierres charriées dans le lit des glaciers ou roulées dans le lit des cours d'eau, les grains de sable qui ont subi une érosion éolienne, ont une forme souvent proche de celle d'un ellipsoïde. Il en est de même pour les galets et les pelotes fibreuses rejetés sur les grèves sous l'effet du balancement des marées et du ressac des vagues. Différentes théories ont été proposées pour expliquer la forme ellipsoïdale des galets de pierre. Platon dans Le Timée écrivait : « Alors la terre, comprimée par l'air de manière que l'eau ne peut la dissoudre, forme une pierre…. ». David.Hilbert et S.Cohn-Vossen rappellent l'importance des théories probabilistes dans les phénomènes d'érosion, dans leur ouvrage: « Geometry and the Imagination », traduction anglaise, p.13, Chelsea pub.1952. L'analyse des mécanismes de formation de la feuille de papier ainsi que des considérations énergétiques développées au chapitre IV-2, justifient le formalisme elliptique de structuration des ensembles d'objets lorsqu'ils sont soumis à des sollicitations aléatoires.

**30-** Le principe de la moindre action est l'équivalent de la condition de Lagrange, suivant laquelle dans la durée du déplacement, la valeur moyenne de la différence entre l'énergie cinétique et l'énergie potentielle des particules est extrémale, ou suivant Euler : que la variation d'énergie potentielle des particules au cours de leur transfert est minimale. Voir dans la bibliographie: http://webinet.cafe-sciences.org/articles/le-principe-de-moindre-action-un-bijou-de-la-physique/
A masse volumique constante pour un fluide qui s'écoule dans une texture poreuse la circulation du fluide s'effectue avec une moindre action pour aller d'un point à un autre.

**31-** En thermodynamique statistique l'entropie mesure le degré de désordre d'un système au niveau microscopique. Plus l'entropie du système est élevée, moins ses éléments sont ordonnés, liés entre eux. Ludwig Boltzmann a exprimé l'entropie statistique, S, en fonction du nombre d'états microscopiques, N(x), définissant l'état d'équilibre d'un système donné au niveau macroscopique. On a  S = - ∫ p(x) log[p(x)] dx, où p(x)= N(x)/∑ N(x), 0‹p(x)‹1. L'équilibre d'un système thermodynamique est réalisé quand son entropie a la valeur maximale compatible avec les contraintes auxquelles il est soumis.
Dans le cas de l'écoulement d'un fluide dans un milieu poreux, les orientations des interfaces des éléments du fluide qui s'écoulent au voisinage des parois des pores, pour des faibles valeurs du nombre de Reynolds, sont conformes en moyenne aux orientations des éléments de la surface du pore équivalent du milieu poreux. La distribution en orientation des rayons de courbure d'une ellipse, normée c'est-à-dire rapportée au périmètre de l'ellipse, est très proche d'une fonction

Gaussiène tronquée circulaire. Dans ces conditions l'entropie du fluide en écoulement dans un milieu poreux dont le pore équivalent est ellipsoïdal est maximale.

**32-** Pour l'étude des analogies hydrodynamiques on peut consulter notamment : S.P.Timochenko, J.N. Godier, « Theory of elasticity », Third edition, Mc Graw-Hill, Hydrodynamical Analogies, 114, pp.325-328. M. Paschoud, « Le problème de la torsion et l'analogie hydrodynamique de M. Boussinesq », Bulletin technique de la Suisse Romande, N° 23, 7 Nov. 1926, pp. 277-284. E.Guyon, J-P. Hulin, L. PETIT , « Hydrodynamique Physique », inter éditions / éditions du C.N.R.S, 1991, Analogies avec l'électromagnétisme, 3.3, pp.122-123. J. Ferrandon, « Les lois de l'écoulement de filtration », Le génie civil, tome CXXV N° 2, pp. 24-28, 15 janv. 1948.

**33-** Envisageons le mouvement d'un fluide incompressible sur la surface d'un cylindre droit de base elliptique. En tous les points de la surface du cylindre le mouvement d'un élément de volume dans un petit intervalle de temps peut être décomposé d'une part en un mouvement hélicoïdal qui correspond à une translation et à une rotation en bloc sans changement de forme de l'élément, sur la surface d'un cylindre de section circulaire, d'axe orthogonal au plan des sections circulaires du cylindre elliptique et d'autre part, de manière concomitante, en une déformation iso volume de l'élément de fluide.
Le mouvement hélicoïdal peut être incrémenté dans la direction de l'axe des sections obliques circulaires du cylindre en fonction d'une variable liée à la durée de l'écoulement sur la surface. Le pas réduit de l'hélice est défini par le rapport de la hauteur du cylindre elliptique projetée sur une droite de direction celle de l'axe des sections circulaires du cylindre et du périmètre du contour de la section oblique circulaire du cylindre. La déformation iso volume résulte d'une part d'une affinité par rapport au grand axe de la section droite elliptique du cylindre, qui a pour effet de réduire l'aire de la section oblique circulaire en celle de la section droite elliptique dans le rapport cos (V) et d'autre part d'une dilatation de l'épaisseur de la tranche de base circulaire de l'élément de volume suivant l'axe de la section elliptique du cylindre dans le rapport : 1/cos(V). V, est l'angle entre les normales aux plans des sections droites elliptiques et obliques circulaires du cylindre. Par construction, cos (V), est égal au rapport des aires des sections droites elliptiques et obliques circulaires du cylindre.
Cette analyse du mouvement d'un élément de volume à la surface d'un cylindre elliptique peut s'envisager dans le cas du fluide qui s'écoule à l'interface des pores dans la texture d'un milieu poreux. Le but de cette étude est de construire de manière globale les lignes de courant du fluide d'un point à un autre de la texture poreuse par sommation des déplacements et des déformations élémentaires du fluide suivant cette analyse.

**34-** Ce raisonnement correspond à une conjecture énoncée par Henri. Poincaré à l'issue du congrès de Solvay en 1911, et qui est citée par Jean-Paul Auffray: « le quantum élémentaire d'action constitue un véritable atome, un atome de mouvement, dont l'intégrité vient du fait que les points qu'il contient sont équivalents d'un point à un autre du point de vue de la probabilité ». Ce physicien théoricien reprend cette conjecture pour la définition de l'espace fait des i points qui est l'objet de ses recherches. Voir : « The string, string theorists forgot to notice ». (La corde que les physiciens des cordes ont oublié de voir) : jpauffray@yahoo.fr, November 2007.

**35-** Selon la définition donnée par E.Cartan, « La théorie des groupes », Revue du palais de la découverte, 17, 15, 1989, « un groupe peut être considéré comme formé de toutes les opérations d'une nature donnée qui conservent certaines propriétés des objets auxquels elles sont appliquées, ou certaines relations entre ces objets ».
Un groupe possède une loi associative pas forcement commutative, avec un élément neutre chaque élément ayant un symétrique, cf. G. Lochak, « La géométrisation de la physique », Ed. Champs, pp.184, 192. M.Lachieze-Rey, « Au dela de l'espace et du temps », pp.250. G.Dhont , B. Zhilinskii, « Symetrie dans la nature », PUG, 2011, pp.29-30.

**36**-Compte tenu de la taille des pores, qui est petite par rapport à l'extension des interfaces, la géométrie des filets du fluide sur la surface du pore équivalent correspond à celle d'un fluide qui s'écoule entre des cloisons parallèles et relativement de faible espacement. La vitesse du fluide est parallèle au plan des interfaces et ses variations sont faibles dans les directions parallèles à ce plan par rapport à ses variations dans la direction perpendiculaire avec une pression constante dans cette direction. Ces conditions correspondent aux conditions des écoulements de Hele-Shaw pour les fluides visqueux entre des cloisons parallèles. Voir à ce sujet les analyses de : R. Comolet, « Mécanique expérimentale des fluides », tome.II, pp. 100-102, ainsi que : E. Guyon, J-P. Hulin, L.Petit, « Hydrodynamique physique », Ed. du C.N.R.S,opus.cit. pp. 382-384.

**37**-L'ensemble des lignes de courant du fluide constitue ainsi un tube de courant de très petite section en forme de couronne cerclant la surface du pore équivalent. L'homogénéité du domaine poreux étant assumée par hypothèse au niveau de l'étendue du volume élémentaire représentatif, la continuité de la surface du pore équivalent et l'absence de singularités rendent possible cette transposition et les évaluations analytiques sur sa surface. L'équilibre dynamique du fluide est ainsi réalisé d'un point de vue holiste, à l'échelle macroscopique du volume représentatif de l'ensemble de la texture qui est également celui du fluide en écoulement.

**38**- L'indétermination de l'orientation des plans méridiens de la sphère est à l'origine de l'instabilité observée au cours de l'élaboration de la texture d'un ensemble lorsque celui ci approche l'isotropie ce qui correspond à un désordre maximum pour la distribution des éléments dans la texture. Ce cas se produit par exemple pour la répartition des fibres en suspension aqueuse, à la formation de la feuille de papier, lorsque le gradient de vitesse approche la valeur zéro dans l'épaisseur de la veine fluide en mouvement. Cet état très instable rend moins homogène la suspension fibreuse. Le papetier dit que « l'épair » de sa feuille est nuageux, une caractéristique physique qui peut s'avérer être un défaut ou un avantage suivant les utilisations du papier.

**39**- Dans la modélisation de l'écoulement du fluide sur le pore équivalent la localisation d'une particule est probable à 100 % aux points nodaux qui sont les points d'intersection de l'ensemble des trajectoires. Par contre la direction de la vitesse de la particule est indéterminée en ces points qui sont des ombilics sur la surface du pore équivalent ellipsoïdal.

**40**- Lorsque la vitesse de formation de la feuille est élevée, plusieurs centaines de m/mn, voire 2000 m/mn, sur la machine à papier, le branlement dans la direction transversale de la table de filtration de la suspension fibreuse n'est pas possible mécaniquement. Cependant le mouvement transversal peut être réalisé au moyen de règles dont le profil est crénelé ce qui provoque des micros courants transversaux qui déplacent les fibres dans la suspension. Ce procédé a été proposé par Robert Charuel, professeur à l'E.F.P.G dans les années 1980 et est aujourd'hui enseigné par Jean-Claude. Roux, Professeur de Génie Papetier à Grenoble-I.N.P- Pagora.
Comme mentionné au chapitre I-2, la feuille de papier est constituée de flocs qui sont des agrégats de grains fibreux, voir la photographie figure 20. La taille de ces agrégats dépend de celle des micros turbulences dans la suspension des fibres. Leur distribution dans la feuille parait régie par le même principe organisationnel qui structure les grains dans les flocs à l'échelle millimétrique. L'autosimilarité qui a souvent été constatée pour la distribution en orientation des flocs anisotropes dans la feuille relativement à celle des fibres à l'intérieur des grains, se vérifie en comparant les mesures d'orientation des éléments fibreux par des échantillonnages effectuées à des échelles différentes : d'une part à l'échelle millimétrique dans les grains fibreux et d'autre part à l'échelle décimétrique dans des échantillons de papier de dimension : 6,6 cm x 6,6 cm, qui constituent une surface représentative pour évaluer la répartition des flocs.
Il est possible de définir l'anisotropie des flocs ainsi que leur orientation et leur taille dans la feuille de papier par la mesure optique du gradient des niveaux de gris de l'échantillon éclairé en transmission. L'interprétation des mesures est faite suivant le concept de la surface conforme équivalente, voir les travaux de recherche de : A. P. Alves Da Costa: « Contribution à l'étude du facteur de formation de la feuille de papier » pp.99-129, thèse de doctorat, nov. 2001, Unversidade da Beira Interior, Covilha, Portugal, et Institut National Polytechnique de Grenoble, France.

**41-** Un énoncé du mouvement des corps a été édicté par Epicure puis repris par Lucrèce (v.98-v.55 av. J.-C.) qui l'a versifié dans son poème : « De natura rerum », (De la nature des choses).
Au Chant II, versets 214-241, il est écrit :
   «Parvenu en ce point, nous brûlons de te faire
   encore savoir ceci : tandis que, par le vide,
   les corps tombent tout droit sous l'effet de leur poids,
   ils font, en un moment tout indéterminé
   et en des lieux tout aussi indéterminés,
   un écart dans leur course ; oh petit, juste assez
   pour que leur mouvement puisse être dit changé….»
Traduction par Bernard Pautrat, en alexandrins non rimés, dans la collection du livre de Poche, 4677, p. 189.
Ce concept du mouvement des corps met en exergue l'importance du « clinamen », c'est-à-dire la déviation des trajectoires, argumenté et théorisé de nos jours par les physiciens et les philosophes en tant que principe d'évolution. Voir à ce sujet les recherches sur la flèche du temps de Maurice Solovine, « Epicure, doctrines et maximes », traduites et accompagnées d'une note sur le « clinamen », pp.255, ainsi que l'ouvrage de J.M.Serres, « La naissance de la physique », pp : 11-14, 26-27, 74-75, et l'ouvrage : « Démocrite et l'atomisme ancien, Fragments et témoignages, » p. 27 et 65, texte traduit par Maurice Solovine, introduction et commentaires de Pierre-Marie Morel AGORA , collection : les classiques.

**42-** À la fois cyclo-ellipsoïdale et cyclo-cylindrique la courbe définie par les équations (4) est l'intersection de deux surfaces quadratiques. Cette courbe ellipsoïdo-cylindrique pourrait être considérée comme une variété de Clélie par extension à l'ellipsoïde et au cylindre elliptique de cette variété de courbes dont la dénomination fût choisie par Luigi Guido Grandi en 1728 pour l'étude des courbes cyclo-sphériques.
Cette courbe pourrait également s'orthographier « Clé-Lie », compte tenu de son appartenance à un groupe de Lie et de sa potentialité pour dénouer le « nœud Gordien » des trajets tortueux des particules du fluide dans un milieu poreux, voir au chap. IV.
Les circonvolutions des courbes ellipsoïdo-cylindriques des équations (4) peuvent être illustrées dans un tout autre domaine par le mouvement qu'effectue un garçon de café, bien entraîné, en portant son plateau en équilibre au dessus de son épaule par une élévation et une torsion de $2\pi$ de son bras puis en l'abaissant tout en le détordant par une deuxième rotation de $2\pi$, dans le même sens, pour revenir à sa position initiale. C'est ainsi que Georges Lochak, Président de la Fondation Louis de Broglie, a illustré ce mouvement apparenté à celui du spin des particules, sur la figure 7 de son ouvrage: « L'objet quantique », écrit en collaboration avec Simon Diner et Daniel Fargue et publié aux Editions Champs Flammarion, 1991, Vol. 250, pp. 62-64.
Des variations synchronisées de la variable, u, et du paramètre, w, dans les équations (4), permettent de décrire sur l'ellipsoïde une variété de courbes cycliques en forme d'hélice.

**43-** Pour une aire constante de l'ellipsoïde, la longueur du périmètre de la courbe ellipsoïdo-cylindrique varie peu en fonction des anisotropies de l'ellipsoïde. Dans le cas de la sphère la courbe sphero-cylindrique se différencie d'un grand cercle de la sphère qui est une ligne géodésique au sens du parcours le plus court ou le plus long pour un mobile qui irait d'un point à un autre de la sphère. L'étude de la courbe définie par les équations (4) et de ses propriétés peut s'effectuer en utilisant des algorithmes de calcul complémentaires de ceux de la bibliothèque NAG utilisée pour cette étude en 1993 au Centre Interuniversitaire de Calcul de Grenoble.

**44-** Les paramètres descripteurs du groupe des cylindres elliptiques sont dans un repère d'axes orthonormés : la direction de l'axe neutre de symétrie du groupe qui est la direction des axes de translation et de rotation du mouvement des particules fluides, la longueur de la génératrice des cylindres, l'orientation et la longueur de chacun des axes principaux de l'ellipse directrice des cylindres. Ces paramètres peuvent être définis de manière équivalente en fonction des ellipticités et de l'aire, S, de la surface du pore équivalent de la texture poreuse.

La courbe ainsi définie par les équations (4), synthétise tel un stick déformable, l'espace embrassé par les particules du fluide en écoulement à travers la texture d'un ensemble poreux désordonné aléatoire, d'un point de vue global et dans des conditions d'équilibre dynamique.

**45-** Les écoulements des fluides visqueux suivant la loi de Darcy et dans une cellule de Helé-Shaw sont analysés dans les ouvrages de mécanique des fluides. On peut consulter par exemple : E. Guyon, J-P. Hulin, L. Petit, « Hydrodynamique Physique », Inter Editions/Editions du CNRS, 7.3, pp. 379-384 ainsi que R.Comolet, « Mécanique expérimentale des fluides », Tome II, pp. 98-102, Masson, 1976.

Dans le cas de l'écoulement permanent et conservatif d'un fluide incompressible, suivant la loi de Darcy et selon Hele-Shaw la vitesse moyenne de débit est proportionnelle négativement au gradient du potentiel de pression. Le rotationnel de la vitesse est nul dans ces conditions. La divergence de la vitesse étant nulle le Laplacien du potentiel est nul et vérifie une fonction harmonique. Ces propriétés de l'écoulement sont établies d'un point de vue global pour la vitesse moyennée, la texture du milieu poreux étant estimée homogène par hypothèse.

A l'échelle microscopique la rotation de l'élément de volume du fluide nécessite une dépense d'énergie cinétique minimisée dans les conditions ou l'écoulement s'effectue en régime laminaire. Cependant lorsque le nombre de Reynolds de l'écoulement dans les pores est supérieur à 1, la dépense d'énergie devient non négligeable et elle apparaît pour une part significative dans la perte de charge. Le formalisme des lois de l'écoulement dans ces conditions a fait l'objet de nombreuses études notamment par : P.C. Carman, « Some physical aspects of water flow in porous media », Discussions of the Faraday Society, N° 3, 1948 pp. 72-77. S. Irmay, « On the theoretical derivations of Darcy and Forcheimer formulas », Transactions, American Geophysical, Union, Vol. 39, N°. 4, August 1958, pp. 702-706. R. Ben Aim, « Les écoulements des fluides à travers les milieux poreux : lois générales », Centre de perfectionnement des industries chimiques, 1980, Nancy, France. C. C. Mei et J.-L. Aurillaut, « The effect of weak inertia on flow through a porous media », J. Fluid Mech. Vol, 222, pp. 647-663, 1991. A. Houpeurt, « Evolution des concepts en mécanique des fluides dans les milieux poreux », Revue de l'institut Français du Pétrole, nov.1992, pp.1293-1301. J. Comiti, Nour-Eddine Sabiri, "Limit of Darcy's law validity in packed beds", E.G.G.E-1, Florence , Italy, May 4-7, 1997, pp.1863-1866. R.Salvado,« Relationship between spundbond process, Structure and properties of nonwovens for hygiène applications », Doctoral Thésis, 2002, (UBI) et (UHA), Chap. 3, Fluid permeability, pp.53-67. R. Salvado, J. Silvy, et J-Y.Drean," Fluid flow in spunbonded nonwovens", Fiber society, Spring conference, May, 2005, St.Galle , Switzerland.

**46-**La terminologie qui est préconisée dans la normalisation de l'I.S.O pour les propriétés d'écoulement des fluides dans les papiers est la perméance qui associe le facteur de perméabilité, K, à l'épaisseur du milieu filtrant.

**47-** Les milieux poreux qui font l'objet de ces études correspondent à une valeur de la constante « k » qui est en moyenne égale à : 4,5 +/- 1. L'échantillonnage de ces milieux peut être considéré comme aléatoire étant donné que, ni la structure des milieux poreux, ni la direction de la perte de charge évaluée dans l'écoulement du fluide à travers leur texture ne sont explicitées par rapport à un référentiel commun à ces différentes recherches.

**48-**Lorsque les courbes sont appariées en doublets, l'échange des polarités dans le volume de l'ensemble peut être réalisé localement en modifiant la valeur du paramètre, w, dans les équations(4). Ce « tour de main » qui permet d'homogénéiser un ensemble est la base du fonctionnement de certains mélangeurs malaxeurs industriels et de robots culinaires.

La bipartition des éléments dans un ensemble et leurs intrications dynamiques ont été postulées par Lao-tzu pour définir la cosmologie dans la philosophie bouddhiste Taoïste. Francois Cheng, de l'Académie Française, qui en a fait l'analyse dans son essai : « Vide et plein, le langage pictural chinois », cite l'extrait suivant du Tao-te-ching de Lao-tzu :
 « Le Tao d'origine engendre l'Un
 L'un engendre le Deux
 Le Deux engendre le Trois
 Le trois produit les dix mille êtres
 Les dix mille êtres s'adossent aux Yin
 Et embrassent le Yang
 L'harmonie naît au souffle du Vide médian ».

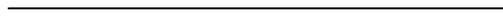

**II. Un concept de caractérisation des ensembles à texture désordonnée aléatoire : le pore équivalent.**

**II-1.**

**II-2.**

**III. Validation et applications du concept de pore équivalent.**

. **III-4**

**III-5.**

**V. L'équipartition bipolaire des principes du Yin et du Yang.**

**VI- Conclusions**.

**Résumé**

Cette étude concerne les propriétés des ensembles d'objets associés dans une structure qui résulte de processus multiples où intervient le hasard, comme c'est le cas dans les matériaux dont la texture est non ordonnée, aléatoire. L'auteur fait référence à la feuille de papier, un ensemble fibreux qui peut être considérée comme un archétype pour de nombreuses structures existantes dans la nature ou élaborées par l'homme. Les propriétés d'usage du papier sont mises en corrélation d'avec sa texture en montrant l'incidence des effets du hasard qui intervient au cours de son procédé de fabrication. Les développements théoriques, le formalisme et les méthodes d'application présentés dans cette étude ont une portée générale au delà du seul domaine du matériau papier.

Une propriété discriminante des ensembles d'objets distribués dans l'espace de manière désordonnée aléatoire est la répartition en orientation de leurs interfaces. Cette répartition s'obtient généralement par l'analyse d'images échantillonnées dans les ensembles. Dans une texture poreuse fibreuse la densité de la probabilité en orientation des fibres ou de l'orientation des interfaces de la texture, pondérée par leur longueur ou leur aire peut s'interpréter en tant que rayon de courbure d'un contour ou d'une surface qui caractérise, d'un point de vue statistique et global, la géométrie de la texture en deux ou en trois dimensions. Cette figure dénommée par l'auteur : le pore équivalent, a dans de nombreux cas une forme elliptique ou ellipsoïdale et est semblable au pore moyen défini par la corde moyenne évaluée dans les différentes directions entre les interfaces de la texture.

Différentes méthodes de construction du pore équivalent sont analysées : par transformation conforme du réseau des fibres ou des interfaces de la texture, par l'analyse stéréométrique des images et de coupes tomographiques, par des micro tomographies avec des rayons X de haute énergie à l'E.S.R.F. de Grenoble, par diffraction et par diffusion de la lumière d'un faisceau laser impactant la texture fibreuse ou la réplique de ses surfaces.

Le concept du pore équivalent permet d'étudier le comportement des ensembles désordonnés aléatoires lorsqu'ils sont sollicités dans des champs de forces, en simplifiant l'analyse. Ainsi un phénomène qui se produit dans un ensemble plan peut s'analyser sur le contour linéaire de son pore équivalent et un phénomène qui se produit en volume dans un ensemble en trois dimensions peut s'analyser sur la surface gauche bidimensionnelle de son pore équivalent. Ce concept a été appliqué pour l'étude des propriétés physiques, mécaniques, optiques et de conduction ionique de matériaux tels que les papiers et les cartons, les feutres, les voiles textiles non-tissés, les mousses de polymères, les alliages métalliques avec des joints de grains, les structures géologiques, la rugosité des matériaux, leur degré de brillant, leur abrasion et le relief des ensembles naturels.

L'ellipse et l'ellipsoïde ainsi que leurs compositions multimodales sont les figures les mieux adaptées pour représenter le pore équivalent des matériaux à texture désordonnée aléatoire. Qu'une loi qui définit la courbure d'une configuration géométrique déterministe, elliptique, s'impose pour représenter la répartition en orientation des interfaces d'éléments dont la distribution dans l'espace est statistiquement probabiliste, est un fait remarquable qui nous interroge.

Cette assertion est corroborée par l'analyse de l'écoulement des fluides dans un milieu poreux. L'énergie dissipée dans l'écoulement est répartie en fonction des composantes de déplacement, en translation et en rotation ainsi que de déformation du fluide, sur le pore équivalent dont la surface est conforme à celle de l'espace tangentielle des interfaces de la texture. Le milieu poreux étant homogène et les particules du fluide indiscernables les unes par rapport aux autres, compte tenu des échanges stochastiques continuels d'un élément de volume dans un autre, on conclut que la quantification de leurs mouvements est invariante d'un point de vue statistique en chaque point de la surface du pore équivalent. De manière générale un groupe de cylindres elliptiques permet de représenter cette quantification. Les intersections de la surface du groupe de cylindres d'avec la surface du pore équivalent définissent ainsi les trajectoires probabilistes des particules du fluide, de manière virtuelle et holiste dans le milieu poreux.

Une configuration géométrique particulière du groupe de cylindres elliptiques et du pore équivalent ellipsoïdal, satisfait les conditions de dissipation d'énergie minimale pour l'écoulement du fluide et d'entropie maximale dans les conditions des sollicitations. Ces trajectoires tapissent l'ellipsoïde par un faisceau de lacets homotopes ellipsoïdo-cylindriques, en boucles ouvertes ou fermées, suivant les appariements possibles en leurs points nodaux de regroupement et/ou de tangence isocline. La laminarité et l'irrotationnalité de l'écoulement sont établies à l'échelle macroscopique de manière globale et aux faibles valeurs du nombre de Reynolds ce qui est en accord avec les résultats obtenus par d'autres méthodes d'analyse.

La courbe ellipsoïdo-cylindrique ainsi définie est stationnaire quand à la moindre action dans le déplacement d'un élément de volume du fluide sur la surface de l'ellipsoïde pour aller d'un point ombilic à l'autre en position antipodale. Cette courbe permet de réaliser une tapisserie de l'ellipsoïde, de manière unicursale, en fonction d'une variable angulaire cyclique, ce qui permet de construire de manière générale un ellipsoïde tri axes. Tel un stick déformable la courbe ellipsoïdo-cylindrique caractérise dans sa globalité l'espace embrassé par le fluide en écoulement dans le milieu poreux, dans des conditions d'équilibre dynamique du fluide.

Lorsque l'ensemble d'éléments désordonné aléatoire, est isotrope la courbe est une hippopède qui fait partie de la variété des courbes sphéro-cylindriques. Le développement dans l'espace de cette courbe, projeté dans un plan permet de décrypter le tai-chi, figure symbolique de l'équipartition de l'énergie vitale suivant les principes du Yin et du Yang dans la philosophie bouddhique Taoïste.

Cette étude des ensembles désordonnés aléatoires, en particulier des milieux poreux fibreux, établit un lien structurel entre le désordre stochastique de leurs éléments à petite échelle et l'ordre qui émerge de ces ensembles à une plus grande échelle dans des conditions d'équilibre dynamique vis-à-vis de leurs sollicitations. Le hasard le plus grand qui soit est la variable nécessaire et suffisante qui permet d'optimiser au mieux dans sa globalité leur comportement suivant les lois déterministes et probabilistes qui conditionnent leur évolution.

Le vocabulaire de cette étude est celui du langage commun le plus souvent adapté à la classe des matériaux ; il est transposable dans d'autres domaines d'intérêt. Des notes annexes ainsi qu'une bibliographie sur des travaux effectués suivant les concepts présentés ou proches du domaine de l'étude complètent le texte.

**Mots clés**

Ensemble aléatoire, hasard, ordre, désordre, harmonie, échelle macroscopique, échelle microscopique, matériau stochastique, milieu poreux, pétrographie, structures fibreuses, papier, feutre, non-tissé, propriétés physiques, rugosité, stéréométrie, transformation conforme, courbure, pore équivalent, pore moyen, texture, particules, orientation des fibres, anisotropie, analyse d'images, micro tomographie, diffusion de la lumière, diffraction de la lumière, rayons X de haute énergie, E.S.R.F., écoulement des fluides, écoulement laminaire, écoulement irrotationnel, principe de moindre action, entropie statistique, ellipse, ellipsoïde, groupe de cylindres elliptiques, courbe ellipsoïdo-cylindrique, homotopie, tapisserie de l'ellipsoïde, hippopède, tai-chi, Yin, Yang.